\def\etal{{\it et al.\thinspace}}

\def\gcm3{{g cm${}^{-3}$}}
\def\h50{\hbox{$\rm\thinspace h_{50}$}}
\def\h50m1{\hbox{$\rm\thinspace h_{50}^{-1}$}}
\def\etal{{\it et al.\thinspace}}

\def\Fig{Fig.\thinspace}

\def\Figs{Figs.\thinspace}
\def\p3m{P${}^3$M}
\def\ap3m{AP${}^3$M}
\def\-{{\em{---}}}

\documentstyle[emulateapj,psfig,apjfonts]{article}

\received{}
\accepted{}
\journalid{}{}
\articleid{}{} 

\lefthead{Barnes \& Quinn}
\righthead{The Stability of Planetary Systems}

\begin{document}

\title{The (In)stability of Planetary Systems}
\author{Rory Barnes and Thomas Quinn}
\affil{Dept. of Astronomy, University of Washington, Box 351580, Seattle, 
WA, 98195-1580}
\begin{center}E-mail:rory@astro.washington.edu,trq@astro.washington.edu
\end{center}

\begin{abstract} We present results of numerical simulations which examine the
dynamical stability of known planetary systems, a star with two or more
planets. First we vary the initial conditions of each system based on
observational data. We then determine regions of phase space which
produce stable planetary configurations. For each system we perform
1000 $\sim10^6$ year integrations. We examine $\upsilon$ And, HD83443,
GJ876, HD82943, 47UMa, HD168443, and the solar system. We find that
the resonant systems, 2 planets in a first order mean motion
resonance, (HD82943 and GJ876) have very narrow zones of
stability. The interacting systems, not in first order resonance, but
able to perturb each other ($\upsilon$ And, 47UMa, and SS) have broad
regions of stability. The separated systems, 2 planets beyond 10:1
resonance, (we only examine HD83443 and HD168443) are fully
stable. Furthermore we find that the best fits to the interacting and
resonant systems place them very close to unstable regions. The
boundary in phase space between stability and instability depends
strongly on the eccentricities, and (if applicable) the proximity of the
system to perfect resonance. In addition to $10^6$ year integrations,
we also examined stability on $\sim 10^8$ year timescales. For each
system we ran $\sim 10$ long term simulations, and find that the
Keplerian fits to these systems all contain configurations which may
be regular on this timescale.

\end{abstract}

\section{Introduction} 
By September 2003, 13 planetary systems had been discovered (including
our own Solar System). The extra-solar planetary systems (ESPS) are,
possibly due to observational biases, markedly different from our own
in several ways. In our solar system (SS) the Jovian mass planets all
orbit at distances larger than 5AU, and on nearly circular orbits
($e\lesssim 0.05$). ESPS, on the other hand, contain giant planets in
a wide range of distances and eccentricities; some are 10 times closer
to their primary than Mercury, and others orbit with eccentricities
larger than 0.5. In this paper we attempt to categorize these systems
dynamically, constrain the errors of the orbital parameters, compare
our SS to ESPS, explore the long term stability of each planetary
system, and determine the mechanism(s) that maintain stability.

We examine these systems through numerical simulations. The
integrations begin with slightly different initial conditions in order
to probe observationally allowed configurations. This exploration of
parameter space permits a quantitative measure of the stability of
each system, and, hence, predicts which distribution of orbital
elements will most likely result in a stable system. In addition, a
comparison of stability between systems may reveal which elements are
most critical to the stability of planetary systems in general.

A quick inspection of the known systems reveals three obvious
morphological classifications: resonant, interacting, and
separated. Resonant systems contain two planets that occupy orbits
very close to a 2:1 mean motion resonance. A 3:1 system, 55Cnc (Marcy
\etal 2002), has been announced, but its dynamics will not be examined
here. Interacting systems contain planets which are not in mean motion
resonance, but are separated by less than a 10:1 ratio in orbital
period. These systems are not dynamically locked, but the planets
perturb each other. The SS falls into this category. The final
classification is systems in which the (detected) planets' orbits are
beyond the 10:1 resonance. These planets are most likely dynamically
decoupled, however some of these systems warrant investigation.

We examine planetary systems on two different timescales. First, we
explore parameter space in $10^6$ year integrations. For these
simulations we vary initial conditions to determine stable regions
within the observed errors.  Second, we continue several stable
simulations for an additional $10^8$-$10^9$ years. From these runs we
then learn how robust the predicted stable regions are, and we also
determine the mechanisms that lead to stability. Specifically we
evaluate the hypothesis that some stable system require secular
resonance locking.

In many ways this paper performs the direct analysis that is approximated
by MEGNO (Cincotta \& Sim\'o 2000, see also Robutel \& Laskar 2001,
Michtchenko \& Ferraz-Mello 2001, and Go\'zdziewski 2002). MEGNO searches
parameter space for chaotic and periodic regions. Our simulations
show that to first order MEGNO's results do uncover unstable regions. In
general, however, chaotic systems can be stable for at least $10^9$ years,
as shown below. Our SS also shows chaos on all timescales (for a complete
review, see Lecar, \etal 2001), therefore only direct N-body integration
can determine stability.

This work represents the largest coherent study of planetary system
dynamics to date. Our simulations show that the true configurations of
most planetary systems are constrained by just a few orbital elements
(or ratios of elements), and that stable regions can be identified
with integrations on the order of $10^6$ years. We also find that
stability, as well as constraints on stability, are correlated with
morphology. Resonant system stability depends strongly on the ratio of
the periods, interacting systems on eccentricities, and separated
systems are stable. 

This paper is structured as follows. In $\S$2 we describe the
generation of initial conditions, integration technique, and introduce
the concept of a stability map. In $\S$3 we analyze the results of the
resonant systems HD82943 and GJ876. In $\S$4 we examine the
interacting systems of 47UMa, $\upsilon$ And, and the SS. In $\S$5 the
separated systems, specifically HD168443 and HD83443, are discussed
briefly. In $\S$6 we summarize the results of $\S$3-5, In $\S$7 we
discuss possible formation scenarios and inconsistencies between this
work and other work on planetary systems. Finally in $\S$8 we draw
general conclusions, and suggest directions for future work.

\section{Numerical Methods}
In this section we outline the techniques used to perform and analyze
these simulations. This paper follows the example of Barnes and Quinn
(2001, hereafter Paper I). First we describe how the initial
conditions of each of the short term simulations are determined. In
$\S2.2$ we describe our integration method. Finally in $\S2.3$ we
describe a convenient way to visualize the results of these
simulations, the stability map.

\subsection{Initial Conditions}
In order to explore all of parameter space we must vary 5 orbital
elements, (the period $P$, the eccentricity $e$, the longitude of
ascending node $\Omega$, the longitude of periastron, $\varpi$, and
the inclination i) and every mass in the system. The period and the
semi-major axis $a$ are related by Kepler's third law. The argument of
pericenter, $\omega$ is the difference $\varpi$ and $\Omega$.

For each system we perform 1000 simulations, each with different
initial conditions. Orbital elements that are easily determined via
the Doppler method (P,e,$\varpi$) are varied about a Gaussian centered
on the nominal value, with a standard deviation equal to the published
error. We do not permit any element to be more than 5$\sigma$ from the
mean. For the $i$ and $\Omega$ elements, flat distributions in the
ranges $0^o < i < 5^o$ and $0 < \Omega < 2\pi$ were used. Note that
this distribution of $i$ and $\Omega$ permits a maximum mutual
inclination of $10^o$. These randomized orbital elements are relative
to the fundamental plane.

The Doppler method of detection does not produce a normal error
distribution. As Konacki \& Maciejewski (1999) show, the error in
eccentricity can have a large tail toward unity. However their method,
or other statistical methods, such as bootstrapping, require all the
observational data (including reflex velocity errors) to determine the
shape of this error curve. As not all the observations of multiplanet
systems have been published we are forced to use a normal distribution
in order for comparison between systems to be meaningful. We therefore
encourage all the observational data to be published, as some of the
results presented here (specifically comparing the percentage of
simulations that survived) are less meaningful because we are unable
to accurately estimate the error distributions of the orbital
elements. For completeness we also vary the primaries' masses, $M_*$,
based on other observations (generally determined via spectral
fitting). However as the stars are all at least 100 times more massive
than their planetary companions, the slight variations in primary mass
should not affect stability.

The masses of the planets are determined by the following relation
\begin{equation}
m = \frac{m_{obs}}{|sin(cos^{-1}(sin i)cos\Omega) |}
\end{equation}
where $m$ is the true mass of the planet, and $m_{obs}$ is the
observed minimum mass. By varying the mass this way, we account for
all possible orientations, and connect the inclination of the system
in its fundamental plane, to its inclination in the sky. Note that
this scheme requires the azimuthal angles in the planetary systems to
be measured from the line of sight.

\subsection{Integration}
With initial conditions determined the systems were then integrated
with the RMVS3 code from the SWIFT suite of programs\footnote{SWIFT is
publicly available at
http://k2.space.swri.edu/$\sim$hal/swift.html}(Levison \& Duncan
1994). This code uses a symplectic integration scheme to minimize long
term errors, as well as regularization to handle close approaches. The
initial timestep, $\Delta t$ is approximately 1/30th of the orbital
time of the innermost planet. In order to verify the accuracy of the
integrations, the maximum change in energy, $\epsilon$, permitted was
$10^{-4}$. We define $\epsilon$ as
\begin{equation} 
\epsilon= \frac{max|E_i - E_0|}{E_0}, 
\end{equation} 
where $E_i$ represents an individual measurement of the energy during
the simulation, and $E_0$ is the energy at time 0. There are two
reasons for using this threshold in $\epsilon$. First, as the
integration scheme is symplectic, no long term secular changes will
occur, so high precision is not required. Second, the simulations
needed to be completed in a timely manner. If a simulation did not
meet this energy conservation criterion, it was rerun with the
timestep reduced by a factor of 10. The minimum timestep we used was
$P_{inner}$/3000. Despite this small timestep, a few simulations did
not conserve energy, and were discarded, except that they were
incorporated into the errors. Errors and error bars include
information from unconserved simulations. Simulations which fail to
conserve energy would most likely be labeled as unstable, as the
failure of energy conservation undoubtedly results from a close
encounter between two bodies, which usually results in an ejection.
Therefore the estimates for stability in the systems presented here
should be considered {\it upper} limits.

Throughout this paper we adopt the nomenclature of the discovery
papers (planets have been labeled b, c, d, etc, with order in the
alphabet corresponding to order of discovery). We will also introduce
another scheme based on mass. Planets will be subscripted with a 1, 2,
3, etc, in order of descending mass. This new scheme is more useful in
discussing the dynamics of the system.

The short term simulations are integrated until one of the following
criteria is met. 1) The simulation ejects a planet. Ejection is
defined as the osculating eccentricity of one planet reaching, or
exceeding unity.  2) The simulation integrates to completion at time
$\tau$. For these simulations, $\tau$ is defined as
\begin{equation} \tau \equiv 2.8 \times 10^5 P_1,
\end{equation} 
or 280,000 times the period of the most massive planet. This choice
corresponds to $10^6$ years for the $\upsilon$ And system, as was
simulated in Paper I.

If a system ends without ejection, then the stability of the system must
be determined. There are several possible definitions of stability. In
Paper I, a system was stable if the osculating eccentricity of each
companion remained below 1 for the duration of the simulation. The most
obvious flaw in this definition is that a planet could suffer a close
approach and be thrown out to a bound orbit at some arbitrarily large
distance. Such a system would bear no resemblance to the observed system,
and hence should be labeled unstable. Here we adopt a more stringent
definition, namely that the semi-major axes of all companions cannot
change by more than a factor of 2. Changes in semi-major axis represent a
major perturbation to the system, therefore this second cut is
conservative, and only eliminates systems which expel a planet to large
distances without fully ejecting it.

In addition to these short term simulations, we completed simulations
to explore longer term stability ($\sim 10^8$ yrs). For each system we
ran $\sim 10$ simulations, chosen to cover a wide range of stable
parameter space. For these runs we started with the final conditions
of stable configurations and continued them.  These simulations
therefore give us a handle on how the system is likely to evolve on
timescales closer to its age ($\sim 10^9$yrs). Those systems which
survived these longer runs are the best comparisons to the true
system. Hence they are the best simulations for determining the
factors that lead to planetary system stability.

There are two notable problems with this methodology. First, we ignore
the effects of general relativity, which may be important in some
systems, specifically GJ876, and $\upsilon$ And. General relativity
was included in our treatment of $\upsilon$ And in Paper I. In Paper
I, the innermost planet had a negligible effect on the system, and we
presume that general relativity will continue to be unimportant for
the systems studied here. Second, we treat all particles as point
masses. This is again especially troublesome for GJ876 and $\upsilon$
And, because of their proximity to their (presumably) oblate
primaries. The sphericity of the stars also prohibits any tidal
circularization of highly eccentric planets (Rasio \etal 1996). This
may artificially maintain large planetary eccentricities, and increase
the frequency of close encounters.  However the eccentricities must
become very large for this effect to become appreciable. We therefore
assume that this phenomenon will not adversely affect our
results. Ignoring these two issues should impact the results minimally
while speeding up our simulations considerably.

\subsection{Stability Maps}
When analyzing the simulations, it is useful to visualize the results
in a stability map. In general a stability map is a 3 dimensional
representation of stability as a function of 2 parameters. In resonant
systems, we find that several parameters determine stability. The most
important is the ratios of the periods of the two resonant planets,
which we will call R:
\begin{equation}
R \equiv \frac{P_{outer}}{P_{inner}}.
\end{equation}
In coupled systems, $e_1$ and $e_2$ are the relevant parameters. The
advantage to this visualization is that boundaries between stability
and instability are easily identified.

The disadvantage of this form of visualization is that if the range of
parameter space is not uniformly sampled (as it is here), we cannot
visualize of the errors. It is therefore important to bear this
disadvantage in mind. At the edges of stability maps, the data are
poorly sampled and the information at the edges should largely be
ignored. To aid the visualization we have smoothed the maps. If a bin
contained no data then it was given the weighted mean of all adjacent
bins, including diagonal bins. This methodology can produce some
misleading features in the stability maps. Most notable are tall
spikes, or deep depressions in sparsely sampled regions. We comment on
these types of errors where appropriate.

The procedure as outlined overestimates the size of stable regions in
two important ways. First, the integration times are generally less
than 0.1\% of the systems' true ages. As is shown throughout this
paper, instability can arise at any timescale. Therefore, the
stability zone will continue to shrink as the system evolves. Second
we have chosen a very generous cut in semi-major axis space. Other
studies permit $\Delta a$ to be no larger than 10\% (Chiang,
Tabachnik, \& Tremaine 2001). Lowering this variation would
undoubtedly also constrict stability zones.

\section{Resonant Systems} 
Two systems with orbital periods in 2:1 mean motion resonance have
been detected:
HD82943\footnote{http://obswww.unige.ch/$\sim$udry/planet/hd82943syst.html}
and GJ876 (Marcy \etal 2001a). Table 1 lists the orbital elements and
errors for the resonant systems. For now we do not examine the 3:1
55Cnc system. The current best Keplerian fit to the observations put
HD82943 and GJ876 just beyond perfect resonance. These planets all
occupy high eccentricity orbits and hence have wide resonance
zones. Simulations of these systems show that stability is highly
correlated with the ratio of the periods, $R$, and to a lesser degree
on $e_1$. These systems are the least stable as less than 20\% of
simulations survived to $\tau$.

\begin{figure*}
\psfig{file=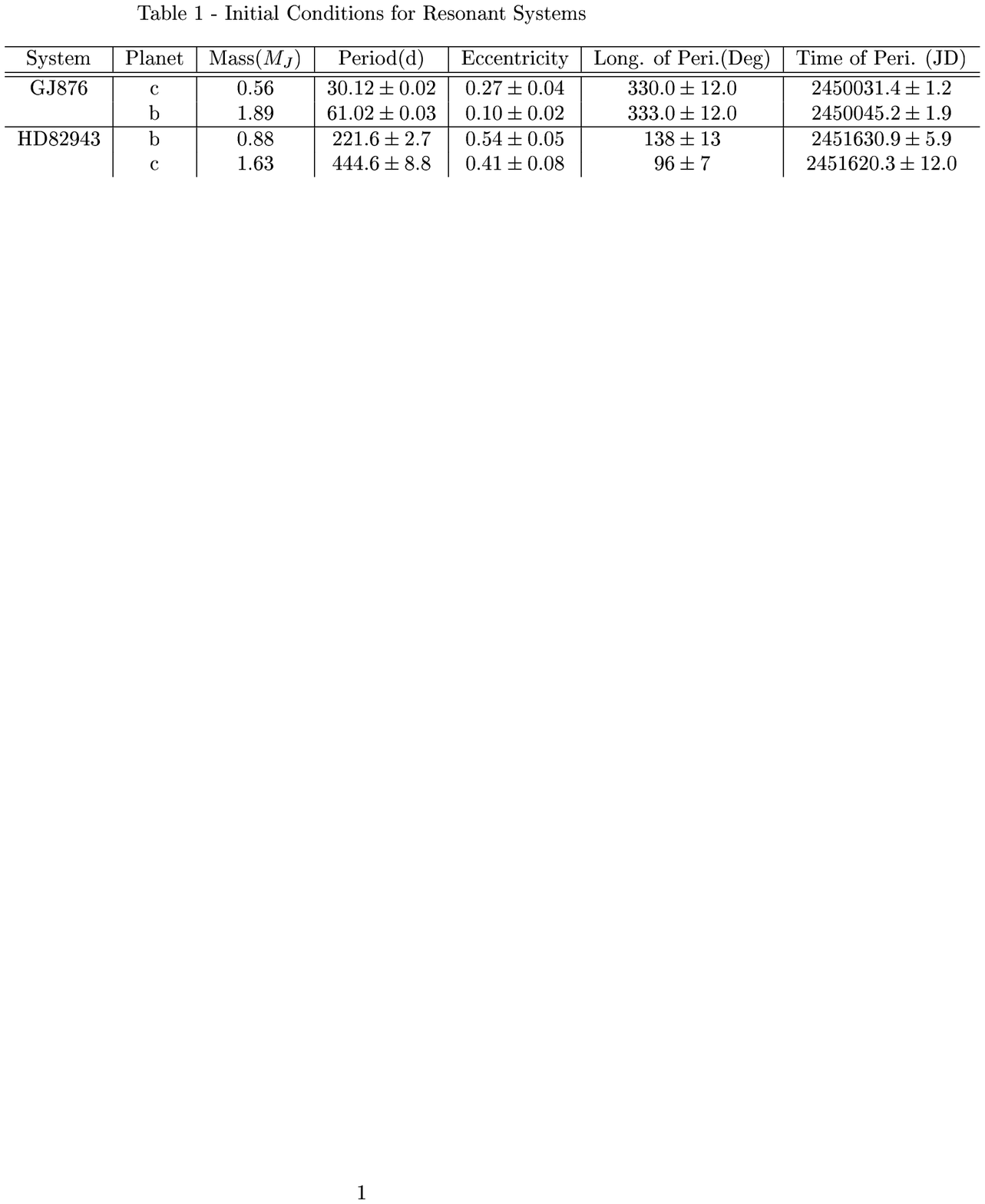,width=19.truecm}
\end{figure*}

\subsection{HD82943} 
Two planets orbit the $1.05 \pm 0.05M_{\odot}$ G0 star (Santos \etal
2000) HD82943 at semi-major axis distances of 0.73 and 1.16AU. Planet
b is the inner and less massive, c, the outer and more
massive. Go\'zdziewski \& Maciejewski (2001) also examined this
system. They found that the system is most likely to be stable in
perfect resonance. These simulations must run for 340,830 years,
$\tau_{HD82943}$. The long term simulations were run for 100 million
years.

For this system we find that $18.8\% \pm 4.3\%$ of the trials were
unperturbed to $\tau_{HD82943}$, and 19.3\% survived for $10^6$ years,
and 4.5\% failed to conserve energy. \Fig 1 shows the instability rate
to $10^6$ years. Most unstable simulations break our stability
criterion within $10^4$ years, but others survived more than 900,000
years before being ejected. The asymptotic falloff to $10^6$ years
implies we have found most unstable configurations. In 91.3\% of the
trials, planet b, the inner and less massive planet, was
ejected/perturbed. In order to check the simulations, \Fig 2 plots the
rate of survival as a function of energy conservation. From this
figure it appears that our limit of $10^{-4}$ is reasonable, as there
appears to be no trend in stability as a function of energy
conservation.  The spike in survivability at $10^{-8}$ corresponds to
regular orbits that were stable for our initial choice of $\Delta
t$. The lack of a trend with energy conservation (specifically
survival probability increasing with decreasing $\epsilon$), implies
our cutoff value of $\epsilon$ is stringent enough.

Stability in this system is correlated with $R$, $e_c$, $\Delta M$ and
$\Delta \varpi$, where $\Delta M$ is the difference in mean anomaly,
and $\Delta \varpi$ is the difference in initial longitude of
periastron. Slightly beyond perfect resonance is the preferred state
for this system. This system also requires the eccentricity of planet
c to remain below 0.4. These features are shown in \Figs 3 and 4. In
these greyscale images, black represents unsampled regions, darkest
grey marks regions in which no configurations survived, grades of
stability are denoted by lighter shades of grey, and white is fully
stable. As with most stability maps in this paper, the outer 2-3
gridpoints should be ignored. In \Fig 3, the $R-e_c$ stability map,
the large ``plateau'' at low e, is therefore poorly sampled, as is the
island at $R$=2.15, $e_c$=0.52.  The most striking feature of this
figure is that the best fit to the system, the asterisk, places it
adjacent to stability. If $R$ is changed by less that 5\% the system
has no chance of surviving even 1 million years. This map shows that
the current fit to the system is not correct. However the elements do
not need to change by much (specifically $e_c$ needs to be slightly
lower) for the system to have a chance at stability.

\medskip 
\epsfxsize=8truecm 
\epsfbox{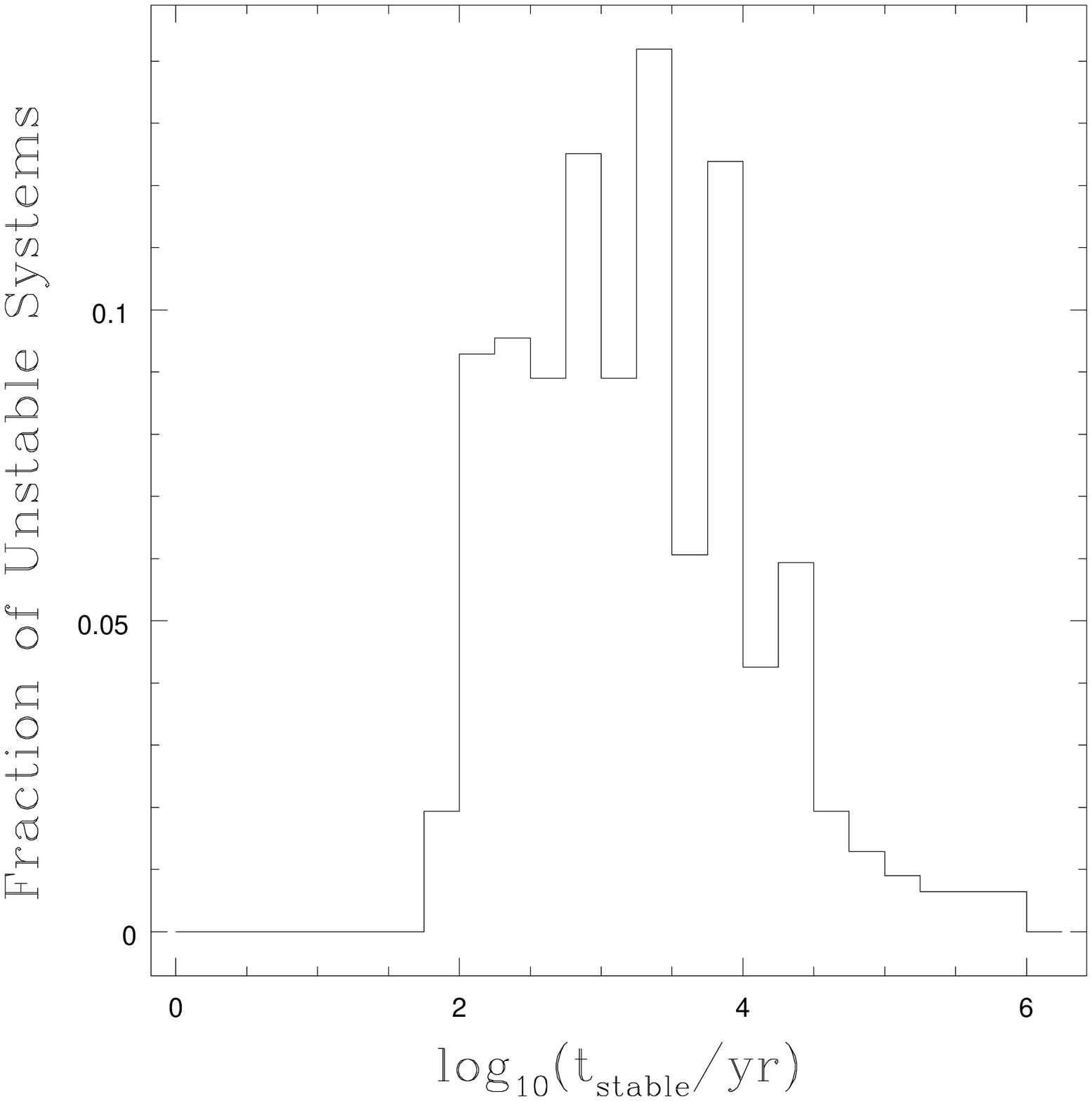}
\figcaption[hd82943.ejrate.ps]{\label{fig:asymptotic} \small{The
distribution of instability times for unstable configurations of
HD82943. Most unstable systems survive for just $10^2$-$10^4$ years
before perturbations change a semi-major axis by more than a factor of
2.}}
\medskip

\medskip 
\epsfxsize=8truecm 
\epsfbox{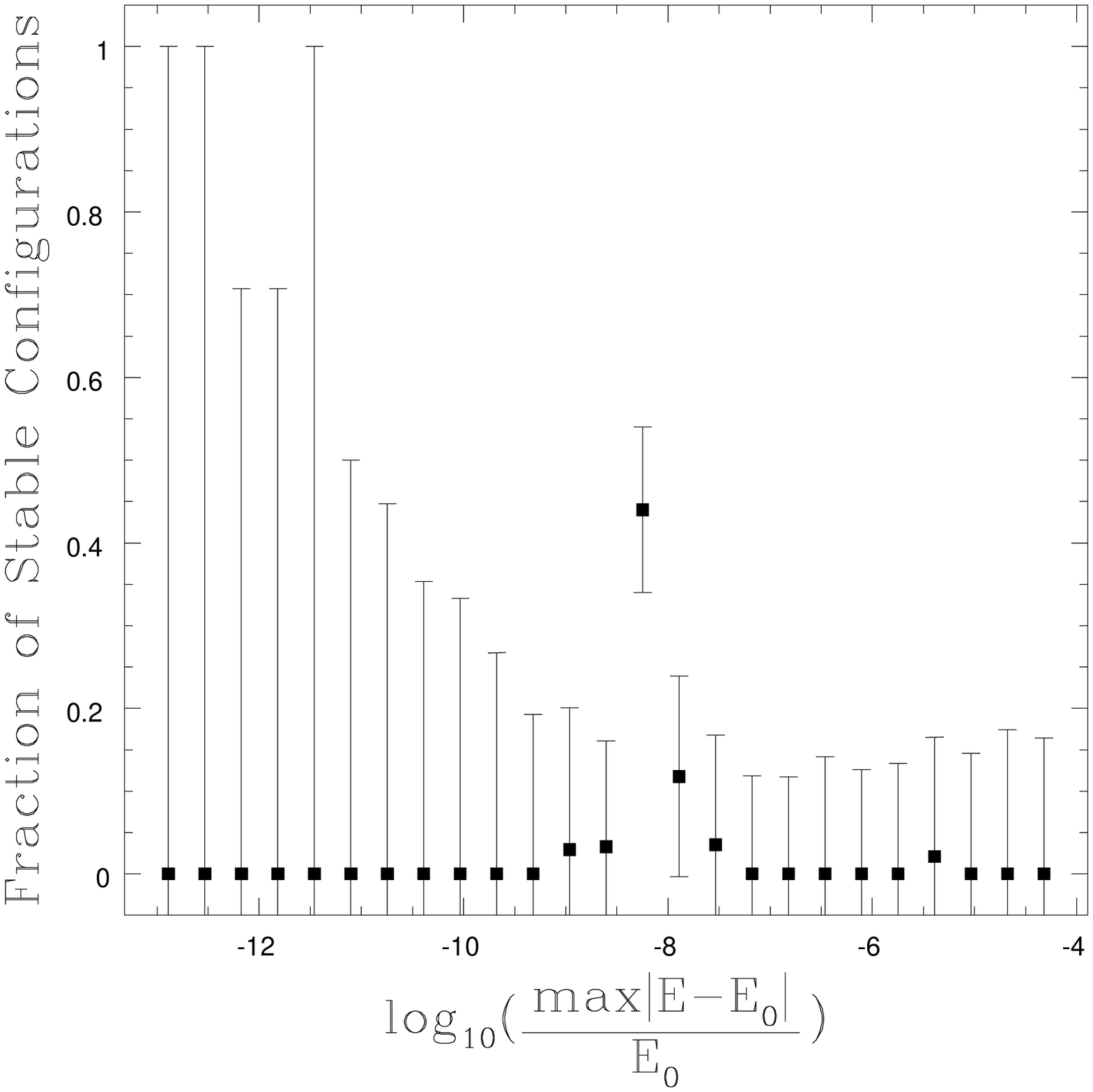}
\figcaption[hd82943.energy.ps]{\label{fig:asymptotic} \small{Survival rate
as a function of energy conservation for HD82943. The lack of a trend
implies the results for the system are accurate.}}
\medskip

\medskip 
\epsfxsize=8truecm 
\epsfbox{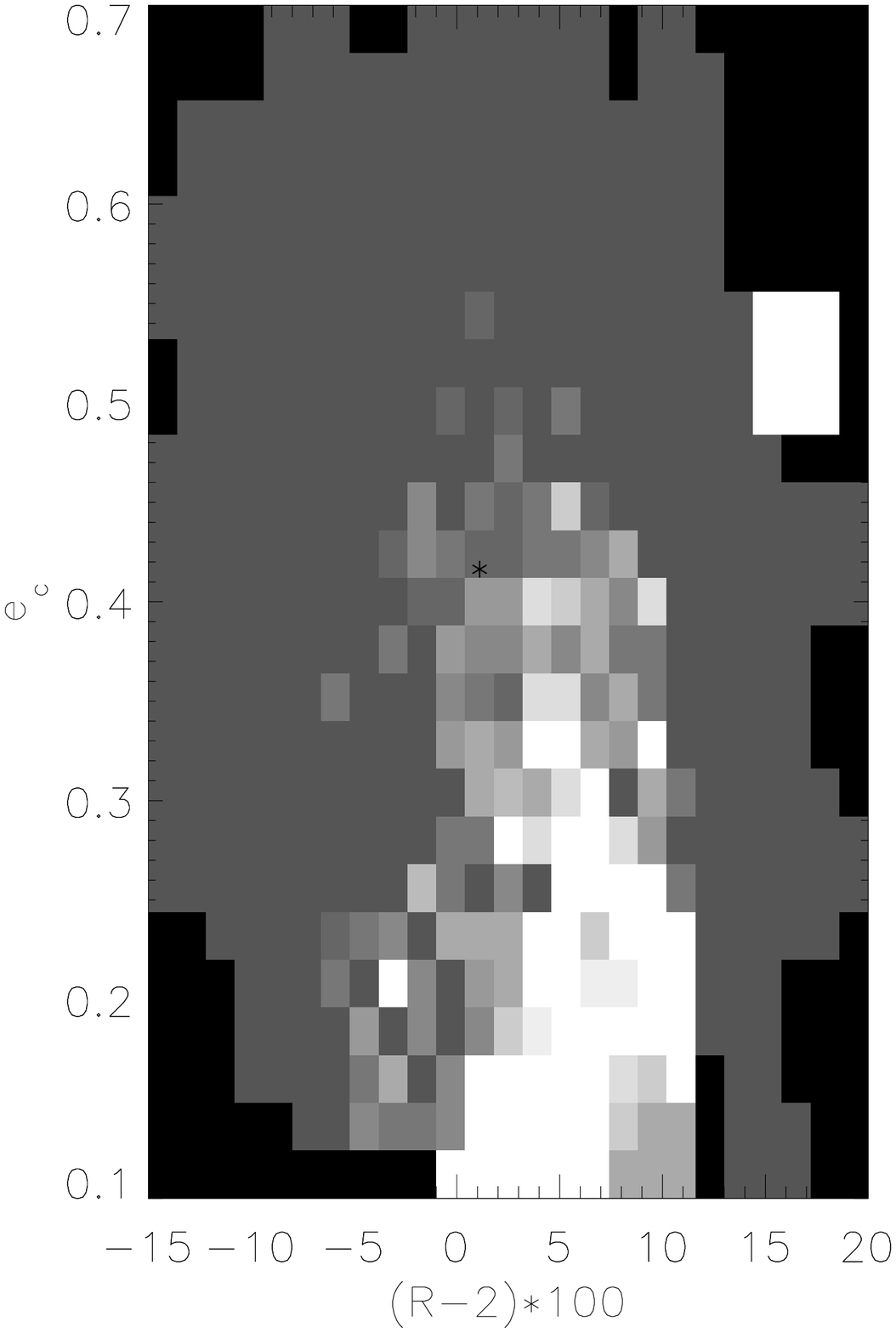}
\figcaption[hd82943.ecc-res.grey.ps]{\label{fig:asymptotic} \small{The $R-e_c$
stability map for HD82943. The asterisk represents the best fit to the
system as of 31 July 2002. The data are most accurate closer to the
asterisk. The system shows a clear boundary in phase space between
unstable (dark grey) regions and stable (white) regions. Black
represents unsampled data. The stable region at $R=2.15,e_c=0.52$ is a
bin in which 1 out of 1 trials survived.}}
\medskip

The system also shows dependence on mean anomaly and longitude of
periastron. Because of the symmetry of ellipses we will introduce a
new variable, $\Lambda$, defined as:
\begin{equation}
\Lambda \equiv \left\{ \begin{array}{ll}
               |\varpi_1 - \varpi_2|, & \Lambda < \pi\\
               360 - |\varpi_1 - \varpi_2|, & \Lambda > \pi
	       \end{array} \right.
\end{equation} 
where the subscripts merely represent 2 different planets, b and c for
this system. The order is unimportant, as the we are only concerned
with the magnitude of this angle. In \Fig 4, the $\Lambda$-$\Delta$M
stability map is presented. The asterisk marks the best fit to the
system. Stability seems to follow a line represented by
\begin{equation}
\Delta M = \frac{4}{3}\Lambda + 120 = \frac{4}{3}(\Lambda + \frac{\pi}{2}).
\end{equation}
This relation is purely empirical. As is shown in the following
sections, this interdependency is unusual for extra-solar planetary
systems. Similar plots of $\Delta M$ or $\Lambda$ versus $R$, show
the same $R$ dependence as in \Fig 3. Therefore $R$ is clearly the
most important parameter in this system, but these other 3 also play
an important role in the system. As more observations of the system
are made, HD82943 should fall into the region defined by $1.95\le R\le
2.1$, $e_c<0.4$, and Eq. 6.

\medskip 
\epsfxsize=8truecm 
\epsfbox{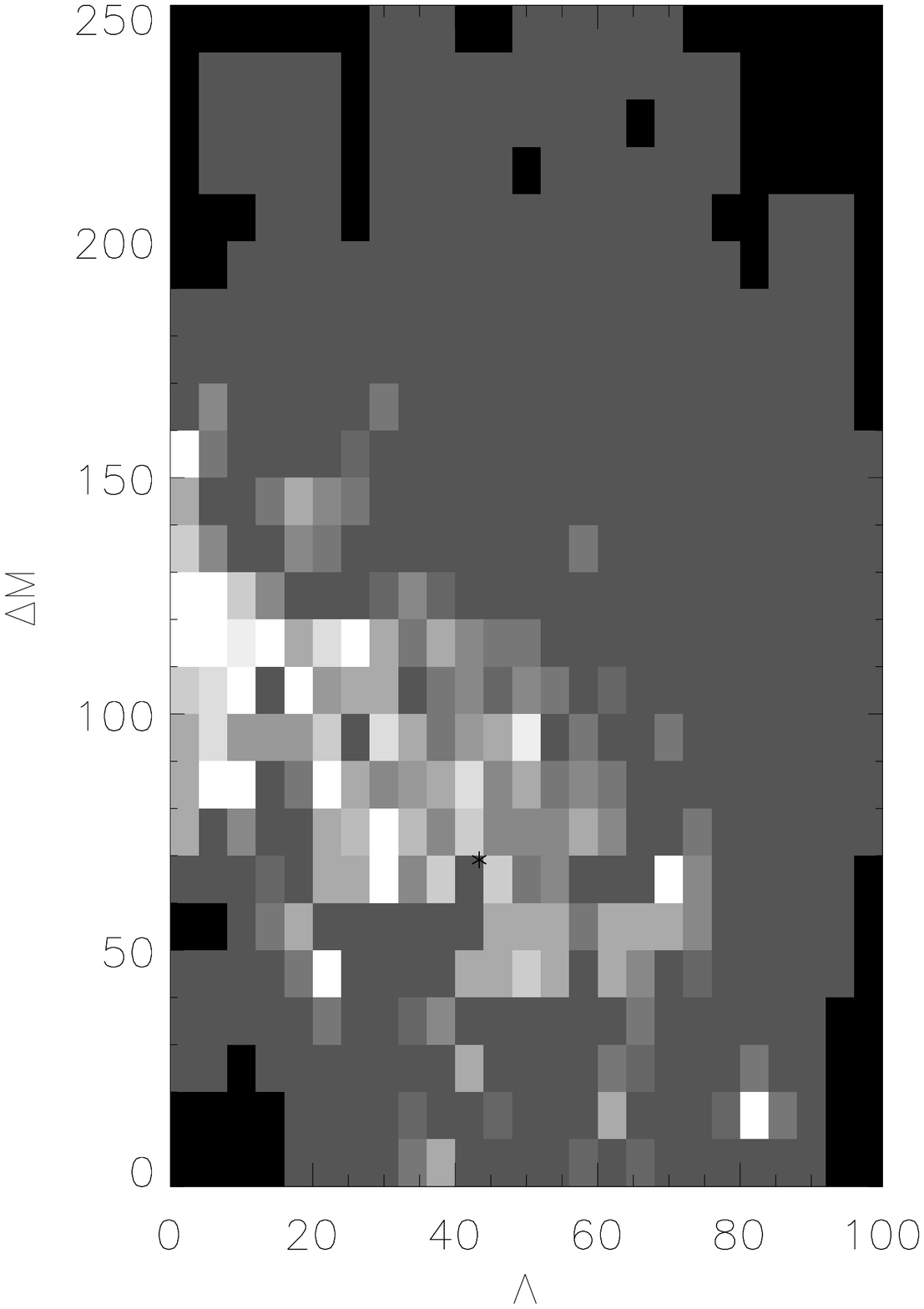}
\figcaption[hd82943.dm-lam.grey.ps]{\label{fig:asymptotic} \small{The
$\Lambda$-$\Delta$M stability map for HD82943. The asterisk represents
the best fit to the system as of 31 July 2002. The data are most
accurate closer to the asterisk. Stability appears to follow the line
represented by Eq 6. Note however that the system is also constrained
to $\Lambda \le 75^o$ and $\Delta M \ge 30^o$. The island at $\Lambda
= 80^o$, $\Delta$M = $20^o$ is a bin in which 1 out of 1 trials
survived.}}
\medskip

HD82943 shows a wide range of dynamics. Some examples of these are
shown in Figs.\ 5-8. The initial conditions of these systems are
presented in Table 2. \Fig 5 shows an example of the evolution of a
regular system, HD82943-348, which shows no evidence of chaos. Note
that instead of $\Lambda$(t), we present the distribution function of
$\Lambda$, P($\Lambda$), the probability of $\Lambda$, vs.\
$\Lambda$. This representation of $\Lambda$ shows that the motion is
like that of a harmonic oscillator; the longitudes of periastron are
librating with an amplitude of $60^o$.

\begin{figure*}
\psfig{file=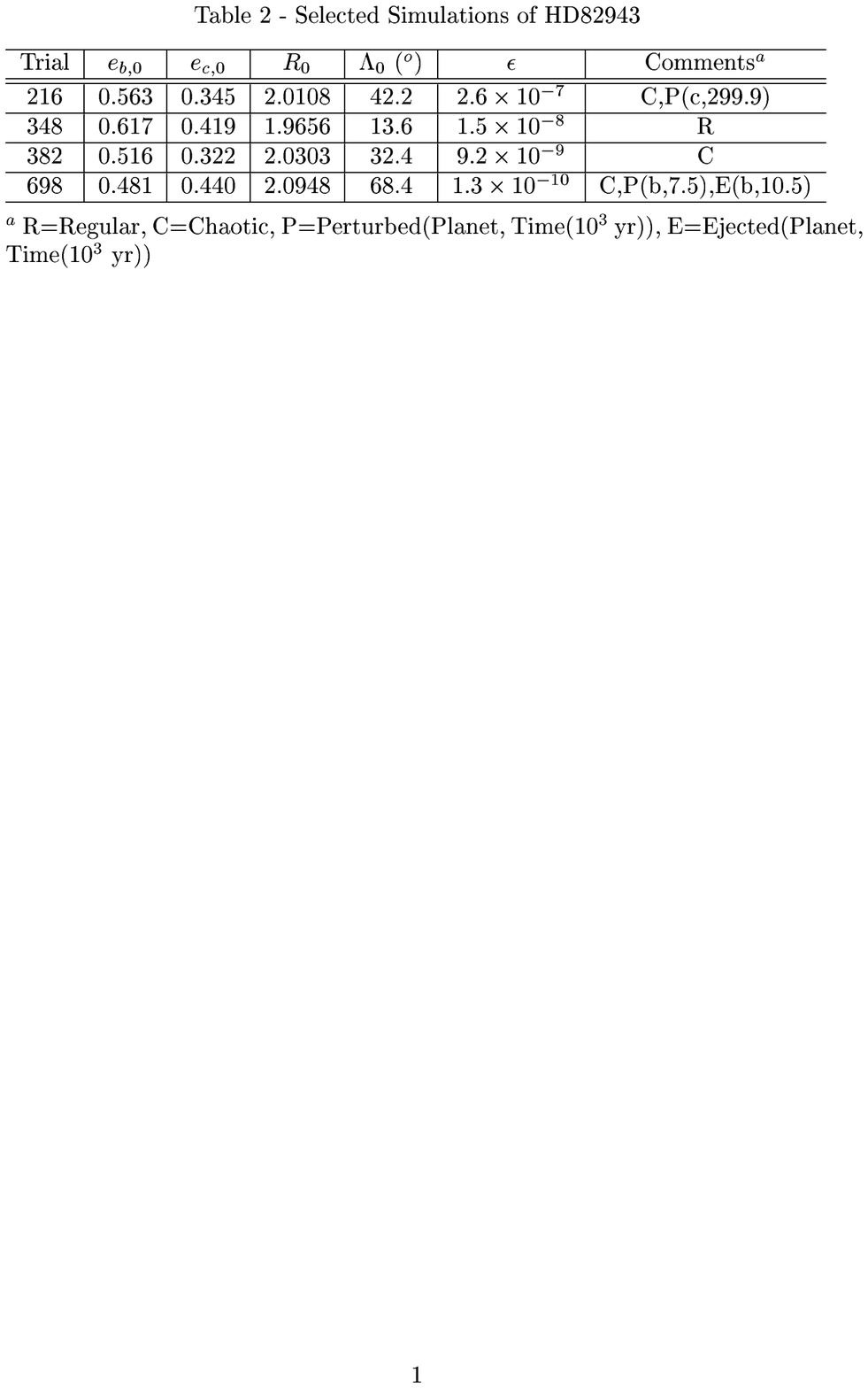,width=19.truecm}
\end{figure*}

\Fig 6 (HD82943-382) is a stable case which is clearly
chaotic. Although the eccentricities remain close to their initial
values, the inclinations jump to large values quickly. Note that
$\Lambda$ never exceeds $75^0$, but its motion is slightly
nonharmonic, another indication of chaos.

In \Fig 7 we plot the evolution of a system which ejects planet
b. This system experiences close approaches very quickly as is
evidenced in the top left panel of \Fig 7. The eccentricities undergo
dramatic fluctuations, and the inclinations nearly reach
$40^o$. Although poorly sampled, $\Lambda$ suggests circulation.

In \Fig 8, the orbital evolution of simulation HD82943-216 is
shown. This system perturbed planet c after 280,000 years, and despite
the high eccentricities the system obtained (>0.75 for both planets),
remained bound for $10^6$ years. The inclinations also show large
growth. Although initially $\Lambda = 41^o$, the system becomes locked
at just over $100^o$. This is misleading as this is the $\Lambda$
distribution for the full $10^6$ years. It is actually the
superposition of 2 modes. Initially, until perturbation at 280,000
years, the system circulates with a slight preference toward
anti-alignment. However after the perturbation the system becomes
locked at $\Lambda \approx 100^o$. Although not shown, after these
initial perturbations the system settles down to a configuration in
which $a_b\approx 0.344AU$, $e_b\approx 0.79$ and $a_c\approx 2400AU$,
$e_c\approx 0.99$.

\medskip
\epsfxsize=8truecm
\epsfbox{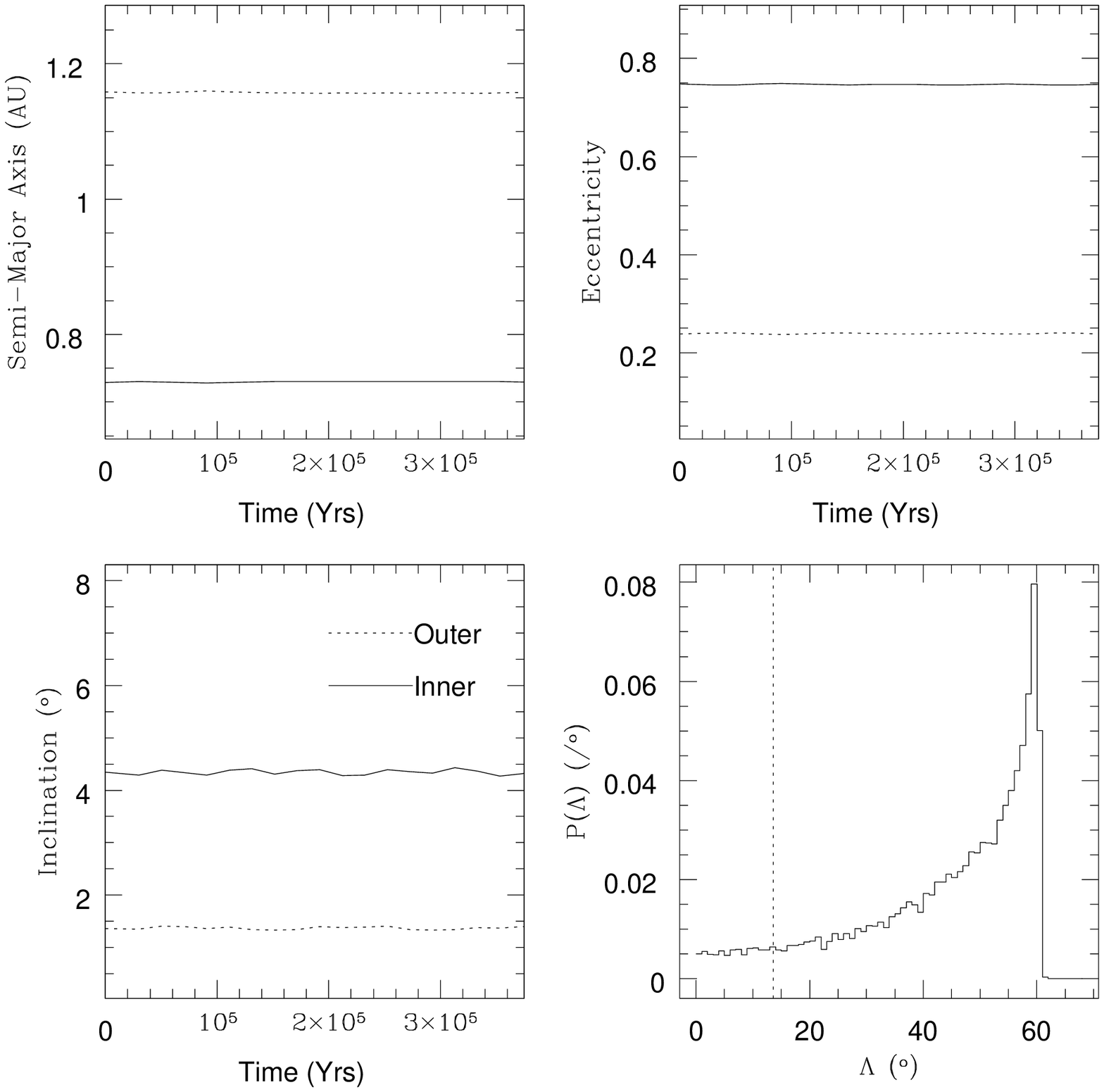}
\figcaption[hd82943_348.smooth.ps]{\label{fig:asymptotic}
\small{Orbital evolution of HD82943-348, a stable, regular
configuration. The data here are smoothed over a 20,000 year
interval. {\it Top Left}: The evolution in semi-major axis. The
planets show slight variations due to resonant interactions. {\it Top
Right}: The eccentricities oscillate with a 700 year period, with
$0.62\le e_b\le 0.85$ and $0.05\le e_c\le 0.45$. This short timescale
is not visible in this plot. {\it Bottom Left}: The inclinations
experience oscillations on 2100 year timescale, again too short to be
visible in this plot. The ranges of inclination are $1^o\le i_b\le
7^o$ and $0\le i_c\le 3^o$. {\it Bottom Right}: From this distribution
function we see the lines of node librate harmonically with an
amplitude of $60^o$. The dashed vertical line represents $\Lambda_0$,
the initial value of $\Lambda$.}}
\medskip

\medskip
\epsfxsize=8truecm
\epsfbox{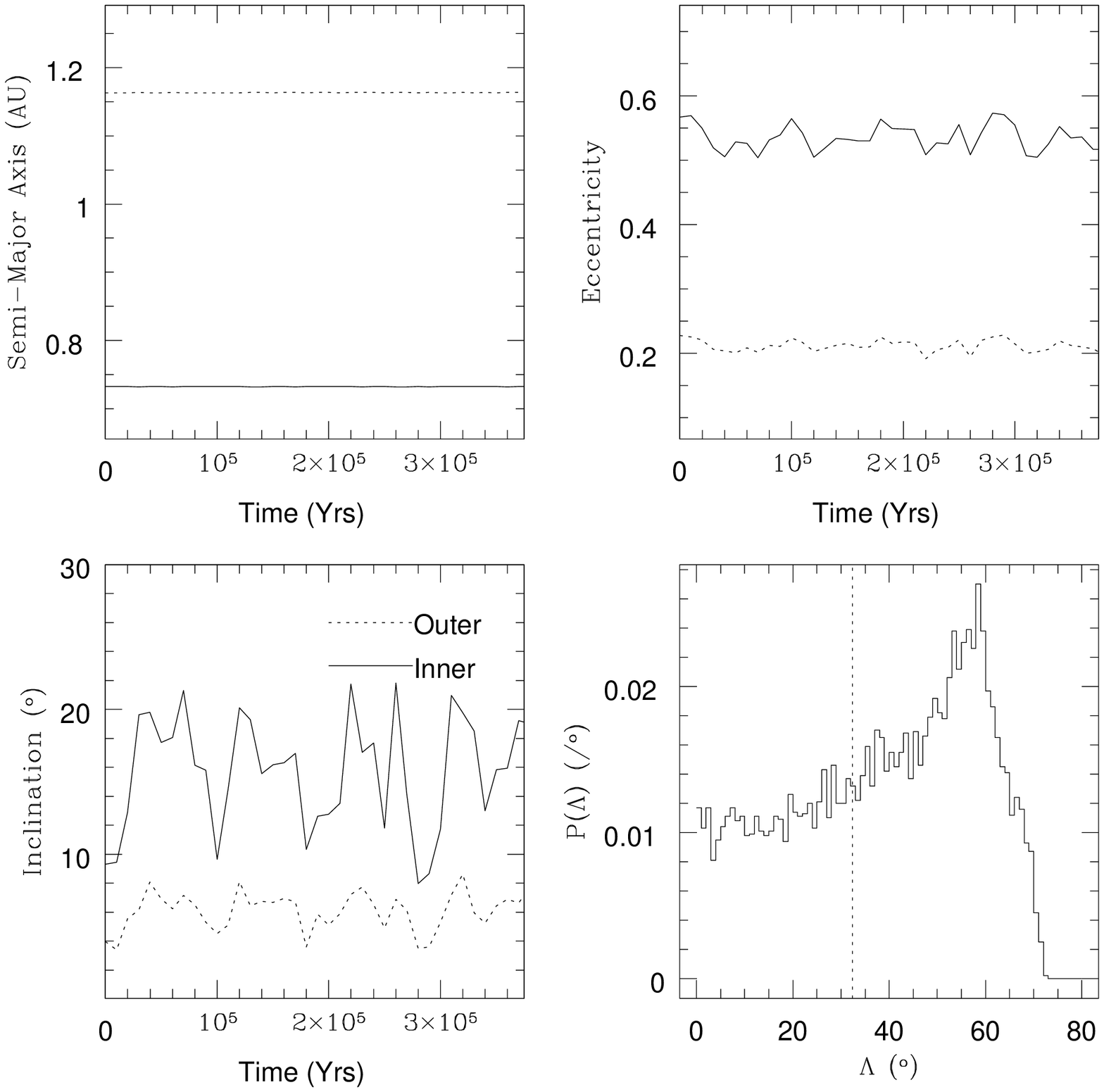}
\figcaption[hd82943_382.bw.ps]{\label{fig:asymptotic}
\small{Orbital evolution of HD82943-382, a stable, chaotic
configuration of HD82943. The data are averaged on a 10,000 year
interval. {\it Top Left}: The evolution in Semi-Major
Axis. The planets clearly never experience a close encounter. {\it Top
Right}: The eccentricities experience low amplitude ($\approx$ 13\%)
chaotic oscillations. {\it Bottom Left}: The inclinations initially
jump to large values and experience large amplitude ($\approx$ 70\%)
fluctuations. {\it Bottom Right}: The $\Lambda$ distribution function
suggests that $\Lambda$ is generally librating, but that there are
chaotic fluctuations superimposed on this motion. The dashed vertical
line is the value of $\Lambda_0$. }}
\medskip

\medskip
\epsfxsize=8truecm
\epsfbox{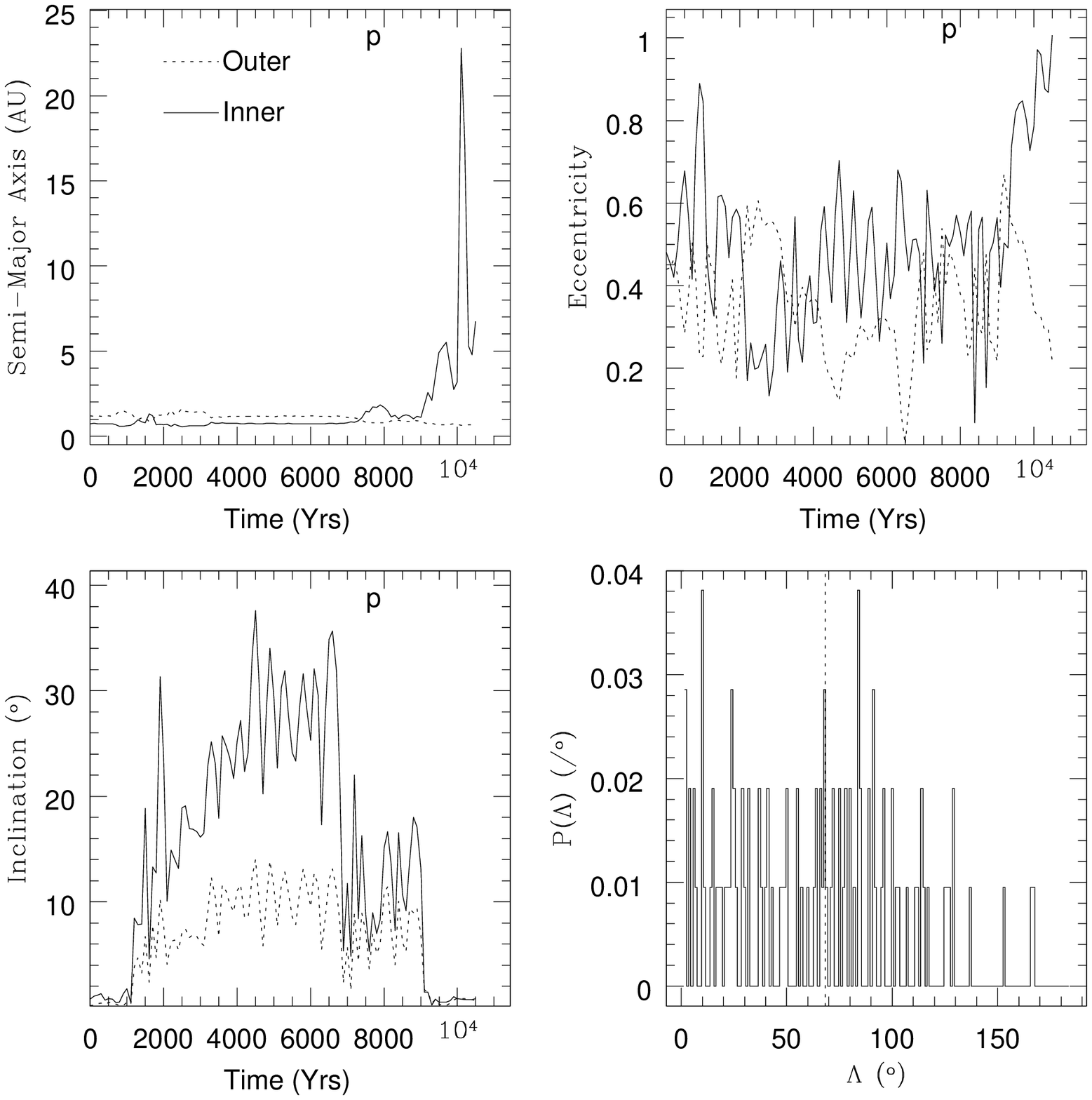}
\figcaption[hd82943_698.bw.ps]{\label{fig:asymptotic}
\small{Orbital evolution of HD82943-698, the perturbation and ejection
of HD82943b. The orbital parameters are sampled every 100 years. {\it Top Left}: The planets experience close encounters
immediately as changes in semi-major axis are visible within 500
years. {\it Top Right}: The eccentricities experience large amplitude
fluctuations, culminating in the ejection of planet b just after
10,000 years. {\it Bottom Left}: The inclinations vary slightly until
a close encounter at 1100 years which sends both inclinations up
substantially. The planets remain in this state until just prior to
ejection, when they return to coplanarity. {\it Bottom Right}: The
$\Lambda$ distribution function suggests that $\Lambda$ remains below
$100^o$, but may occasionally circulate.}}
\medskip

We ran 10 simulations for $10^8$ years. The initial conditions and
results of these simulations are presented in Table 3. The
eccentricity evolution of 4 simulations is shown in \Fig 9. These
simulations, which ran for 100 million years, show some configurations
are regular (top left), some are chaotic (top right, bottom left), and
that instability can arise at any timescale (bottom right). These long
term simulations show that regions exist in phase space in which this
system can survive for billions of years.

\medskip
\epsfxsize=8truecm
\epsfbox{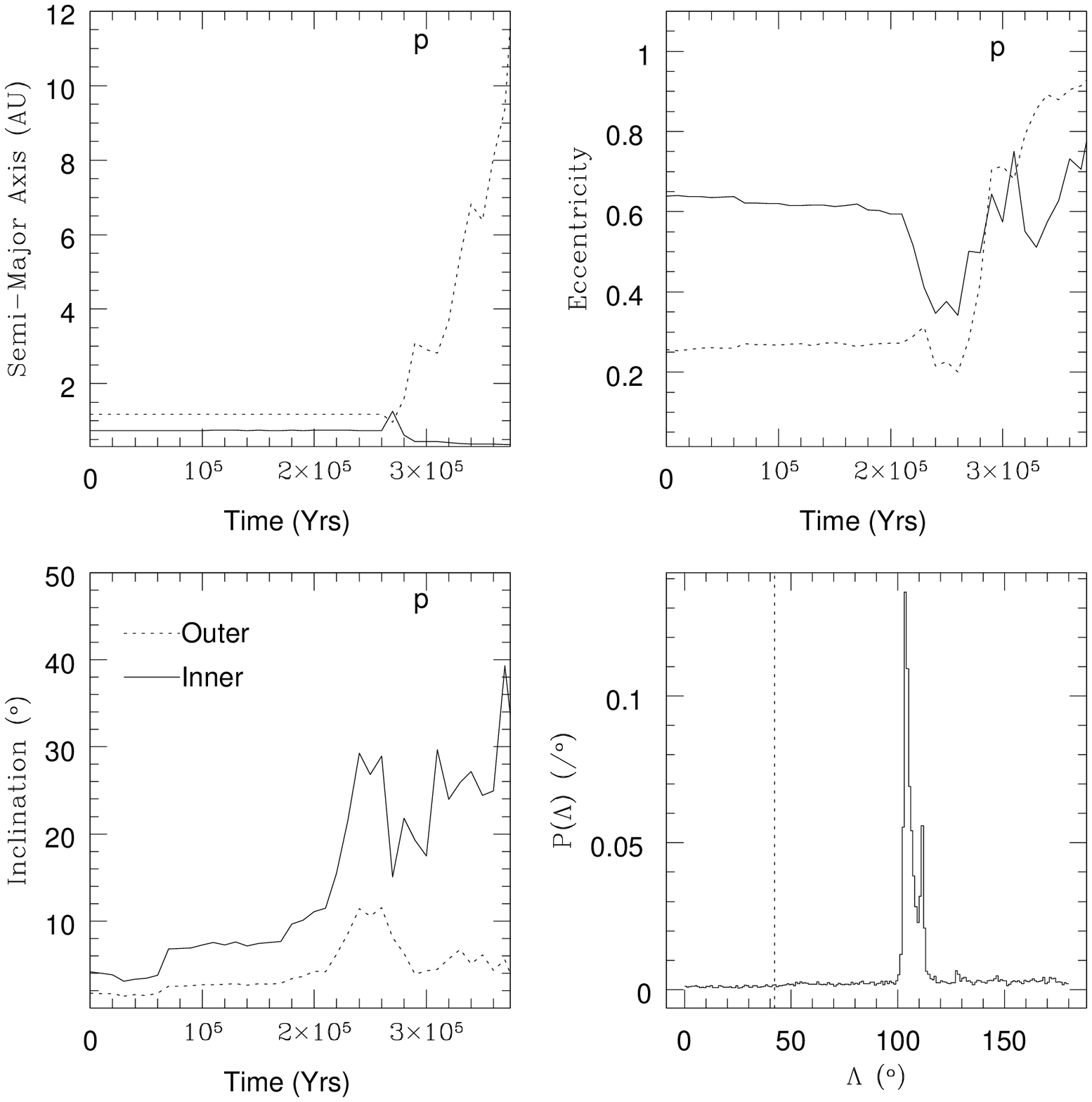}
\figcaption[hd82943_216.bw.ps]{\label{fig:asymptotic}
\small{Orbital evolution of HD82943-216, the perturbation of
HD82943c. The parameters are averaged on 10,000 year intervals. {\it
Top Left}: The planets' semi-major axes are stable and show no signs
of close encounters until 260,000 years. At this point planet c
actually crosses b's orbit. This initial encounter leads to more
encounters as $a_c$ reaches 3AU by 280,000 years, tripping the
criterion for instability. {\it Top Right}: The eccentricities
experience secular change until 210,000 years. The system then moves
into a lower eccentricity state. The eccentricity then grows to large
values and remain at their final values for another 700,000
years. {\it Bottom Left}: As with eccentricity the inclinations show
slow secular change until 210,000 years. The inclinations then leap up
to $30^o$ in the case of planet b.  {\it Bottom Right}: The $\Lambda$
distribution function is the sum of 2 motions: the pre-perturbation
motion is circulation, the post-perturbation motion is fixed close to
$110^o$. The dashed line represent $\Lambda_0$. }}
\medskip

\begin{figure*}
\psfig{file=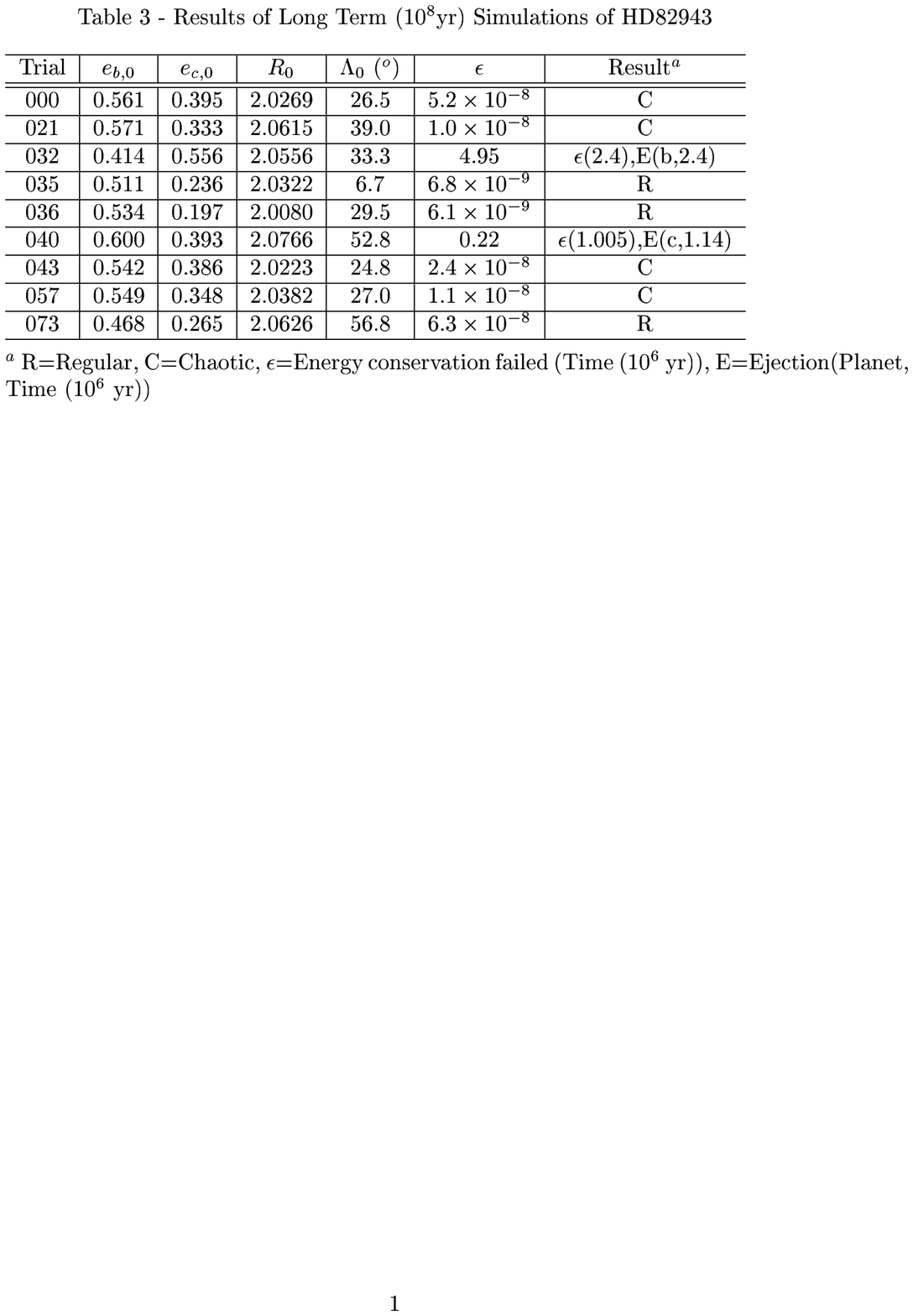,width=19.truecm}
\end{figure*}

\medskip
\epsfxsize=8truecm
\epsfbox{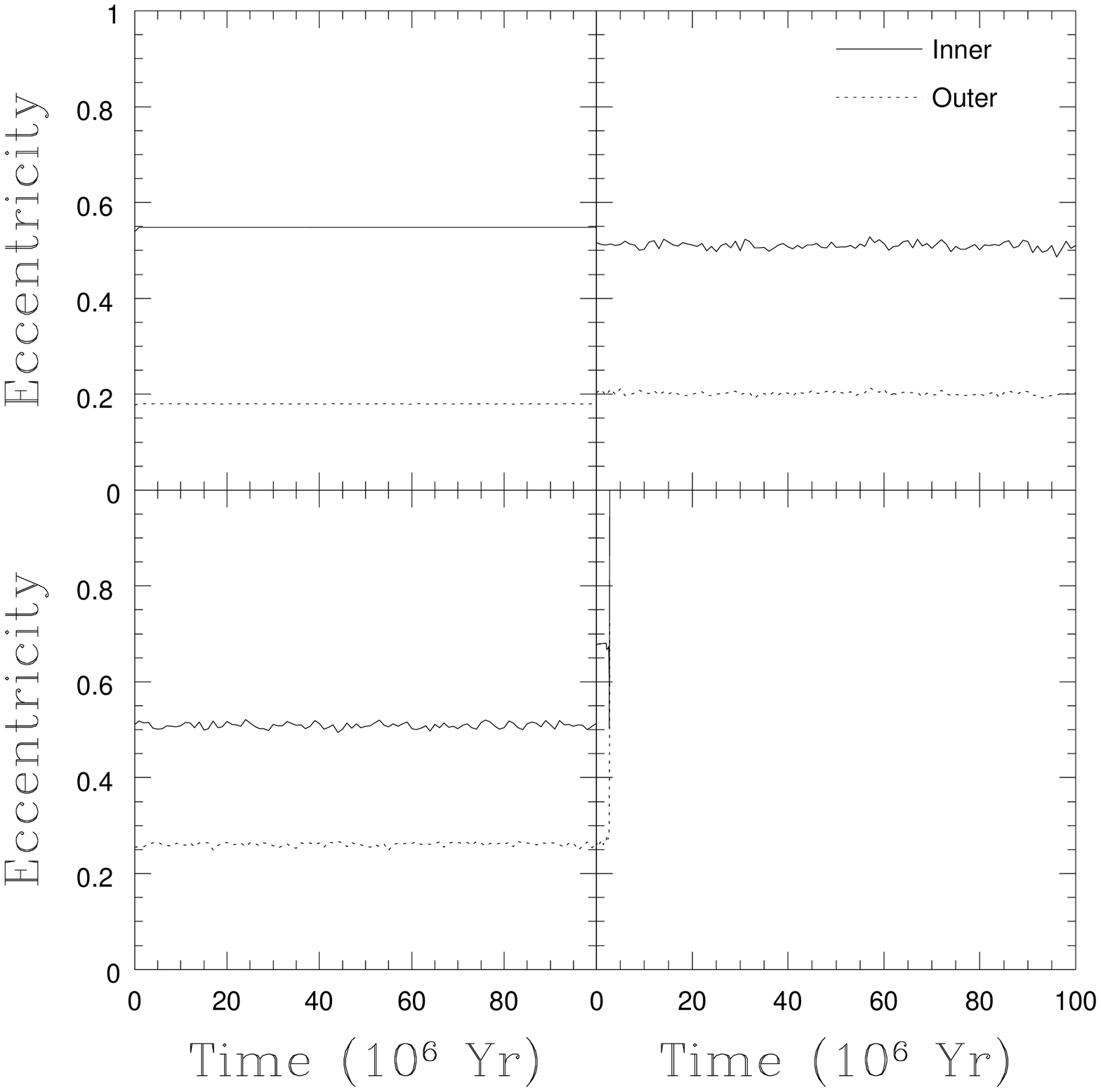}
\figcaption[hd82943.long.bw.ps]{\label{fig:asymptotic}
\small{Long term simulations of the HD82943 system. These data are
averages over 50,000 year intervals. {\it Top Left}: 
Evolution of HD82943-035. An example in which the system is regular. {\it Top Right}:
Evolution of HD82943-021, an example of chaotic evolution. {\it Bottom Right}: Evolution of
HD82943-000, another example of chaotic motion. {\it Bottom Left}:
Evolution of HD82943-032. A chaotic system which ejects planet b after
2.4 million years (7$\tau_{HD82943}$).}}
\medskip

Several papers have suggested that secular resonance locking maintains
stability in ESPS with large eccentricities (Rivera \& Lissauer 2000,
Rivera \& Lissauer 2001a, Chiang, Tabachnik, \& Tremaine
2001). Specifically they suggest that the orientation of the planets'
ellipses should be aligned ($\Lambda\approx$0).  By examining $\Lambda$ in
stable, regular long term simulations we can determine if the
longitudes of periastron remain locked. The probability distribution
of $\Lambda$ for these same four long term simulations is
presented in \Fig 10. $\Lambda$ is sampled once every 100 years. The
top left plot of \Fig 10 is similar to that of an harmonic
oscillator; this configuration is librating about $\Lambda$=0 with an
amplitude of $40^o$.  The other plots are systems which are not
librating, but instead show more random motion. From this figure we
see that regular motion is correlated with libration about alignment,
but chaotic and unstable generally shows chaotic $\Lambda$ behavior.

\medskip
\epsfxsize=8truecm
\epsfbox{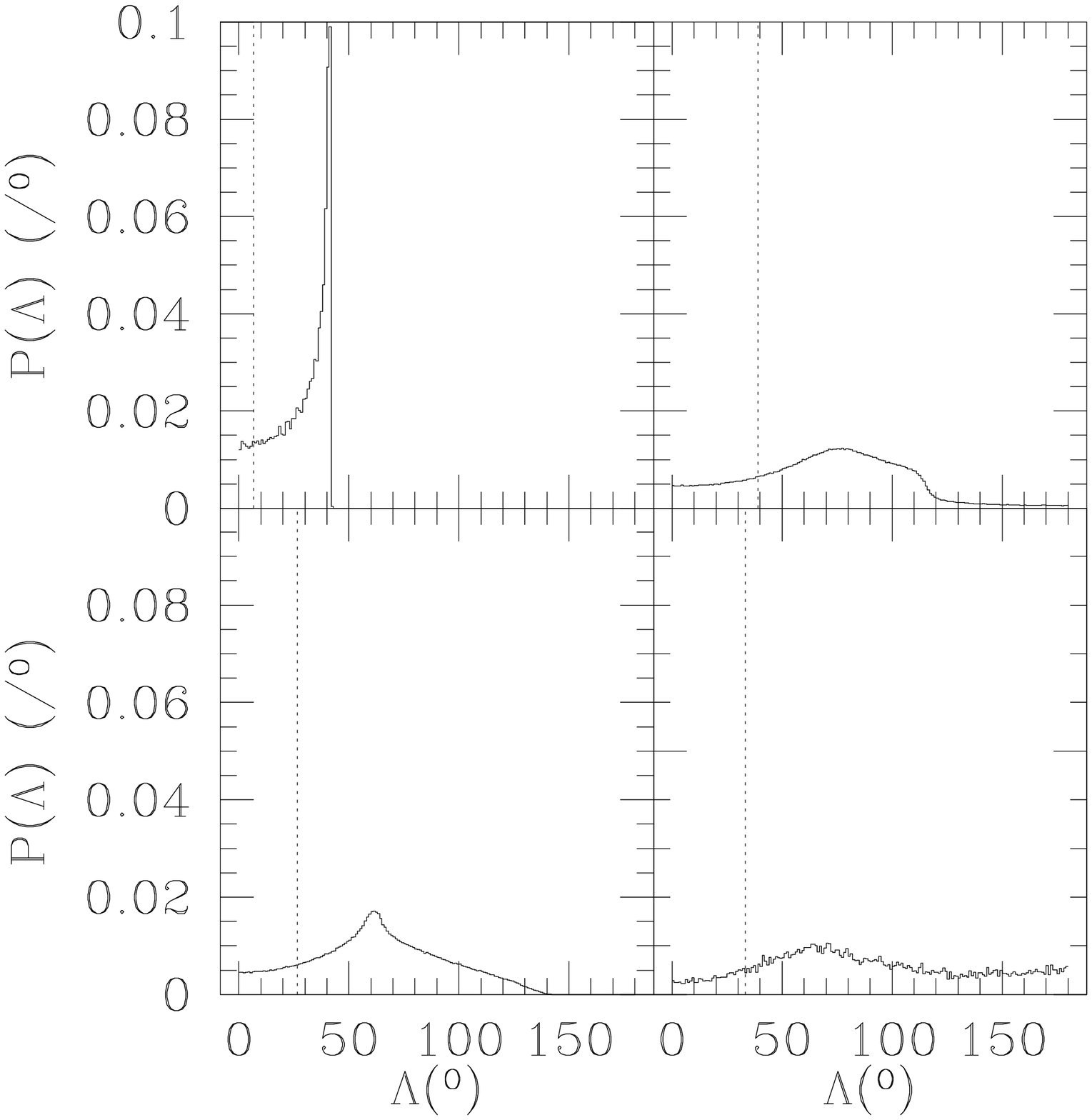}
\figcaption[hd82943.lamprob.ps]{\label{fig:asymptotic}
\small{The distribution function of $\Lambda$ (sampled every 100 years) for four stable cases of
HD82943. These four systems are the same as in \Fig 9. The top left
plots a system librating about $\Lambda$=0. The other plots show that
chaotic motion is usually associated with a circulating $\Lambda$.}}
\medskip

\subsection{GJ876} 
Two planets orbit the $0.32\pm 0.05M_{\odot}$ (Marcy \etal 2001a) M4
star GJ876, also known as Gliese 876. This system is very similar to
HD82943, the major difference being that the planets lie closer to
their primary.  The semi-major axes of these two planets are 0.13 and
0.21AU. Note that in this system planet c is the inner and less
massive planet. A substantial amount of work has already been done on
this system. Notably, in the discovery paper (Marcy \etal 2001) that
stable configurations exist in the system. Rivera \& Lissauer (2001b)
show that Keplerian fitting for this system is not precise enough to
accurately determine the orbits. They suggest, through N-body fitting,
that GJ876 must actually lie in perfect resonance, and that the
orbital elements provided in the discovery paper (which are used here)
suffer from a systematic error. Because $P_1$ is so short for this
system (60 days) the evolution of the orbital elements has been
observed, and corroborated this theory. This section therefore serves
as a check to this hypothesis.

We ran GJ876 for $10^6$ years, but $\tau_{GJ876}$ corresponds to only
47,000 years. On 47,000 year timescales, only $5.6\% \pm 2.8\%$ of
parameter space is stable. On $10^6$ year timescales, 2.4\% of
configurations survived, but only 1.7\% were still unperturbed, and 5\%
failed to conserve energy. In unstable cases planet c, the inner and
less massive, was ejected/perturbed nearly 96\% of the time. \Fig 11
shows the instability rate. GJ876 shows the same trend as HD82943;
most unstable configurations break up in just a few hundred dynamical
times. Again the asymptotic nature of this plot implies most unstable
configurations have been identified. This system shows the same sort
of trend in $\epsilon$ as HD82943, which proves that our results our
correct.

\medskip
\epsfxsize=8truecm
\epsfbox{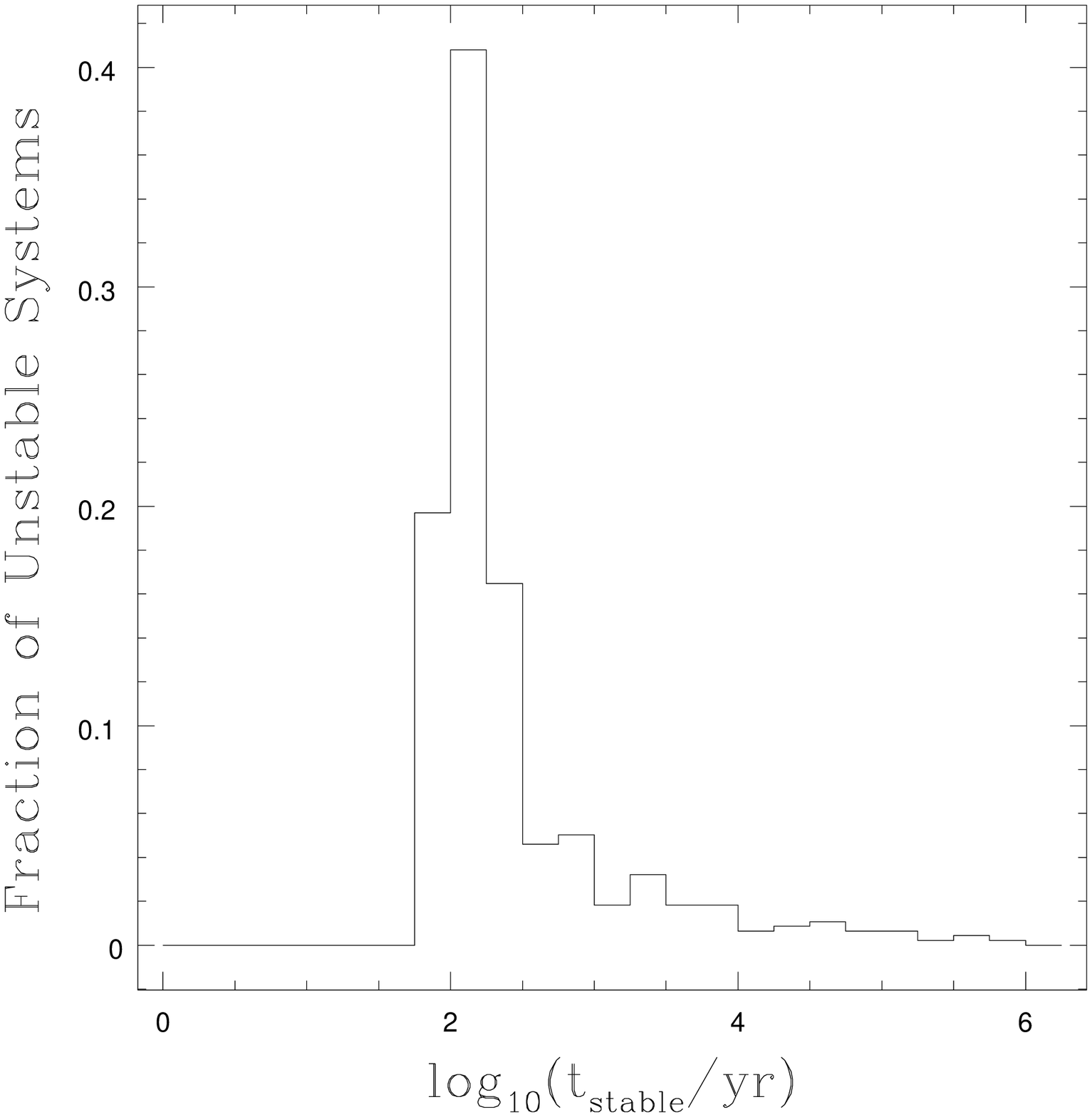}
\figcaption[gj876.ejrate.ps]{\label{fig:asymptotic}
\small{The ejection rate for GJ876. In this system unstable configurations
are usually ejected within 100 years. The rate asymptotically falls to
zero by $10^5$ years.}}
\medskip

Unlike HD82943, there are no obvious zones of stability. We only see
isolated islands in the $R-e_b$ stability map presented in \Fig
12. We choose these parameters for our stability map as they were the
strongest indicators of stability in HD82943, and because of the
suggestion that the system actually lie in perfect resonance. Close to
$R=2.00$ we sampled two simulations near $R=2.02$ and $e_b$=0.7. Both
these were stable. However with such poor statistics, and at such a
large (relative) distance from perfect resonance, we cannot comment on
the likelihood that the system would be more stable in perfect
resonance. We can, however, point out that there are isolated
configurations which may hold stable orbits, and prolonged
observations of this system should demonstrate whether it is indeed in
perfect resonance. However this lack of a large stable region
strengthens the hypothesis that this system lies in perfect resonance.

\medskip
\epsfxsize=8truecm
\epsfbox{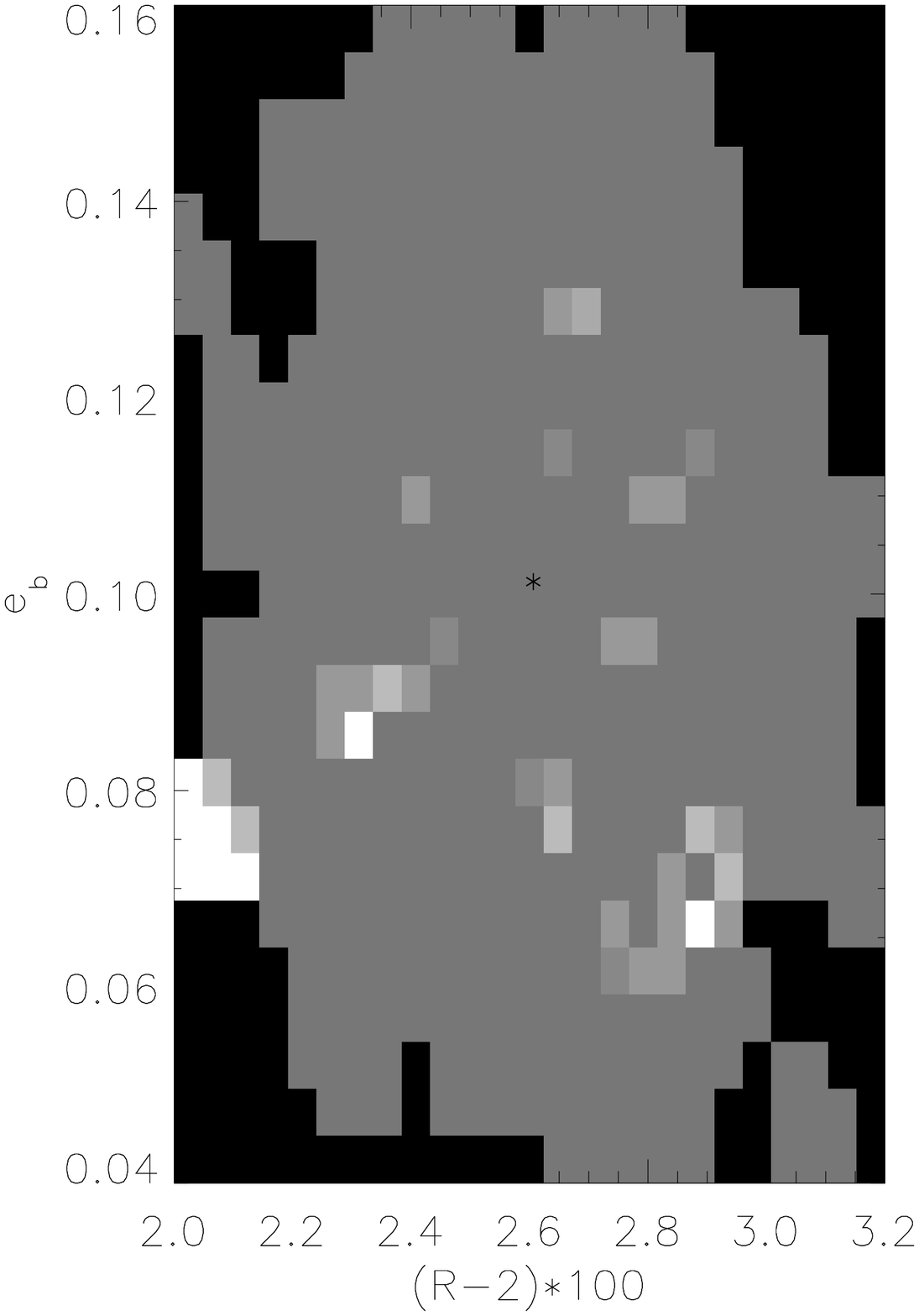}
\figcaption[gj876.ecc-res.grey.ps]{\label{fig:asymptotic}
\small{The $R-e_b$ stability map for GJ876. The asterisk marks the
best fit to the system as of 7 Aug 2002, and the values for stability
are more accurate closer to the asterisk. In this system, as in
HD82943, the two relevant orbital elements are $e_1$ and R. There are
no contiguous regions of stability, only small isolated pockets which
may hold stable zones.}}  
\medskip

This system does, however, lie very close to $\Lambda$=0. However this
proximity to alignment has no bearing on the stability of the system,
in fact, it may actually diminish its chances of stability. In \Fig
13 the probability of stability as a function of initial $\Lambda$ is
shown. Although the values for large $\Lambda$ are poorly sampled,
four data points at 100\% stability does suggest larger values of
$\Lambda$ may be more stable.

Similar dynamics are present in GJ876 as in HD82943. In Figs.\ 14-17,
we show 4 examples of stable and unstable configurations. The initial
conditions for these sample simulations are listed in Table 4. In
\Fig 14 the orbital evolution of simulation GJ876-904 is shown. This
system is one of very few (<10) simulations which show approximately
regular motion. The semi-major axes show resonant
perturbations, and  $\Lambda$ shows a motion typical of chaos; it appears to
be the superposition of libration and circulation.

In \Fig 15 the evolution of a stable, yet chaotic, configuration is
plotted. Although this system was stable for 47,000 years, planet b
was perturbed after 220,000 years. This system however remained bound
for $10^6$ years. The large eccentricity oscillations continue on to
$10^6$ years, and planet c tends to remain in a retrograde
orbit. After $10^6$ years $a_c=0.0515$AU, $0.06\le e_c \le 0.85$,
$i_c\approx 120^o$, $a_b\approx 0.87$AU, $e_b\approx 0.77$, and
$i_b\le 5^o$. The period of $e_c$ oscillations remains at 3300
years. The $\Lambda$ evolution further belies the chaos in this
system, as it tends toward anti-alignment, but also circulates. As
before this distribution is over $10^6$ years, but as the system
remains in approximately the same state from 25,000 years to $10^6$
years, this plot is a fair representation of the motion during the
first $\tau_{GJ876}$.

\medskip
\epsfxsize=8truecm
\epsfbox{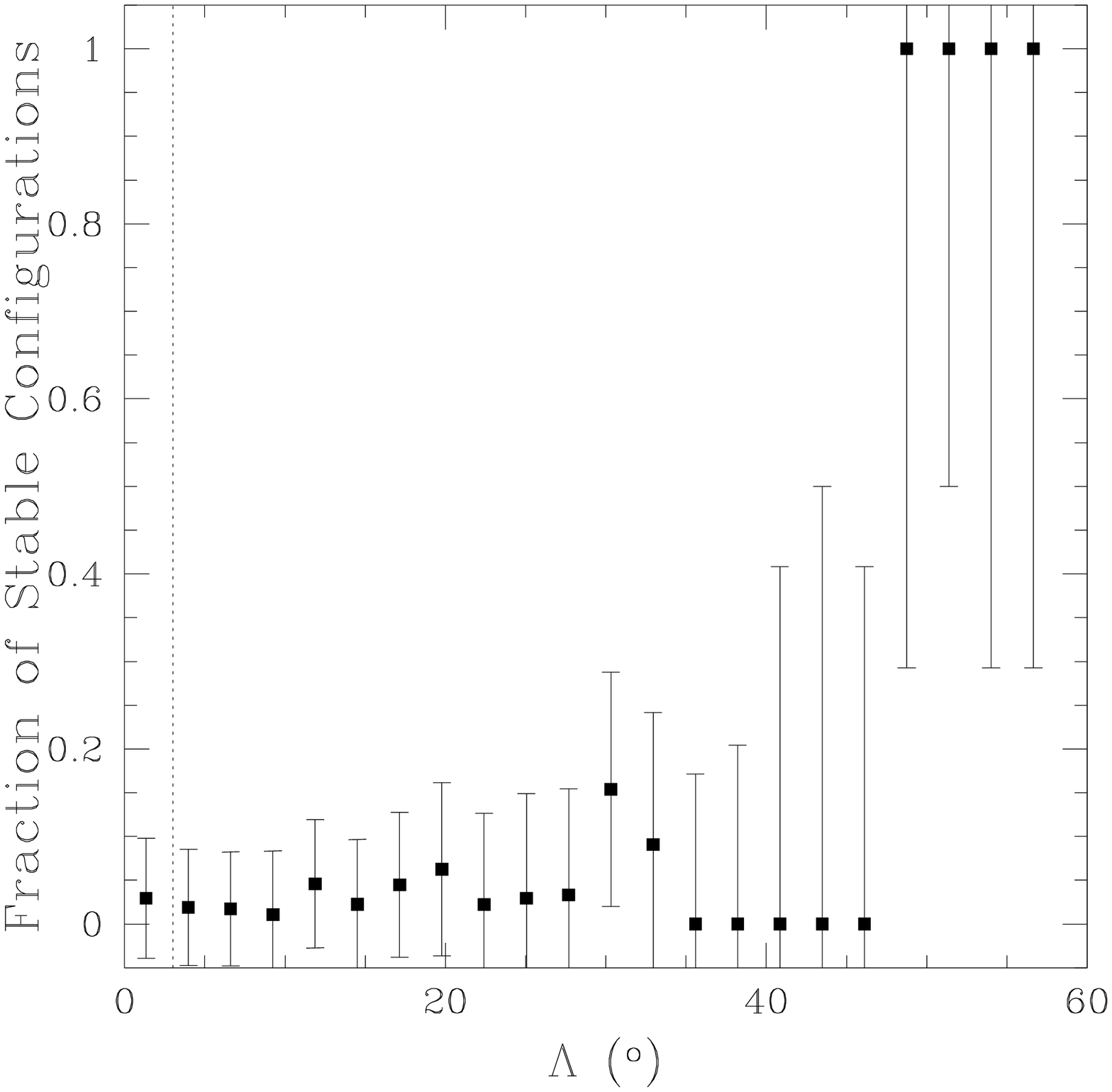}
\figcaption[gj876.lambda.dyn.ps]{\label{fig:asymptotic}
\small{The dependence of stability on initial $\Lambda$. The data
above $50^o$ are poorly sampled, but with such a large difference
between their values and the mean of 5.7\%, they do suggest that
stability may be improved at $\Lambda\gtrsim 50^o$.}}
\medskip

\begin{figure*}
\psfig{file=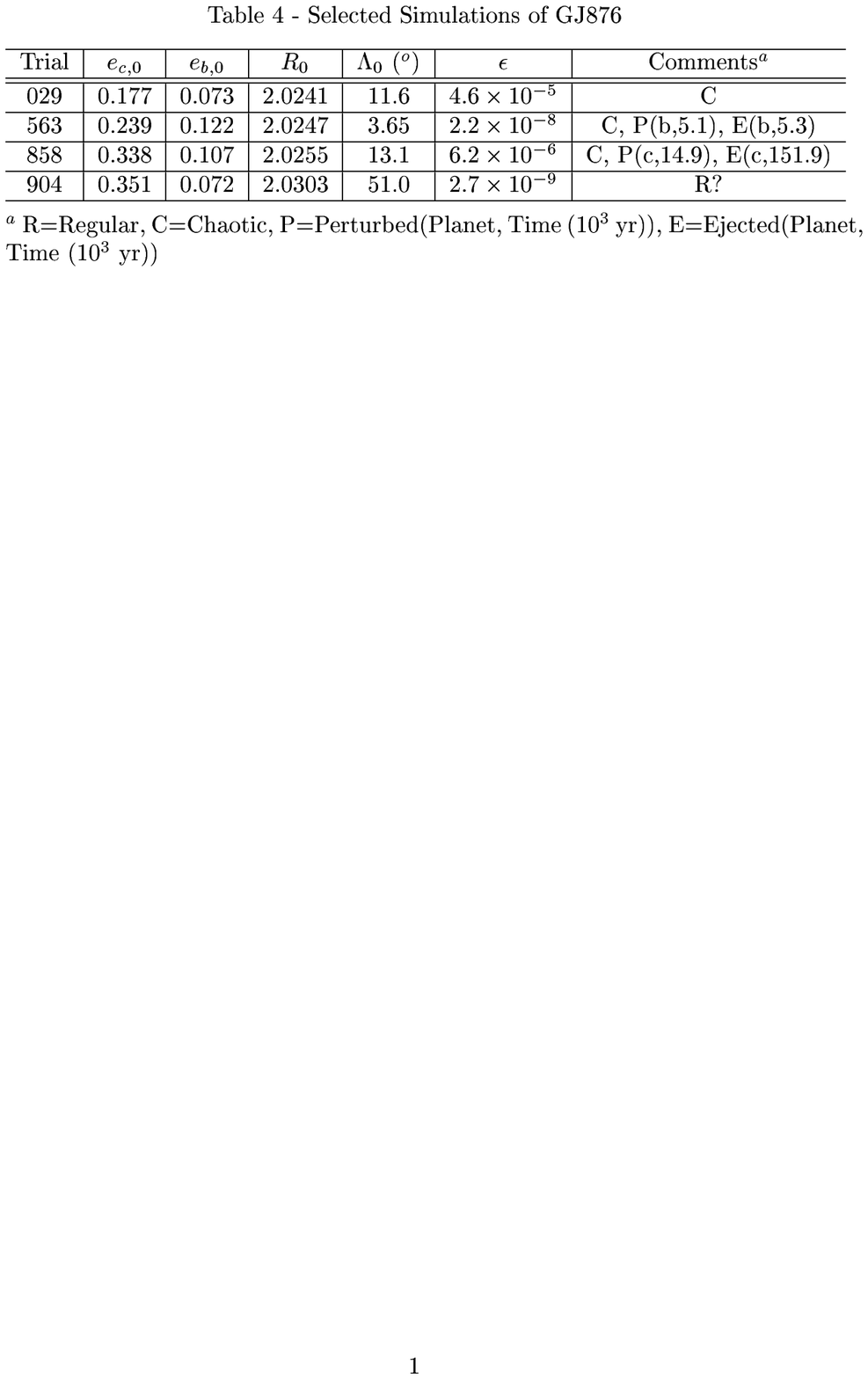,width=19.truecm}
\end{figure*}

\medskip
\epsfxsize=8truecm
\epsfbox{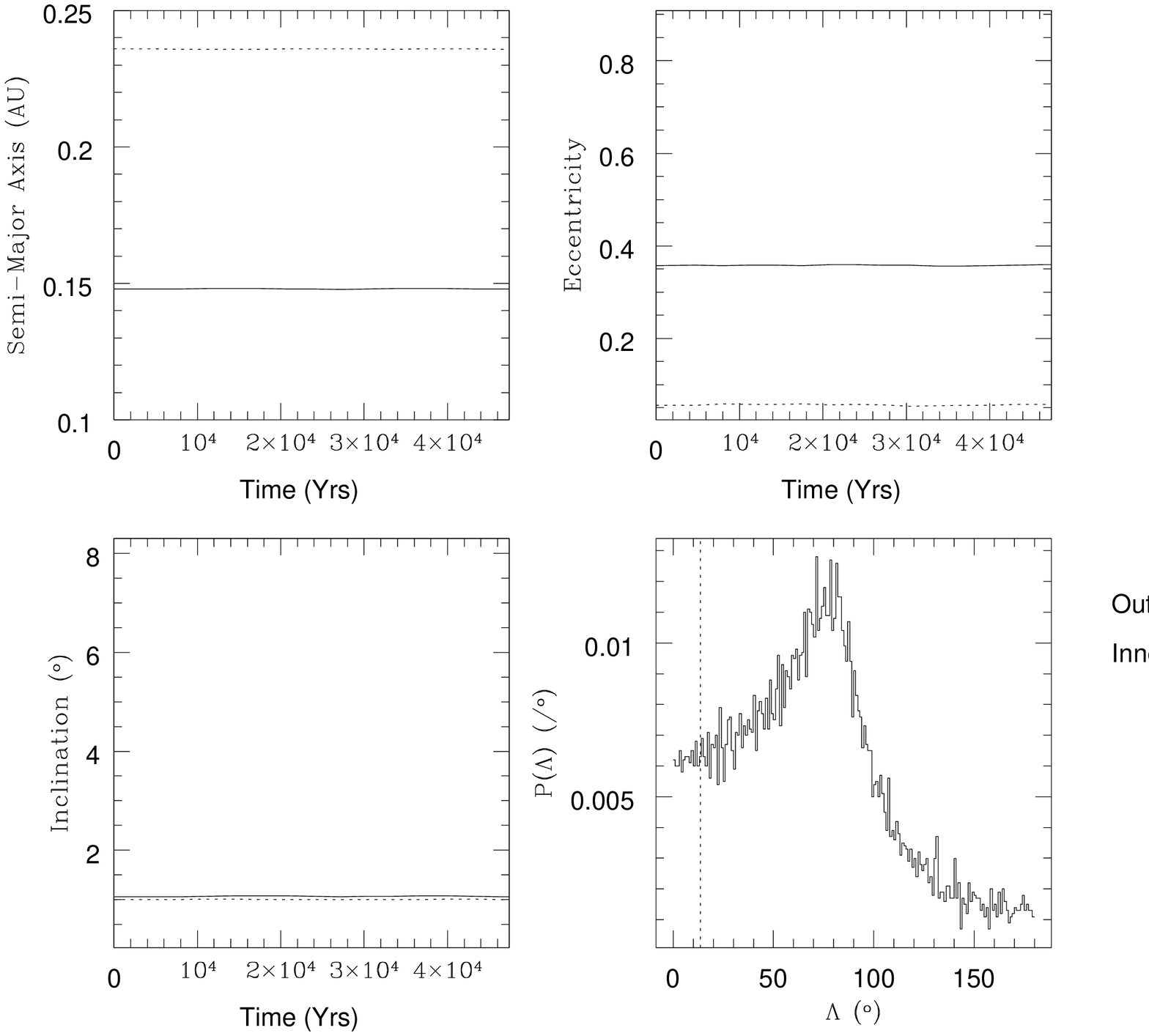}
\figcaption[gj876_904.smooth.ps]{\label{fig:asymptotic}
\small{Orbital evolution of GJ876-904, the stable, regular motion of
GJ876. The data are smoothed on 3000 year intervals, which eliminates
all short period behavior. {\it Top Left}: The low order resonance and large
eccentricities induce low amplitude oscillations in the semi-major
axes of the planets with a period of 2 years. This short frequency is
not visible in this plot due to the smoothing interval. {\it Top Right}: The evolution of eccentricity
for this configuration is that of two eigenmodes, also on a period of 2 years. {\it Bottom Left}: The
inclinations vary slightly with a frequency of 250 years.  {\it
Bottom Right}: The $\Lambda$ distribution function suggests that
$\Lambda$ sometimes librates with an amplitude of $\approx 80^o$, and
sometimes circulates.  The dashed line marks $\Lambda_0$.}}
\medskip

\medskip
\epsfxsize=8truecm
\epsfbox{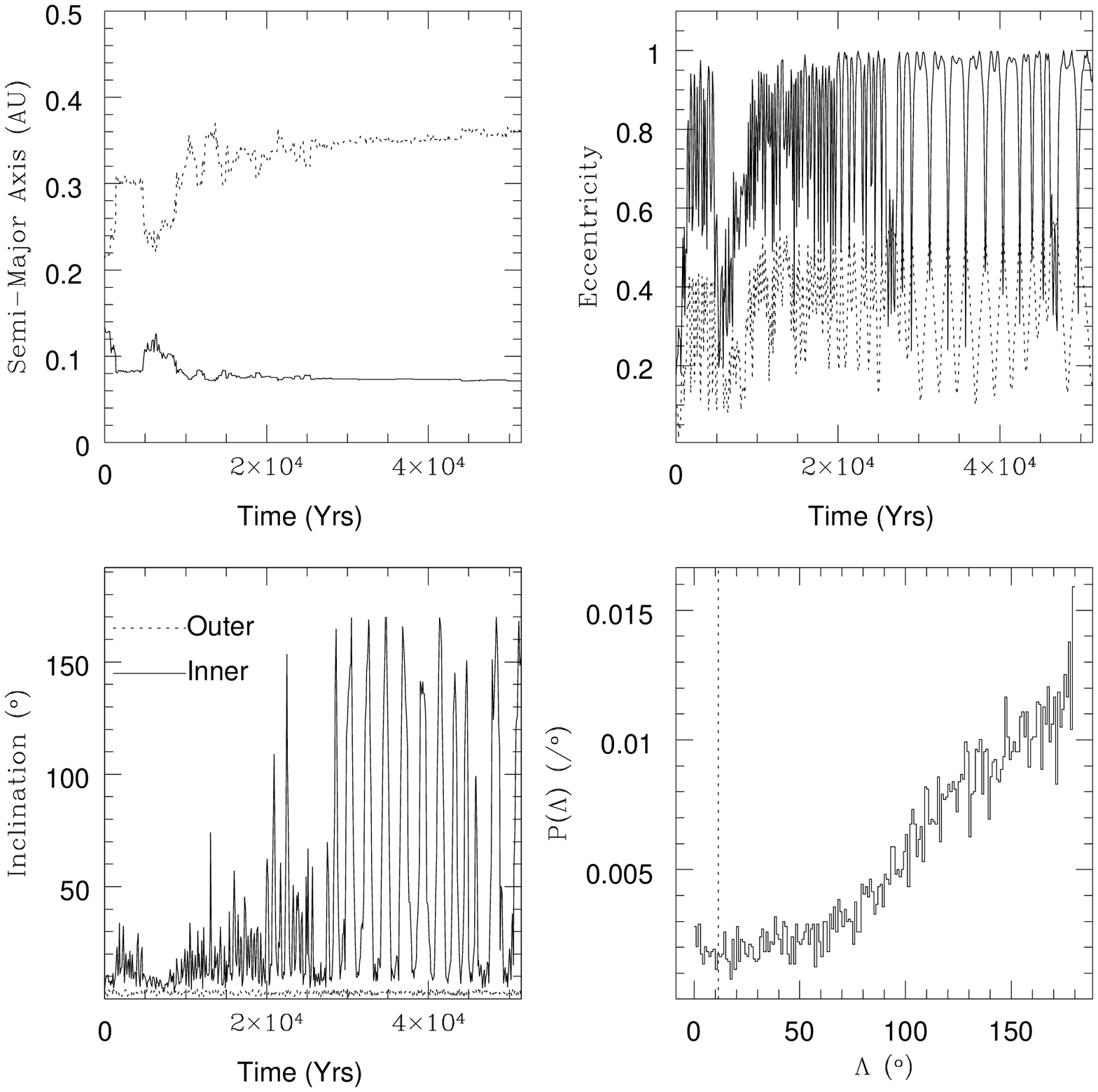}
\figcaption[gj876_029.bw.ps]{\label{fig:asymptotic}
\small{Orbital evolution of GJ876-029, a chaotic stable configuration
of GJ876. {\it Top Left}: Although the semi-major axes vary, they
don't change by a factor of 2 until 220,000
years=4.7$\tau_{GJ876}$. {\it Top Right}: The remarkable eccentricity
evolution of this system. These oscillations persist for $10^6$
years. {\it Bottom Left}: The inclination of planet b varies a small
amount, generally staying below $5^o$. Planet c however experiences
wild fluctuations, however it does eventually settle to $120^o$. {\it
Bottom Right}: This curious $\Lambda$ distribution function suggests
that $\Lambda$  prefers anti-alignment. This implies that a protection
mechanism is keeping the system from breaking apart despite the
extremely large values of  $e_b$.}}
\medskip

In \Fig 16, the evolution of GJ876-563 is shown. This configuration
ejects planet c in just over 5000 years. The system shows evidence for
close approaches right from the beginning as the semi-major axes show
small amplitude, chaotic fluctuations. The inclinations and
eccentricities also experience a large degree of chaos. Although
initially $\Lambda$ is very close to 0, it circulates for the duration
of the simulation.

\medskip
\epsfxsize=8truecm
\epsfbox{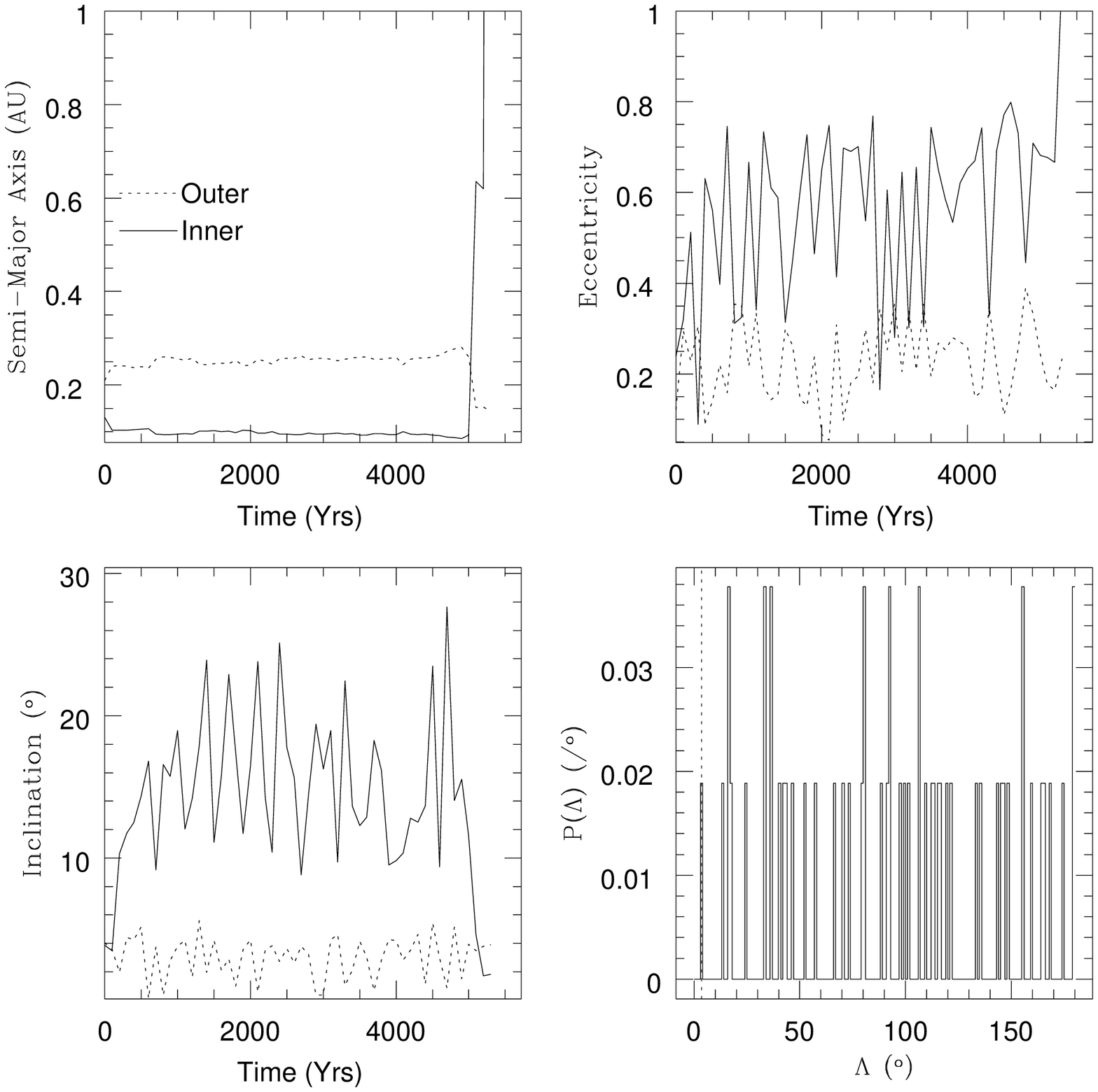}
\figcaption[gj876_563.bw.ps]{\label{fig:asymptotic}
\small{Orbital evolution of GJ876-563, the ejection of GJ876c. {\it
Top Left}: The semi-major axes undergo small fluctuations until a
violent encounter ejects the inner planet after 5000 years. {\it Top
Right}: The eccentricities experience large amplitude fluctuations,
until $e_c$ eventually reaches unity. {\it Bottom Left}: The
inclination of planet b initially jumps and remains at over $30^o$,
but then returns to coplanarity just before ejection. This type of
motion was also seen in \Fig 8. {\it Bottom Right}: As with the
majority of chaotic configurations, $\Lambda$ circulates. Although
poorly sampled, $\Lambda$ appears to have no preferred value.}}
\medskip

In \Fig 17 the evolution of a system which perturbs the outer planet
in just 9000 years is shown. The system ejects planet b in 152,000
years. Before reaching $\tau_{GJ876}$ this system experiences some
remarkable evolution in $a$, $e$, and $i$. Note, too, that $\Lambda$
very quickly moved to an anti-aligned configuration.

\medskip
\epsfxsize=8truecm
\epsfbox{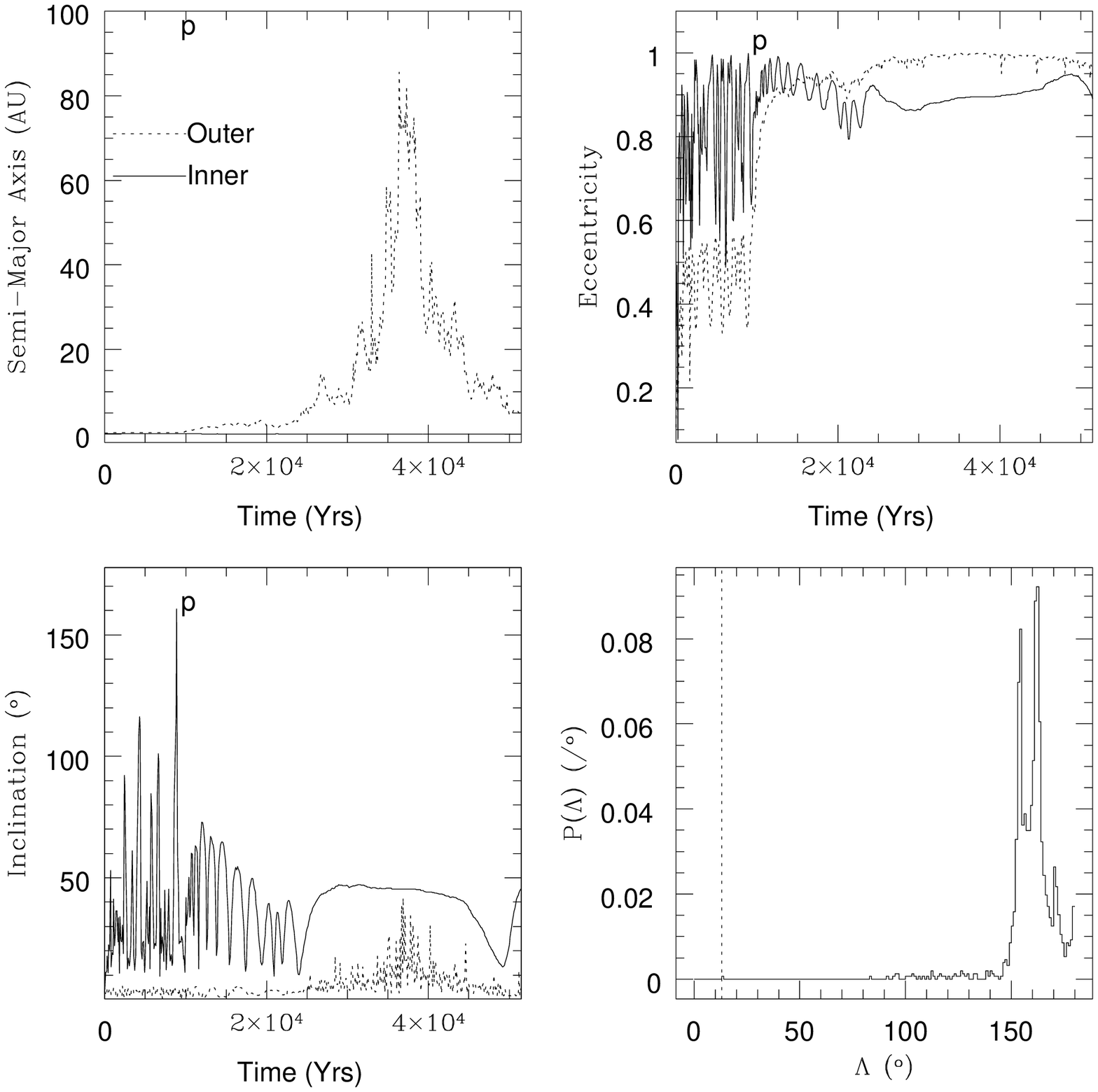}
\figcaption[gj876_858.bw.ps]{\label{fig:asymptotic}
\small{Orbital evolution of GJ876-858, the perturbation of  
GJ876b. Planet b was eventually ejected after 152,000 years. {\it Top
Left}: The semi-major axes evolve quiescently for 9000 years, until a
close approach increases $a_b$, marked by the p. Although $a_b$
returns to its initial value by $\tau_{GJ876}$, just prior to ejection
$a_b$ reached 750AU. {\it Top Right}: The eccentricities immediately
jump to very large values. $e_c$ varies wildly between 0.6 and
0.99. After b is kicked out to large $a$, the oscillations become much
smaller. For nearly 10,000 years $e_b$ remains above 0.98, but it
doesn't reach unity until 152,000 years. {\it Bottom Left}: As with
eccentricity, the inclination of c jumps wildly for 9000 years, even
reaching $162^o$ just prior to perturbation. However it is unclear if
this large inclination produces the perturbation. {\it Bottom Right}:
As with the $e$ and $i$, $\Lambda$ immediately moves from its starting
position. However $\Lambda$ remains very close to anti-alignment for
the duration of the simulation.}}
\medskip

Long term simulations for this system were integrated for 27.5 million
years. A complete summary of the long term simulations for this system
is presented in Table 5. \Fig 18 plots the eccentricity evolution of
4 simulations. Some systems appear regular throughout (top left). Some
configurations are chaotic for the duration of the simulation (top
right, bottom left), and others may eject a planet after an
arbitrarily long period of time (bottom right).

\begin{figure*}
\psfig{file=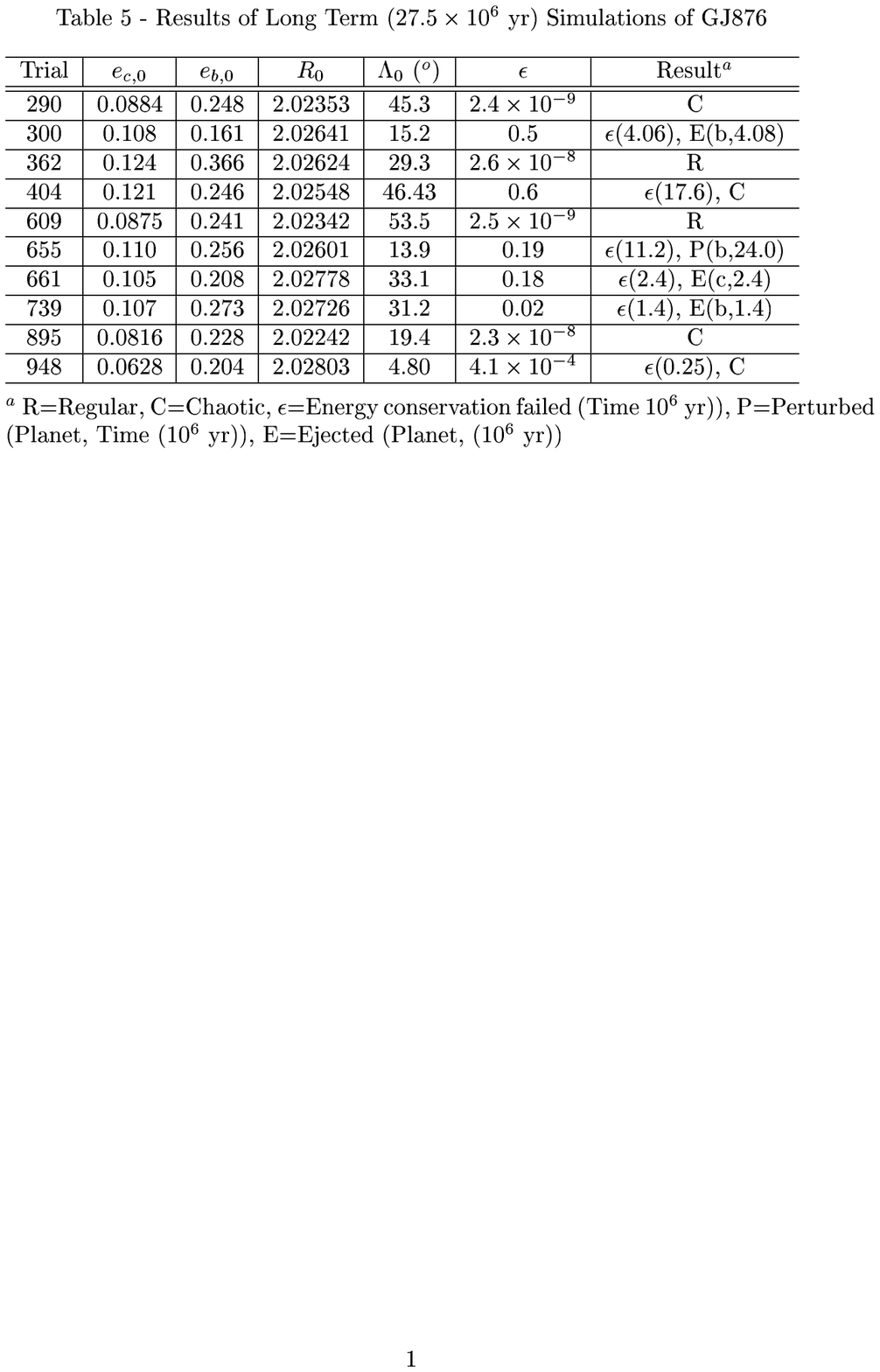,width=19.truecm}
\end{figure*}

\medskip
\epsfxsize=8truecm
\epsfbox{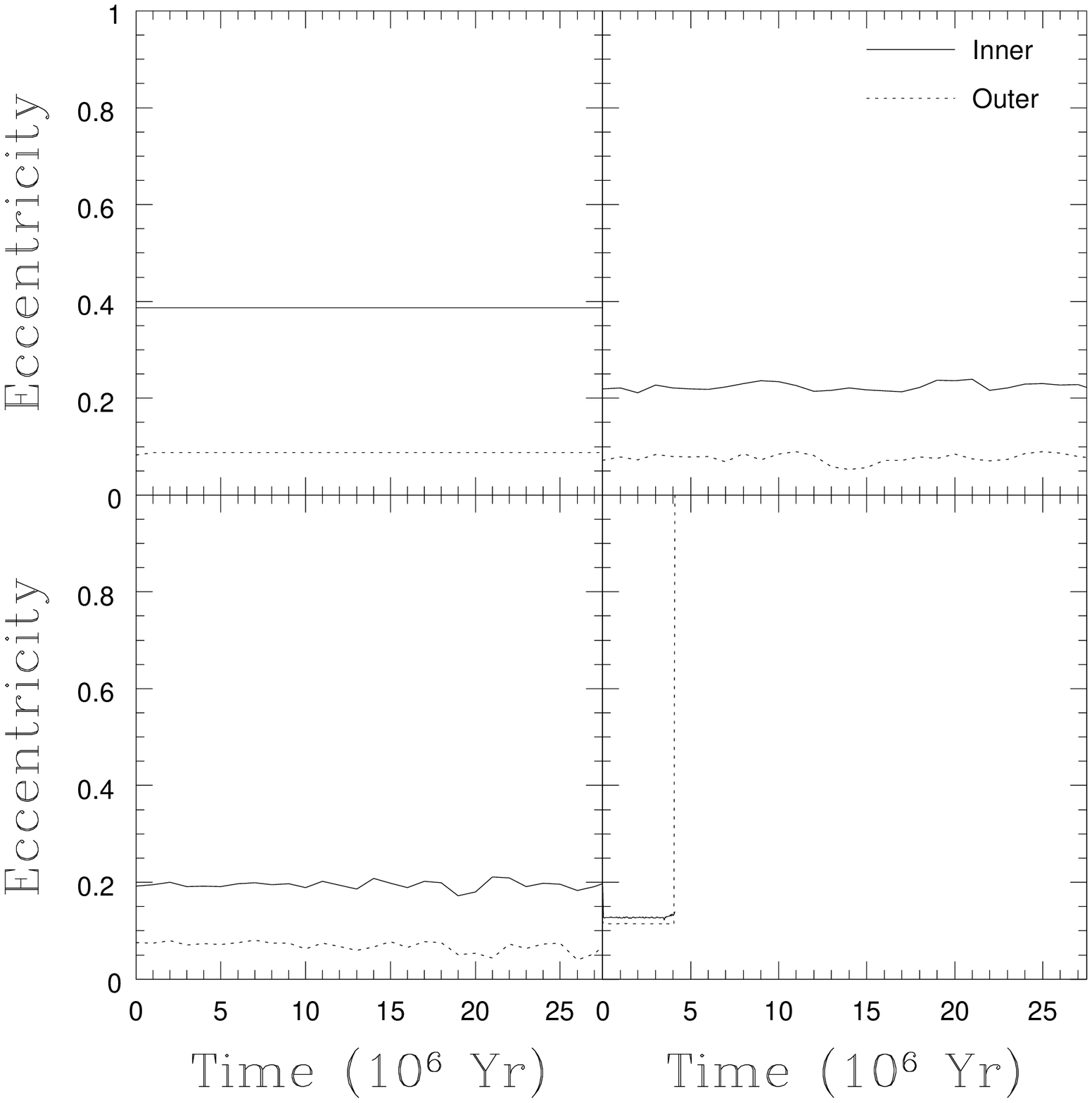}
\figcaption[gj876.long.bw.ps]{\label{fig:asymptotic}
\small{Eccentricity of 4 long term simulations of GJ876. {\it Top
Left}: Evolution of GJ876-362, a regular configuration. {\it Top
Right}: Evolution of GJ876-290, a chaotic, yet stable
configuration. {\it Bottom Left}: Evolution of GJ876-895, a chaotic,
stable configuration. {\it Bottom Right}: Evolution of GJ876-300 which
ejects planet b after 4 million years.}}
\medskip

\Fig 19 shows the distribution function of $\Lambda$ for the same
four systems. The regular system (top left) shows a configuration
which usually librates with an amplitude of $80^o$, but with
occasional circulation. The two chaotic examples (top right, bottom
left) have flat distribution functions. Not surprisingly the unstable
trial shows a very peculiar distribution function. As in HD82943 we
see that regular systems tend to librate, and chaotic configurations
circulate.

\medskip
\epsfxsize=8truecm
\epsfbox{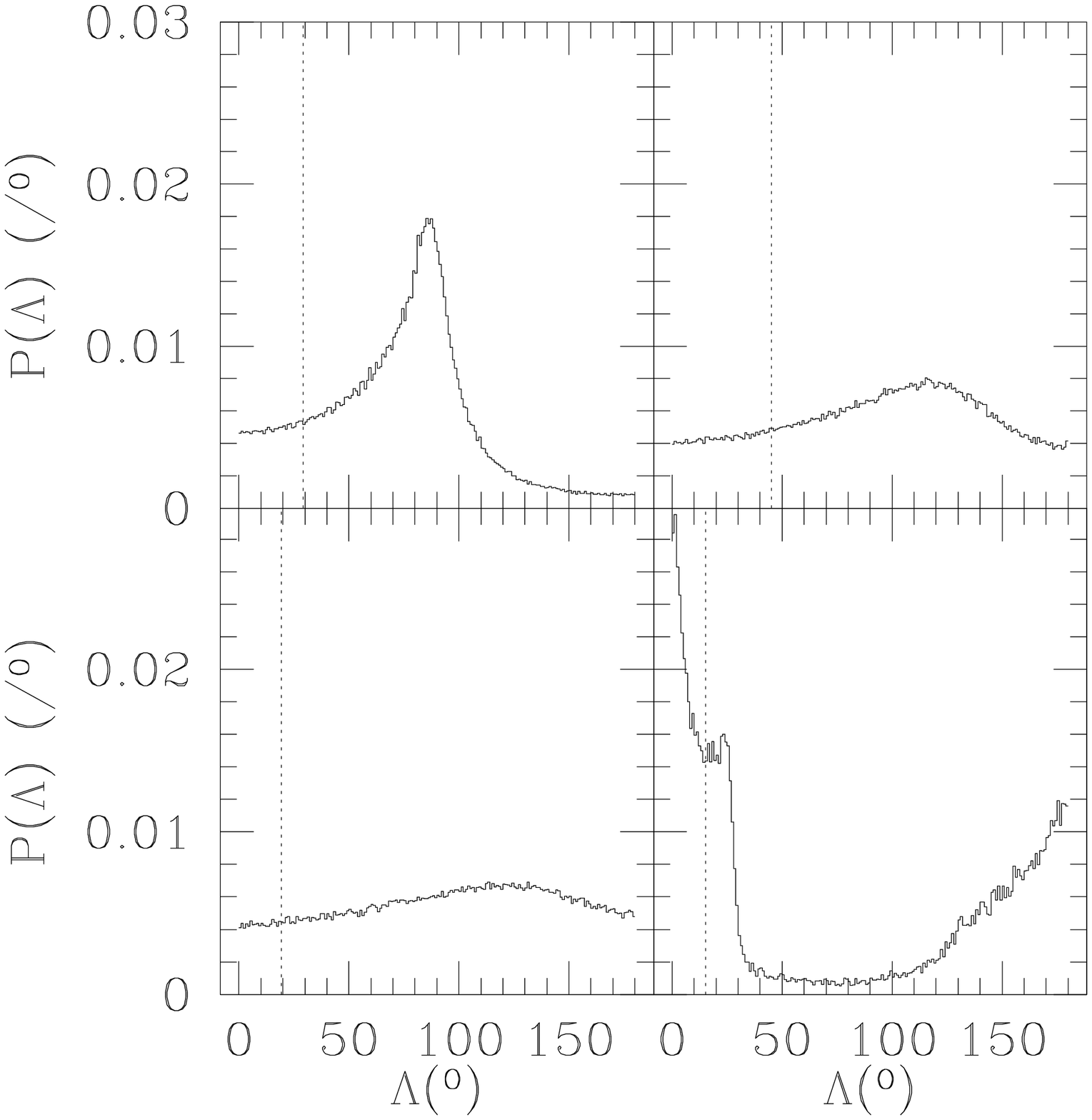}
\figcaption[gj876.lamprob.ps]{\label{fig:asymptotic}
\small{The $\Lambda$ evolution of the same four simulations in \Fig 19.
Chaotic systems (top right, bottom left) show no signs of libration,
while regular systems (top left) are librating, but with occasional
circulation. The unstable example (bottom right) shows a very strange
distribution, which only demonstrates the chaos of this system.}}
\medskip

Although the evidence is compelling that GJ876 does in fact lie in
perfect resonance, our work demonstrates that stable, regular systems
do exist close to the observed Keplerian fit. More observations of
this system will demonstrate if the system is in perfect
resonance. This work clearly demonstrates that stable regions do exist
for a system like GJ876 just beyond perfect resonance. Should this
system lie in perfect resonance, then this work shows that unstable
regions lie very close to its configuration.

New astrometric data has confirmed the mass of the outer planet in this
system (Benedict \etal 2002). This therefore provides the only system with
a known mass. The plane of b's orbit is inclined by $6^o$ to
the line of sight. Benedict \etal confirm the mass and semi-major axis of
this planet to be statistically identical to those presented in Table 1.
However for this paper, the lack of data for planet c precludes any new
insights into the dynamics of the system. At best, if the system is
approximately coplanar, then the variations used here are indeed
representative of the true system, and our results are more robust.

\section{Interacting Systems}
Four known systems meet the interacting system criterion: $\upsilon$
And (Butler \etal 2000), 47UMa (Fischer \etal 2002), the
SS\footnote{http://ssd.jpl.nasa.gov/elem\_planets.html} and HD12661
(Fischer \etal 2003). In this paper we will limit ourselves to the
first three. The number, placement, and size of the planets in each
system are quite different, but all have at least two planets that lie
in between the 2:1 and 10:1 resonances. $\upsilon$ And was the first
known ESPS, and was the subject of Paper I. The experiment in Paper I
is the procedure for this paper, and the simulations have been
performed again. 47UMa was announced in 2001, and, at first glance,
appears more like the SS than $\upsilon$ And. Performing this
experiment on the SS is problematic. The errors in the orbital
elements of the SS are drastically smaller than for the ESPS, therefore
fitting the SS into the procedure requires inflating the SS orbital
element errors to values comparable to those of the ESPS. Essentially
we are asking what would we observe if we took radial velocity
measurements of our sun. We will compare the ESPS to both the gas
giant system ($\S$4.3) and the Jupiter-Saturn system ($\S$4.4). These
three coupled systems' orbital elements are summarized in Table
6. Interacting systems show broad regions of stability which are
correlated with eccentricity.

\begin{figure*}
\psfig{file=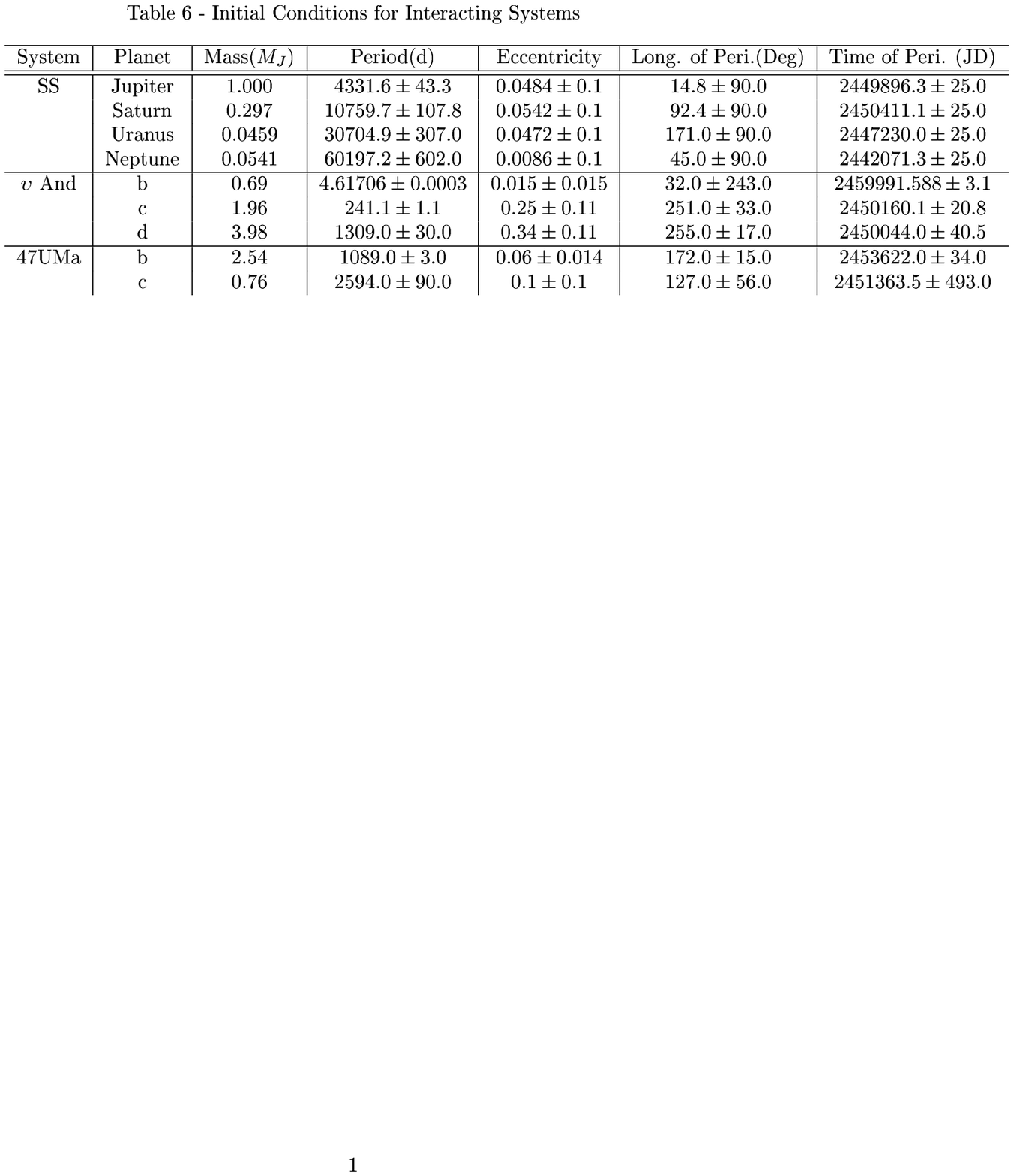,width=19.truecm}
\end{figure*}

\subsection{$\upsilon$ Andromedae} 
The $\upsilon$ And system is a combination of a separated system and
an interacting system. Three planets orbit the $1.02\pm 0.03M_{\odot}$
(Gonzalez \& Laws 2000) F8 star $\upsilon$ And. The inner planet, b
orbits at 0.04AU. The other planets, c and d, orbit at larger
distances (0.8AU and 2.5AU respectively), but are significantly more
eccentric.  The outer planet is the most massive, therefore
$\tau_{\upsilon And}$ corresponds to $10^6$ years.

The $\upsilon$ And system was the subject of Paper I, and has been the
focus of intense research since its discovery. The apparent alignment of
the apses of planets c and d has sparked the most interest with several
groups claiming that the system must be secularly locked, or at least
librate about $\Lambda$=0 (Rivera \& Lissauer 2000; Rivera \& Lissauer
2001b; Chiang, Tabachnik, \& Tremaine 2002), while others suggest this
alignment may be a chance occurrence (Paper I; Stepinski, Malhotra, \&
Black 2000). However, these groups and others (Laughlin \& Adams 1999) all
agree that this system, as presented, can be stable for at least $10^8$
years. 

On a $10^6$ year timescale, 86.1\%$\pm$3.3\% of simulations survived,
and 0.4\% failed to conserve energy. This value is less than 1$\sigma$
from the value published in Paper I, 84.0\%$\pm$3.4\%. \Fig 20 shows
the perturbation rate as a function of time. Once again we see that
most unstable configurations eject a planet immediately, and the rate
falls to 4\% by $10^6$ years. The fact that ejections occur right up
to $10^6$ years implies that we have not detected all unstable
situations, and that the stability map for this system contains more
unstable configurations, and hence the plateau is smaller and/or the
edge is steeper after $10^9$ years.

There is one notable difference between the results of Paper I, and
those reported here: the frequency of ejections of each planet is
different. In Paper I planet b was ejected 10\% of the time, c, 60\%,
and d, 30\%. The SWIFT runs ejected planet b 39\% of the time, c,
54\%, and d, 7\%. \Fig 21 plots the survival probability in this
system as a function of energy conservation. Stability peaks at
$\epsilon=10^{-8}$ but quickly drops.  Although this plot is
qualitatively different from \Fig 2, we again note that this implies
the simulations are valid. This plot is typical for the interacting
systems, confirming our hypothesis that we need only maintain
$\epsilon < 10^{-4}$ for the duration of every simulation.

\medskip
\epsfxsize=8truecm
\epsfbox{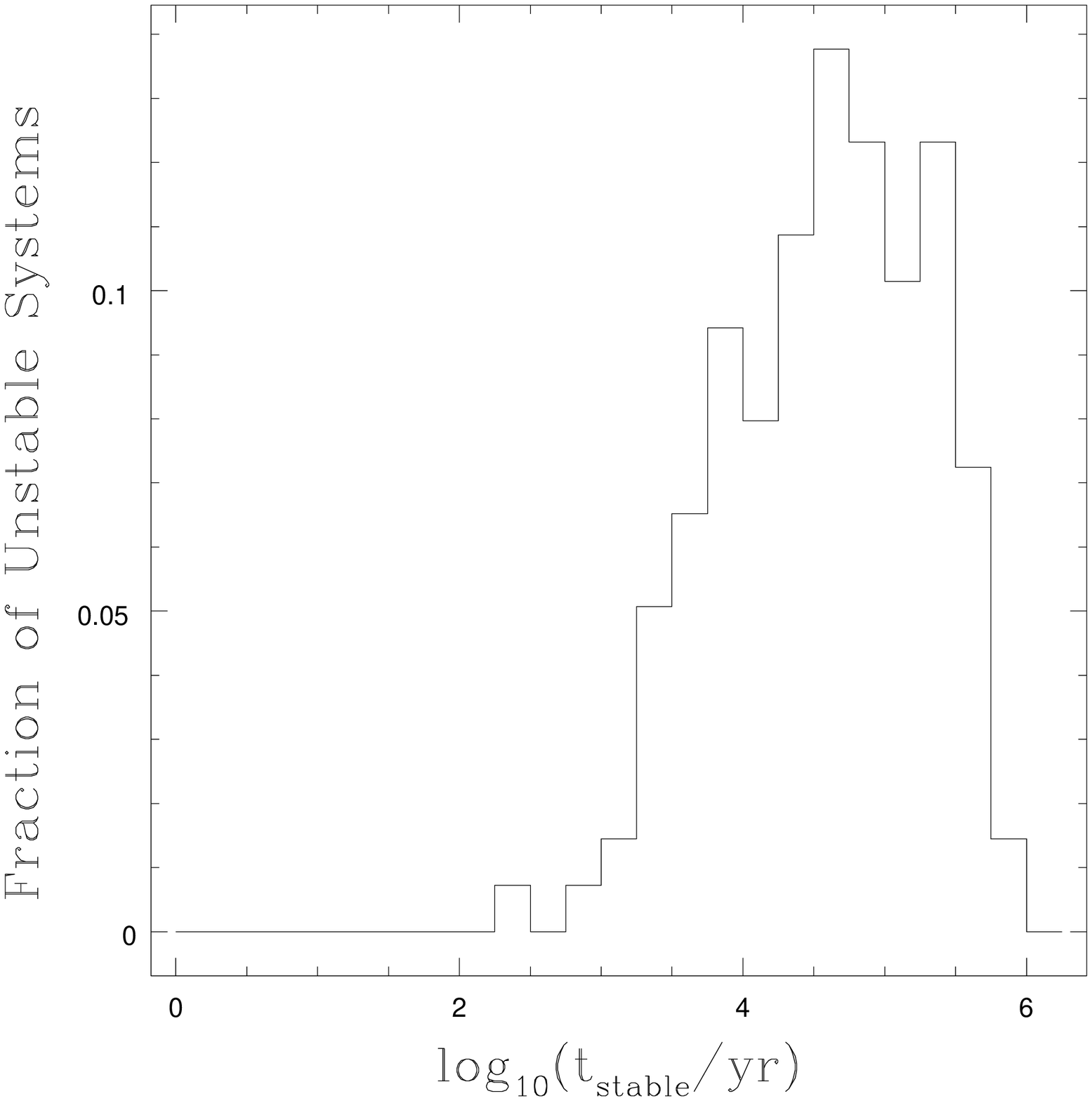}
\figcaption[upsand.ejrate.ps]{\label{fig:asymptotic}
\small{The ejection rate of unstable configurations in $\upsilon$ And.
Configurations may eject planets right up to $10^6$ years. It is therefore
unclear how many more configurations might become unstable.}}
\medskip

In Paper I, a stability map in $e_c$ and $e_d$ was presented in Table
2.  This table showed that $e_d$ and, to a lesser degree $e_c$
determined the stability of the system. In \Fig 22, the $\upsilon$
And stability map is presented. It is nearly an exact match to that in
Paper I. However the best fit to the
system\footnote{http://exoplanets.org/upsandb.html} has changed since
then and the system has moved away from the edge slightly. It is
important to note that two different integration schemes produced the
same results.  From \Fig 22, it is clear that $\upsilon$ And lies
near instability. The edge of stability in $\upsilon$ And, however, is
fundamentally different than in resonant systems. In this interacting
system, a large region of phase space is fully stable, the
``plateau'', but there is a sharp boundary in eccentricity space, the
``edge'', beyond which the system quickly moves into a fully unstable
region, the ``abyss''. Although it appears that both morphologies are
on the edge of stability, they are different types of edges.

\medskip
\epsfxsize=8truecm
\epsfbox{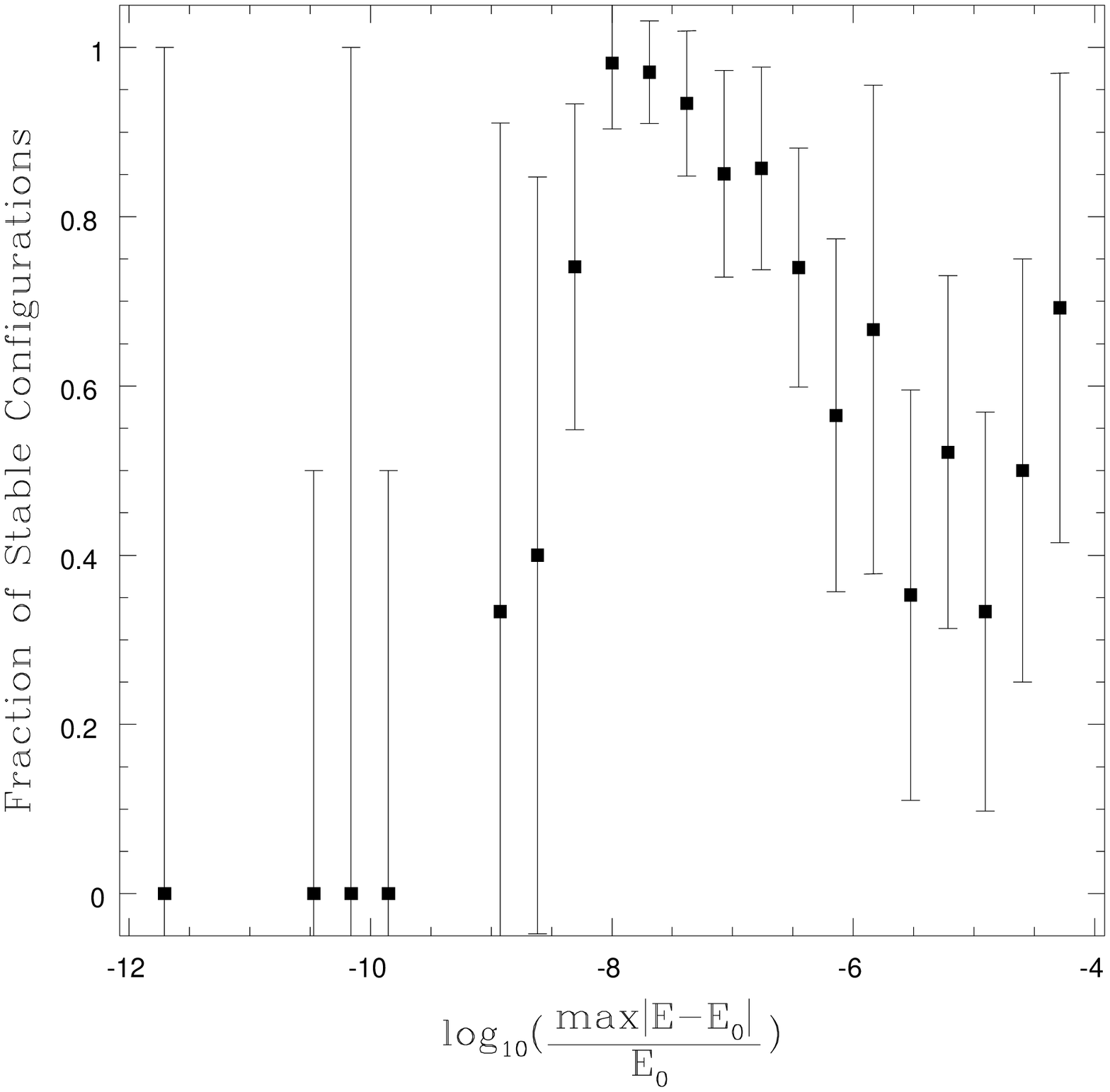}
\figcaption[upsand.energy.ps]{\label{fig:asymptotic}
\small{Survival probability as a function of energy conservation in
$\upsilon$ And. The instability at low $\epsilon$ implies the results for 
$\upsilon$ And are robust.}}
\medskip

\medskip
\epsfxsize=8truecm
\epsfbox{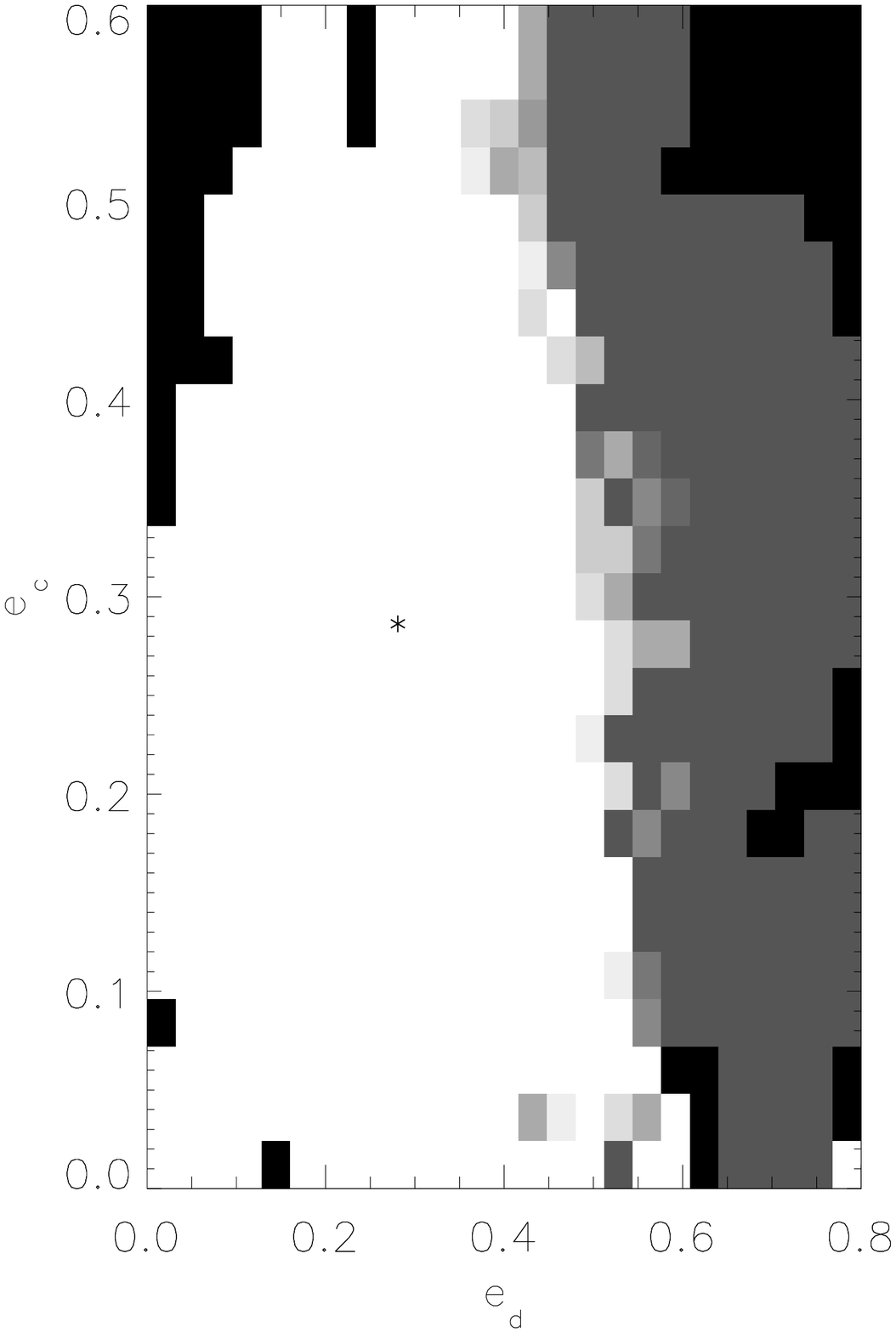}
\figcaption[upsand.ecc.grey.ps]{\label{fig:asymptotic}
\small{Stability map for $\upsilon$ And. Stability in this system depends
on $e_d$ and, to a smaller degree, on $e_c$. As in resonant system
stability maps, precision is correlated with distance from the
asterisk, which marks the best fit to the system as of 24 Sept
2002. $\upsilon$ And lies near the edge of stability.}}
\medskip

As previously mentioned there are some unusual features of this
system; the lines of node are nearly aligned, and the system lies
close to the 11:2 mean motion resonance. This is a weak perturbation,
but between these 2 massive planets, this may be important. However a
quick inspection of plots of stability versus $\Lambda$, \Fig 23, and
$R$, \Fig 24, shows that there is no statistically significant affect
caused by these to (potential) resonances. We do note that our
integration time may not be long enough to detect the importance of
the 11:2 resonance.

\medskip
\epsfxsize=8truecm
\epsfbox{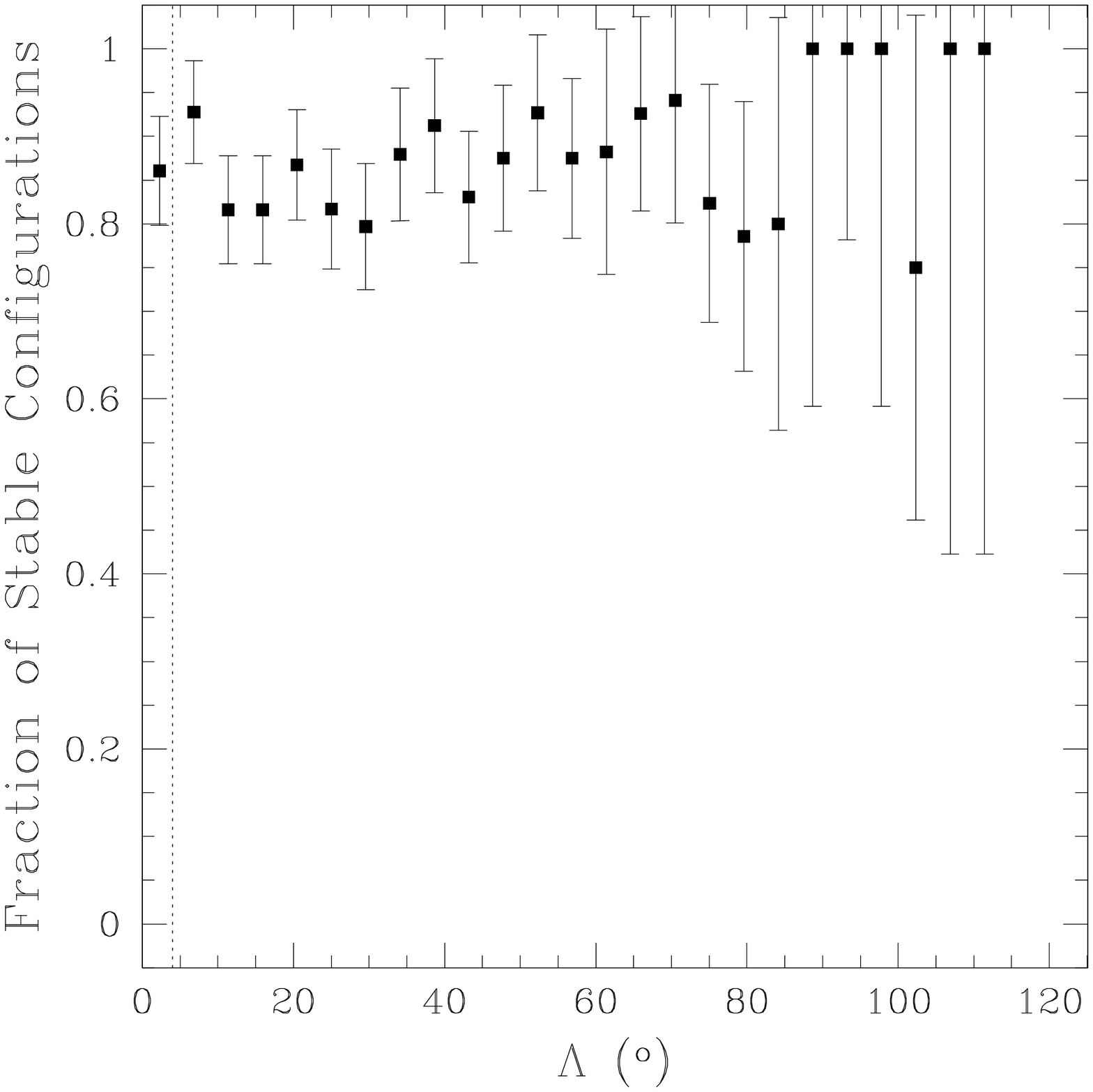}
\figcaption[upsand.lambda.ps]{\label{fig:asymptotic}
\small{Stability as a function of $\Lambda$ in $\upsilon$
And. Although the best fit to the system places it very close to
alignment, marked by the dashed vertical line, there is no significant
trend with $\Lambda$.}}
\medskip

\medskip
\epsfxsize=8truecm
\epsfbox{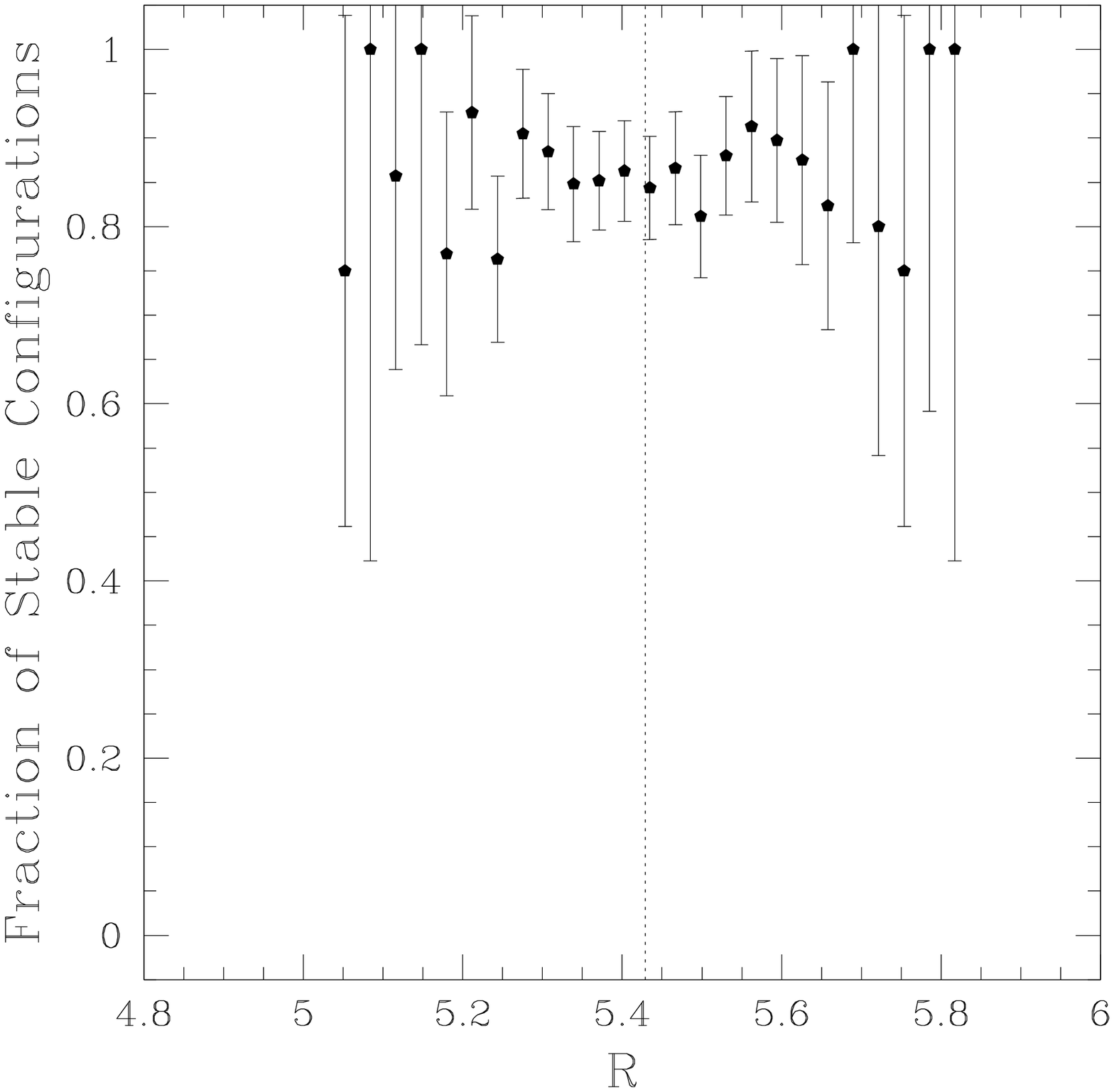}
\figcaption[upsand.res.ps]{\label{fig:asymptotic}
\small{Stability as a function of $R$ in $\upsilon$ And. There is no
evidence that this weak mean motion resonance affects the stability of
$\upsilon$ And. Although there are fully stable points outside of
resonance, all vary by less than 1$\sigma$ of the mean stability rate
of 0.86.}}
\medskip

As has been shown in other papers, this system exhibits both regular,
and chaotic motion. In \Figs 25-29, we present 5 examples of this
system. The initial conditions of these configurations are listed in
Table 7. In \Fig 25, the orbital evolution of a regular, stable
configuration is plotted. However, this plot actually demonstrates the
breakdown of our model. Planet b's eccentricity oscillates with an
amplitude of 0.3 and a period of 120,000 years. Unfortunately,
$\upsilon$ And b is tidally locked by its parent star with a period of
0.011 years. The timescale for tidal circularization is $8\times
10^7$yrs (Trilling 2000). We address this potential inconsistency in
$\S$7. $\Lambda$ for this system librates with an amplitude of $50^o$,
an indicator of regular motion.

\begin{figure*}
\psfig{file=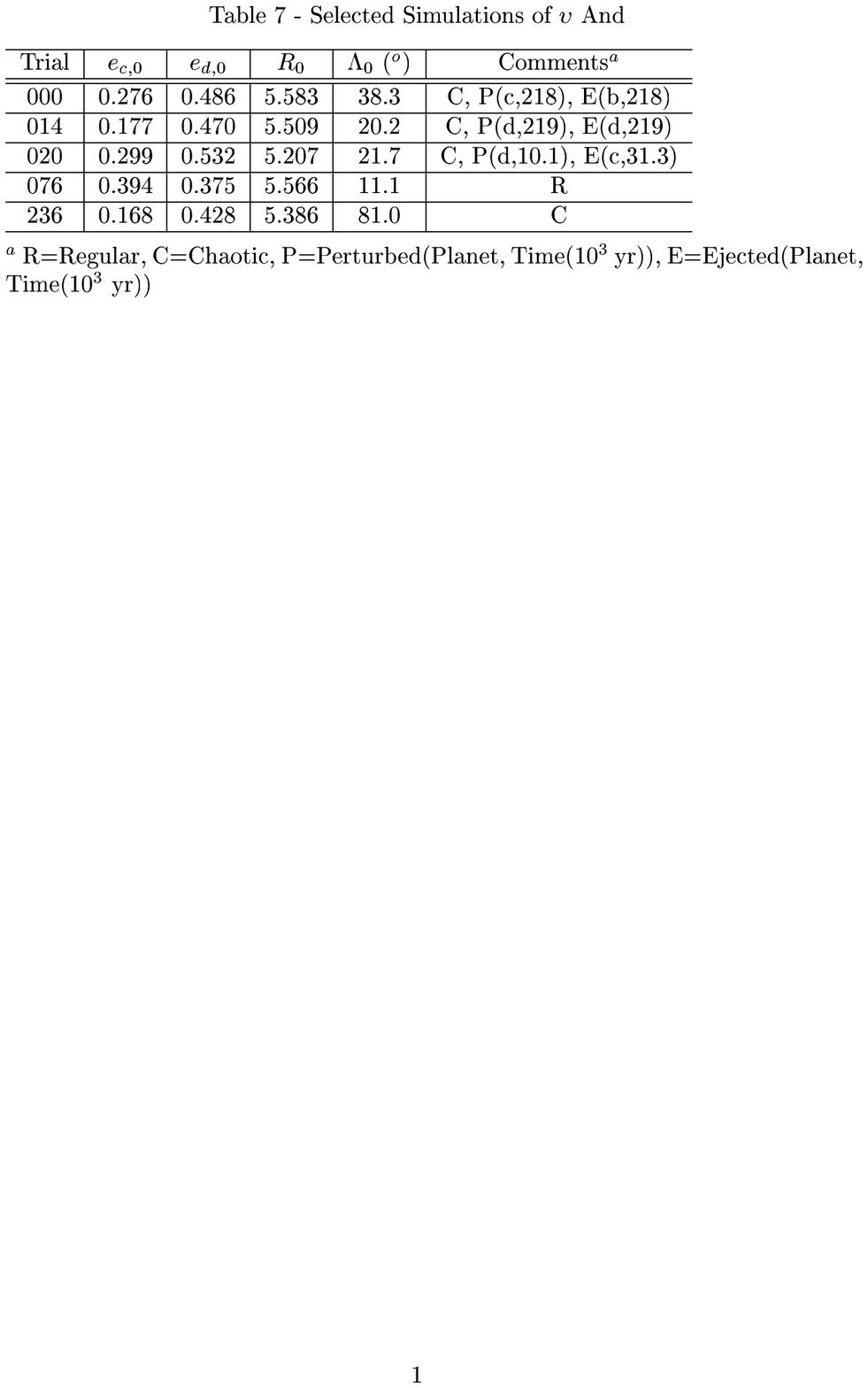,width=19.truecm}
\end{figure*}

\medskip
\epsfxsize=8truecm
\epsfbox{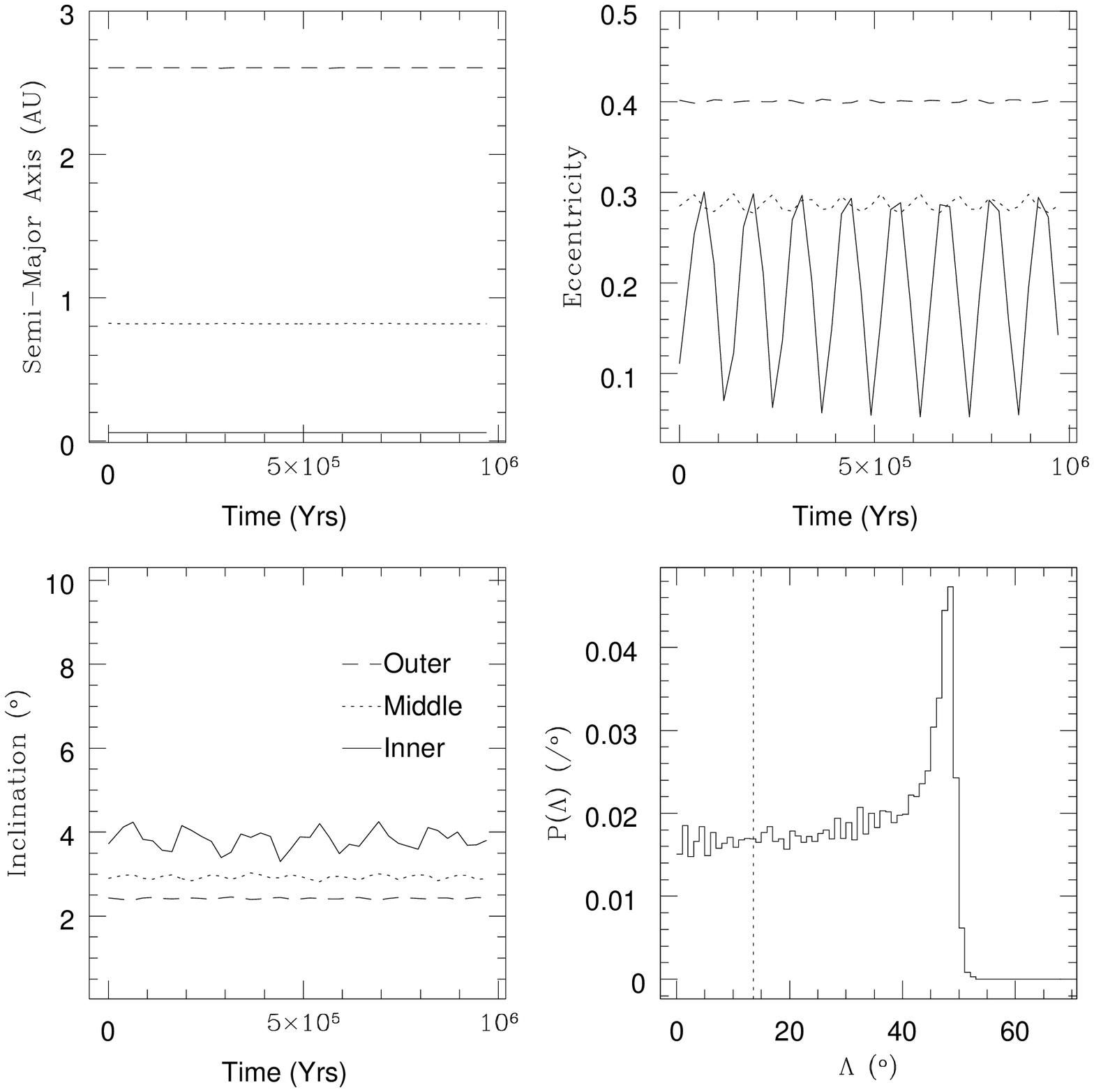}
\figcaption[upsand_076.smooth.ps]{\label{fig:asymptotic}
\small{The orbital evolution of $\upsilon$ And-076, a stable, regular
configuration, smoothed on 25,000 year intervals. {\it Top Left}: The
semi-major axes show no evidence of perturbations. {\it Top Right}:
The eccentricities experience simple sinusoidal variations. The period
of planet b oscillations is 100,000 years, while c and d oscillate on
a 7000 year period. The large amplitude of $e_b$ is most likely
unphysical due to tidal locking with the parent star. The apparently
irregular behavior of $e_b$ is an artifact of its long cycle. {\it
Bottom Left}: The inclinations are also regular, although planet b's
inclination is affected by both planets on its 20,000 year
period. Planets c and d oscillate in inclination on a 4000 year
period. As in eccentricity, the slightly chaotic appearance of
$i_b$ is an artifact of the sampling time convolved with the physical
period. {\it Bottom Right}: The $\Lambda$ distribution librates with
an amplitude of $50^o$.}}
\medskip

In \Fig 26 the evolution of a stable, but chaotic system, $\upsilon$
And-236, is presented. In this configuration $a_d$ is mildly
perturbed, but the eccentricities and inclinations undergo wild
fluctuations. $\Lambda$ also shows evidence of chaos as it librates
and circulates.

\medskip
\epsfxsize=8truecm
\epsfbox{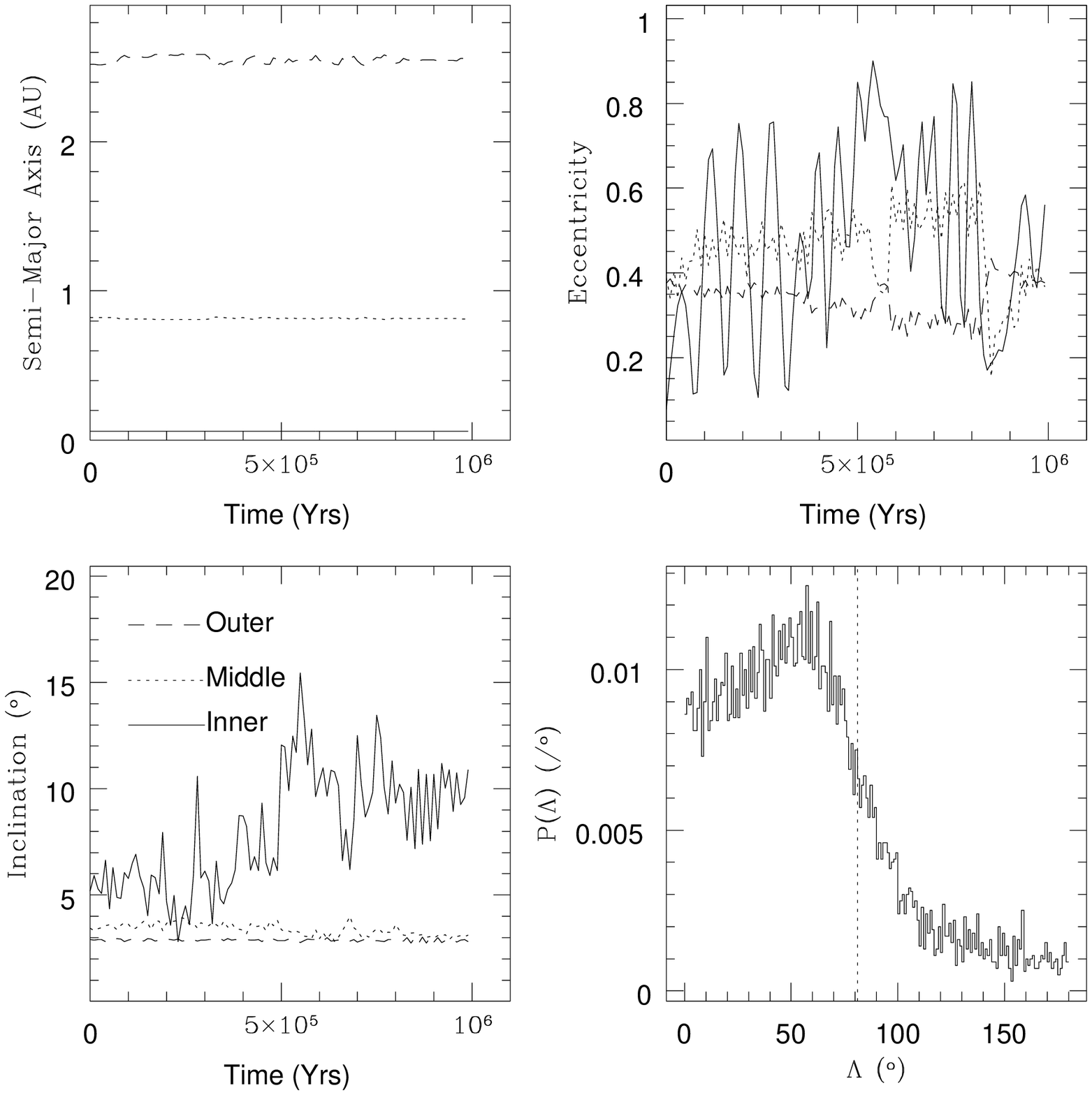}
\figcaption[upsand_236.bw.ps]{\label{fig:asymptotic}
\small{The orbital evolution of $\upsilon$ And-236, a stable, chaotic
configuration. {\it Top Left}: The semi-major axes show perturbations
of planets c and d. {\it Top Right}: The eccentricities are quite
chaotic, especially $e_b$. However, we see very large eccentricities,
and low semi-major axes for planet b. This again suggests that these
large values may be unphysical. However, this does not change the fact
that this system is chaotic. {\it Bottom Left}: The inclinations also
show a large degree of chaos. As with the eccentricities, it is b
which shows the most deviations from regular motion. {\it Bottom
Right}: The $\Lambda$ distribution suggests that this system librates
and circulates. This is the typical behavior of a chaotic system.}}
\medskip

$\upsilon$ And may eject any planet. However, as demonstrated in
\Fig 25, the model breaks down for large $e_b$. The path to ejection
for one such ejection of planet b is shown in \Fig 27. The
eccentricities stay generally low (<0.2) for nearly 200,000 years, but
prior to ejection $e_b$ grows to 0.99 at times, before finally being
ejected.

\medskip
\epsfxsize=8truecm
\epsfbox{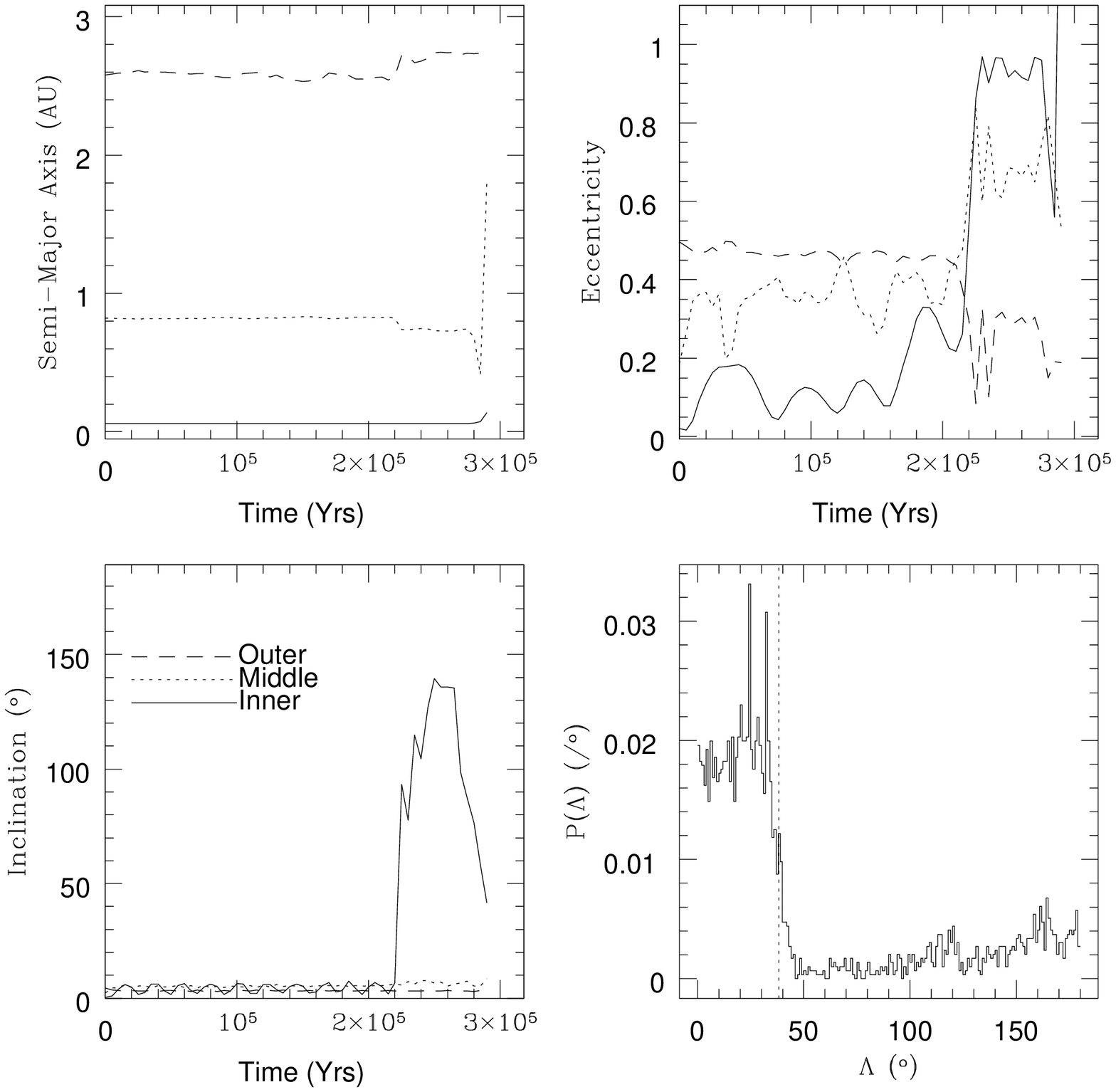}
\figcaption[upsand_000.bw.ps]{\label{fig:asymptotic}
\small{The orbital evolution of $\upsilon$ And-000, the ejection of
$\upsilon$ And b. {\it Top Left}: The system appears to be
experiencing perturbations right from the start as $a_d$ evolves
chaotically. Eventually a close approach between b and c produces an
ejection. {\it Top Right}: Once again we see that tidal
circularization would alter the outcome of this trial. Although it is
unclear if circularization would have been enough to prevent the close
approach that ejects planet b. {\it Bottom Left}: In contrast to
eccentricity, inclination evolves regularly for 200,000 years, when a
close approach sends b into a retrograde orbit, while $i_c$ and $i_d$
continue to evolve sinusoidally. {\it Bottom Right}: This unusual
$\Lambda$ distribution function reveals further chaos in the system,
and, as is seen in other systems, appears to be the sum of a libration
mode, and a circulation mode.}}
\medskip

In \Fig 28 we show the orbital evolution of a system which perturbs
planet d, but ejects planet c. The behavior of the ejecting planet
reaching very large semi-major axis distances, and subsequently
returning, only to be ejected was also seen in Fig 17. Note, too, that
the peak in $e_b$ corresponds with $i_b$.

\medskip
\epsfxsize=8truecm
\epsfbox{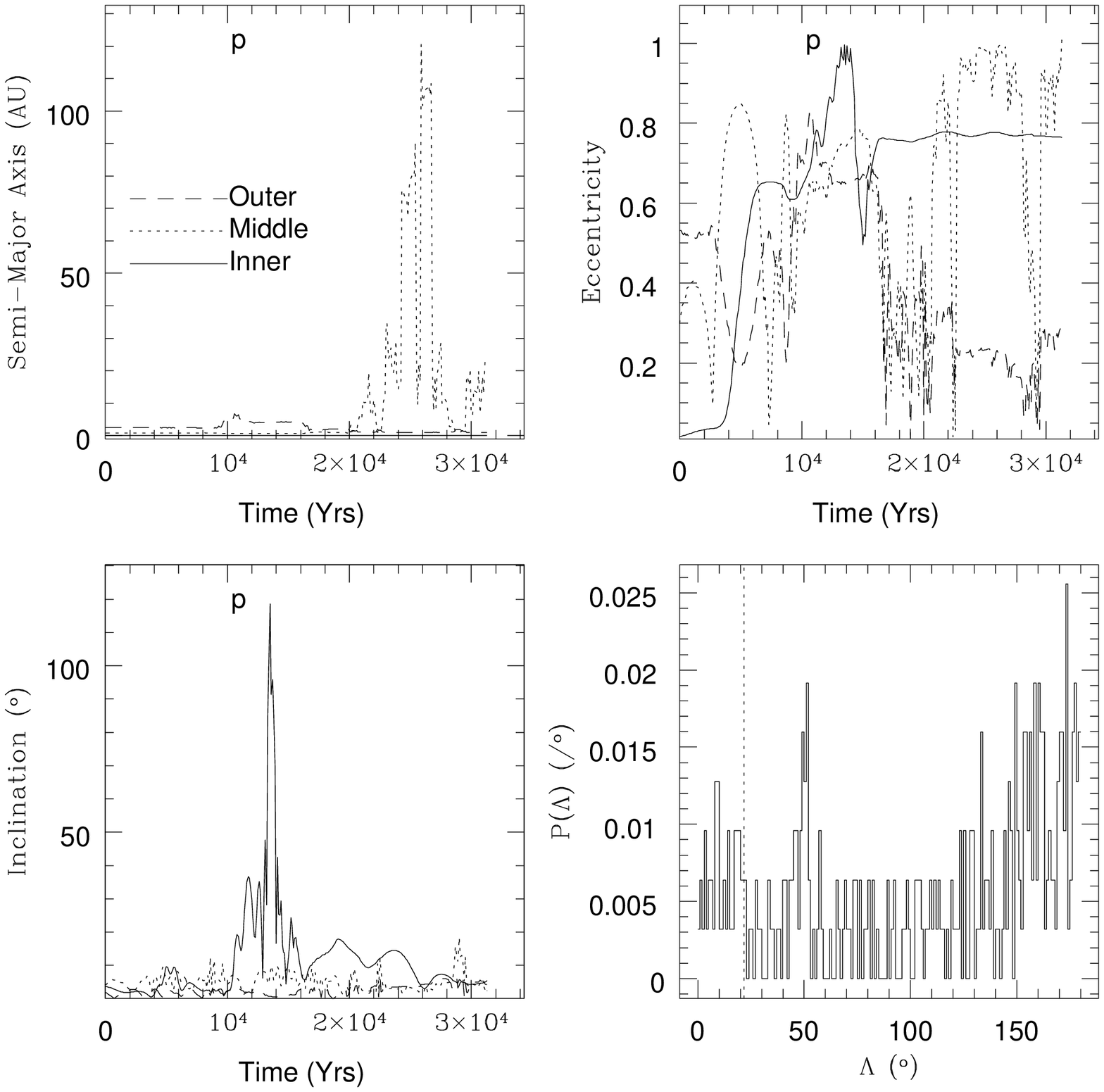}
\figcaption[upsand_020.bw.ps]{\label{fig:asymptotic}
\small{The orbital evolution of $\upsilon$ And-020, the ejection of
$\upsilon$ And c. {\it Top Left}: The semi-major axes evolve
quiescently for 10,000 years before perturbing planet d, marked by the
$p$. 10,000 years later planet c is perturbed to over 120AU. The
planet then returns to low a, but is quickly ejected after another
encounter with planet d. {\it Top Right}: This configuration
experiences wild oscillations from the very beginning.  {\it Bottom
Left}: The inclinations also suffer large, chaotic
fluctuations. Shortly after planet d is perturbed, planet b
experiences a short period of retrograde motion. {\it Bottom Right}:
Although poorly sampled, this graph clearly shows that $\Lambda$
evolves chaotically.}}
\medskip

Finally we show the ejection of planet d, the most massive planet in
the system in Fig 29. This system evolves quasi-regularly for 200,000
years before suddenly ejecting planet d. The only hints of chaos are
in the evolution of $e_c$ and $e_d$, although there are some slight
asymmetries in the sinusoidal evolution of $e_b$. $\Lambda$ is
generally librating, although in the final 20,000 years it does
circulate as the system destabilizes.

Long term simulations run for 100 million years. \Fig 30 is the
eccentricity evolution of 4 simulations. In one case (bottom right panel of
\Fig 30) the inner planet is ejected after 55 million years. The top
left panel shows a system undergoing chaotic evolution. The other two
panels show regular motion.

\medskip
\epsfxsize=8truecm
\epsfbox{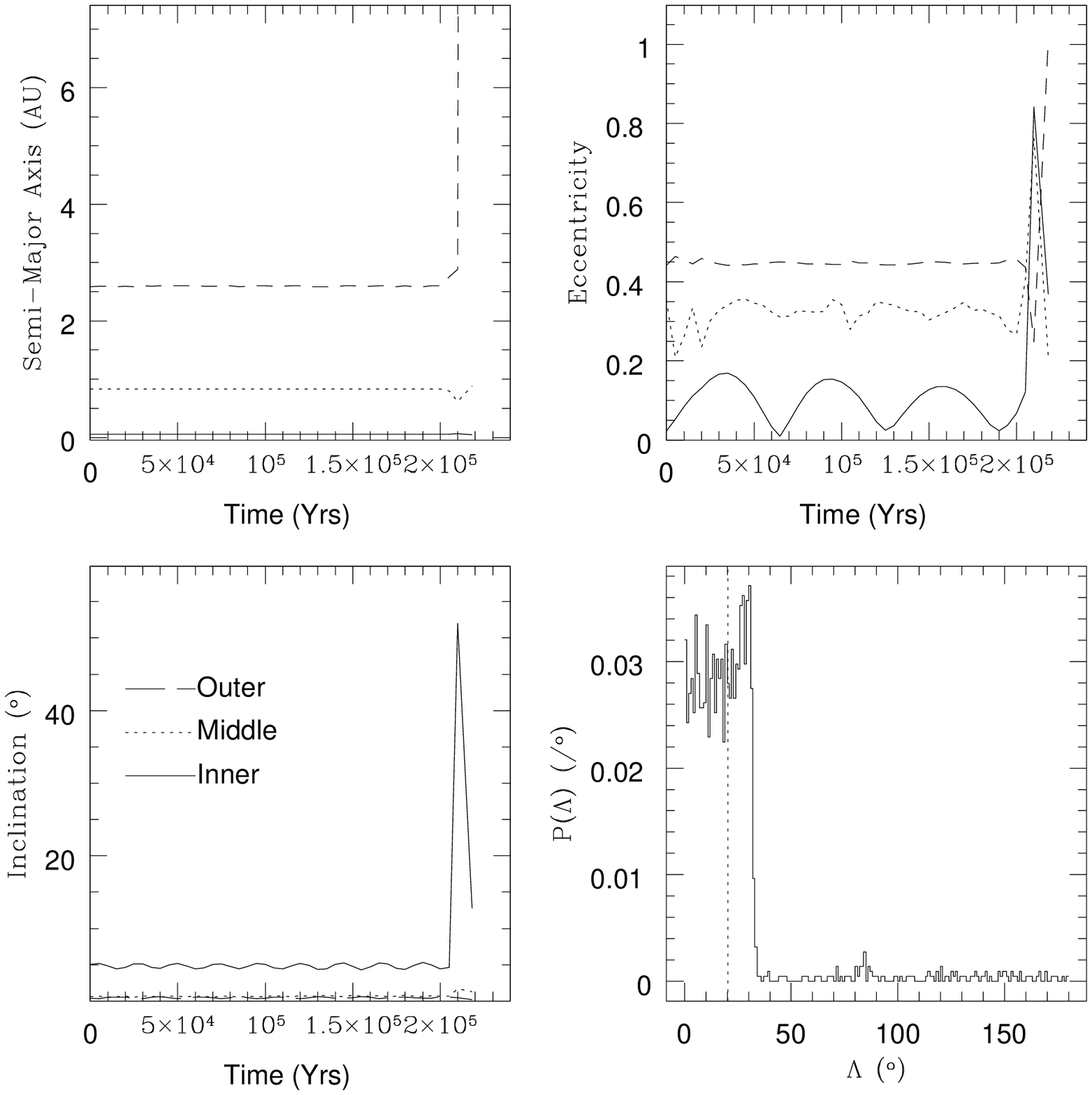}
\figcaption[upsand_014.bw.ps]{\label{fig:asymptotic}
\small{The orbital evolution of $\upsilon$ And-014, the ejection of
$\upsilon$ And d. {\it Top Left}: The semi-major axes evolve without
perturbation for 200,000 years before planet d is suddenly flung from
the system. {\it Top Right}: This configuration experiences low
amplitude chaos in its eccentricity until it suddenly experiences a
close approach at 200,000 years. {\it Bottom Left}: The inclinations,
however, evolve regularly until a close approach sends b into a $45^o$
orbit. A second close approach kicks it back down, but d is ejected
before it returns to the orbital plane. {\it Bottom Right}: This plot
shows the typical behavior of $\Lambda$; it librates while motion is
regular, but circulates after close approaches destabilize the
system. Here, the $\Lambda$ distribution represents 200,000 years of
libration and 25,000 years of circulation.}}
\medskip

\medskip 
\epsfxsize=8truecm 
\epsfbox{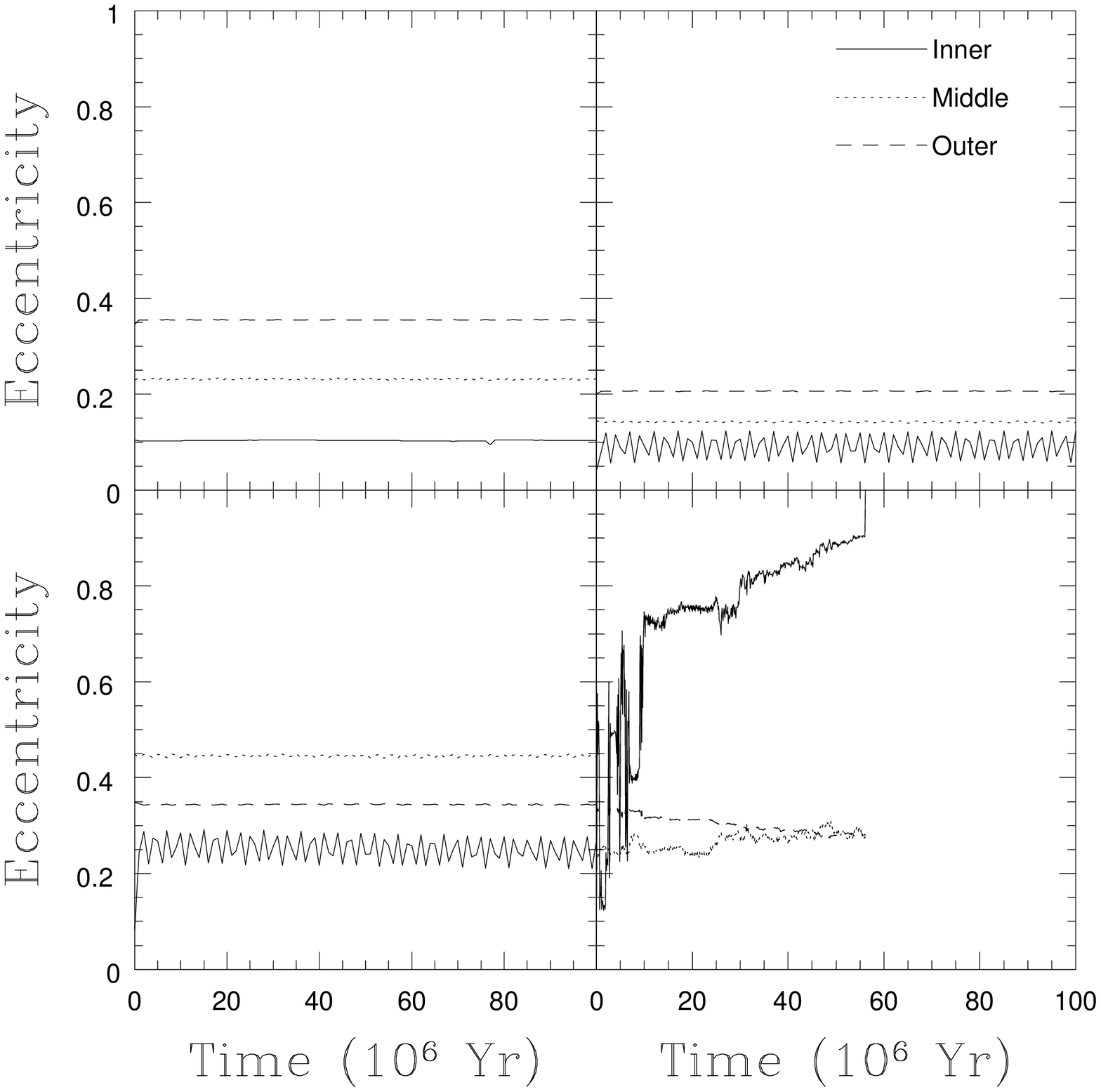}
\figcaption[upsand.long.bw.ps]{\label{fig:asymptotic} 
\small{The long-term eccentricity evolution of 4 simulations of $\upsilon$
And. {\it Top Left:} Eccentricity evolution of $\upsilon$
And-006. This is an example of regular evolution. {\it Top Right:} Eccentricity evolution of $\upsilon$
And-054. This is a chaotic configuration. The chaos has been
mostly smoothed over, though, as the data represent 10,000 year
averages. {\it Bottom Left:}
Eccentricity evolution of simulation $\upsilon$ And-288. Another chaotic
configuration. {\it Bottom Right:} Eccentricity evolution of $\upsilon$
And-192. A chaotic system which ejects the inner planet after 55
million years. This evolution is suspect as the effects of tidal
circularization undoubtedly play a role in the evolution of $e_b$.}}
\medskip

\Fig 31 shows the $\Lambda$ distribution of these
configurations. Secular resonance locking does not occur in the
$\upsilon$ And system; the ellipses tend to be anti-aligned. This is
the same result as in Paper I.  However there do appear to be several
different modes of stability for the apses in this system. The panels
in \Fig 31 correspond to those in \Fig 30, therefore the top left is
the regular case. The chaotic systems (top right, bottom left) show
the $\Lambda$ distribution signature of chaos, as does the example
which ejects planet b (bottom right). In Table 8 we
present a summary of all long term simulations for $\upsilon$ And.

\medskip
\epsfxsize=8truecm
\epsfbox{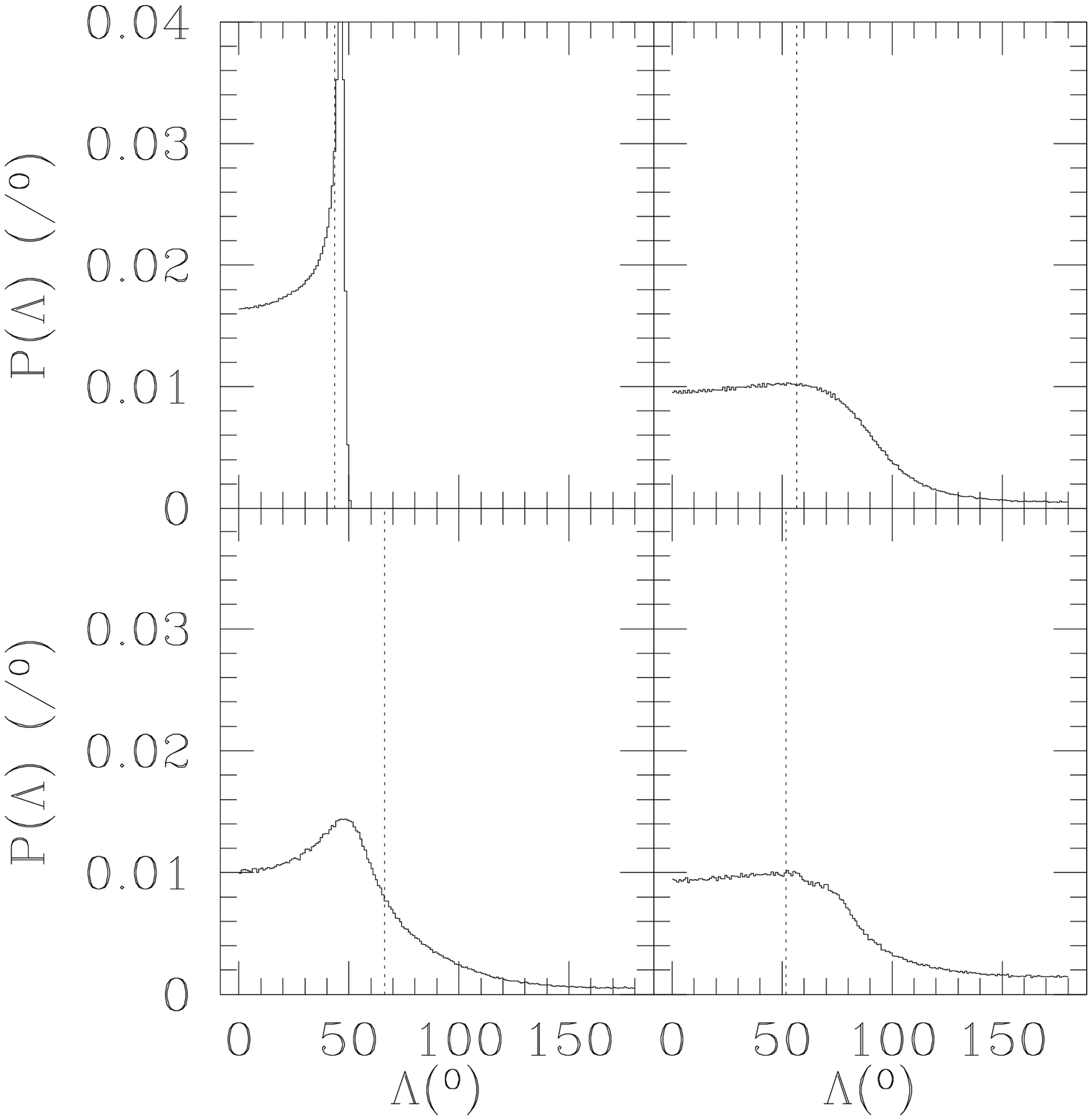}
\figcaption[upsand.lamprob.ps]{\label{fig:asymptotic}
\small{The long-term $\Lambda$ distribution of the same 4 $\upsilon$ And
simulations as in \Fig 30. {\it Top Left:} This $\Lambda$ distribution
is typical of a regular system. {\it Other Panels:} The distributions
are typical of chaotic configurations. From these plots we conclude
the secular resonance locking is important in maintaining stability,
or regularness in $\upsilon$ And.}}
\medskip

\begin{figure*}
\psfig{file=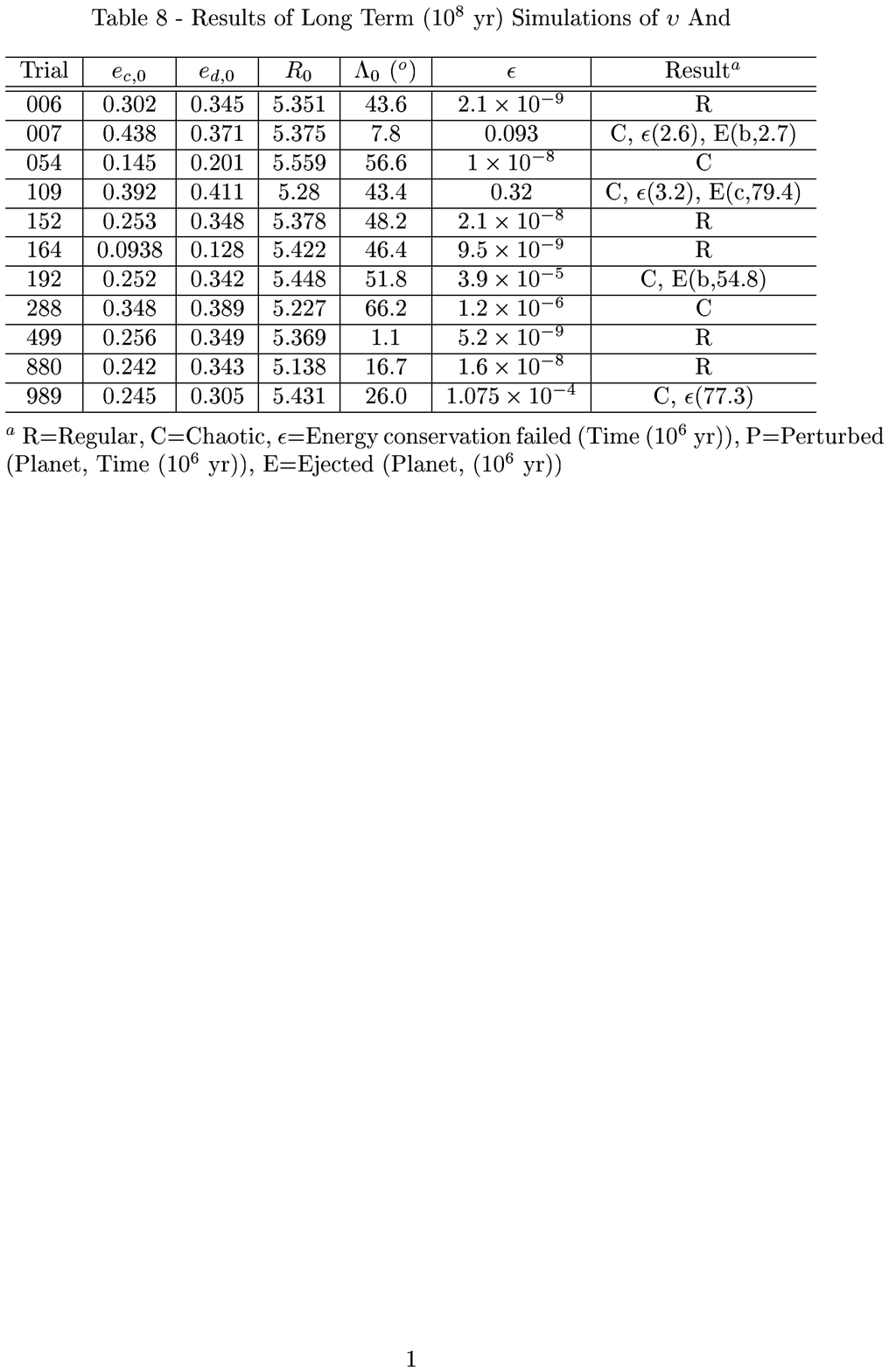,width=19.truecm}
\end{figure*}

\subsection{47UMa}
The 47UMa system consists of a $1.03\pm 0.03M_{\odot}$ (Gonzalez 1998)
star and 2 interacting companions: b and c, at 2.09AU and 3.73AU
respectively. The initial eccentricities in this system are substantially
lower than $\upsilon$ And at 0.06 and 0.1\footnote{http://exoplanets.org}
respectively. More recent measurements place $e_c$ much closer to 0.
However, ``$e_c$=0.3 provides almost as good a fit to the radial velocity
data as does $e_c$=0.005'' (Laughlin, Chambers, \& Fischer 2002).  Should
$e_c \le 0.1$ then, of all the systems examined in this paper, 47UMa most
closely resembles our own. Planet b is the larger planet and hence
determines $\tau_{47UMa}$=840,000 years.

Overall 80.3$\pm$4.7\% of simulations were stable over $\tau_{47UMa}$.
This is less than a 2$\sigma$ difference from $\upsilon$ And. In
unstable configurations planet c, the less massive planet, was ejected
every time. This is similar to $\upsilon$ And in which the most
massive planet perturbed the smaller planets. The instability rate as
a function of time is presented in \Fig 32. The rate for 47UMa is
similar to the other systems in that most unstable configurations
perturb a planet past stability on relatively short timescales, but
with a tail to longer times. The rate does not reach zero, however,
and suggests that more unstable configurations exist.

\medskip
\epsfxsize=8truecm
\epsfbox{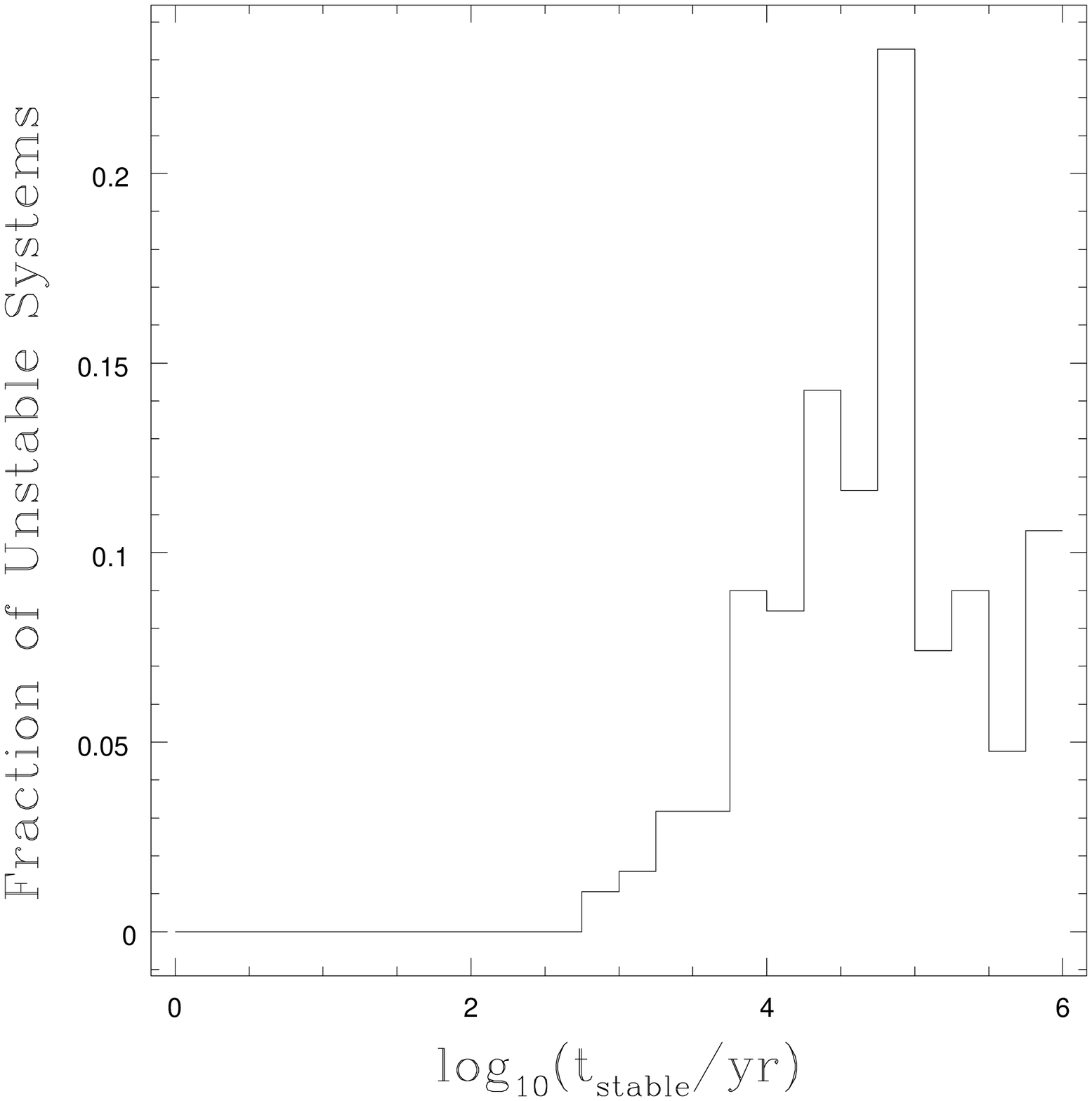}
\figcaption[47uma.ejrate.ps]{\label{fig:asymptotic}
\small{Instability rate in 47UMa. Most ejections occur at 10,000 to
100,000 years. In this system, even unstable configurations generally
survive for at least 10,000 years.}}
\medskip

The 47UMa stability map is presented in \Fig 33. The overall
structure of stability in eccentricity space is qualitatively the same
as in $\upsilon$ And, with one major exception: stability is
correlated with $e_2$ ($e_c$), not $e_1$ ($e_b$). The errors for $e_c$
are substantial. The main difference between 47UMa and $\upsilon$ And
is that the more massive companion is the interior planet. This
configuration makes it more difficult for the exterior planet to
perturb the interior planet which is more tightly bound to the parent
star.

This stability map is in good agreement with other work done on this
system. Using MEGNO, Go\'zdziewski (2002) determined the maximum value for
$e_c$ to be approximately 0.15. A stability analysis in
Laughlin, Chambers, \& Fischer (2002) also shows a similar dependence on
$e_c$. For nearly coplanar systems, such as those presented here, they
determined the maximum value for $e_c$ to be less than 0.2. Although the
exact maximum value for $e_c$ is different for all three studies, it is clear
that the value of $e_c$ determines stability for 47UMa.

When comparing \Fig 33 with \Fig 22, we see that the edge in the
$\upsilon$ And system is much cleaner than in 47UMa. One possible reason
for this is the system's proximity to the 5:2 mean motion resonance. To
examine this possibility, in \Fig 34 we plot stability as a function of
$R$ in this system. Although there appears to be some increase in stability
beyond 5:2, and a decrease inside 5:2, the errors are too large to confirm
that this is a real effect.

\medskip
\epsfxsize=8truecm
\epsfbox{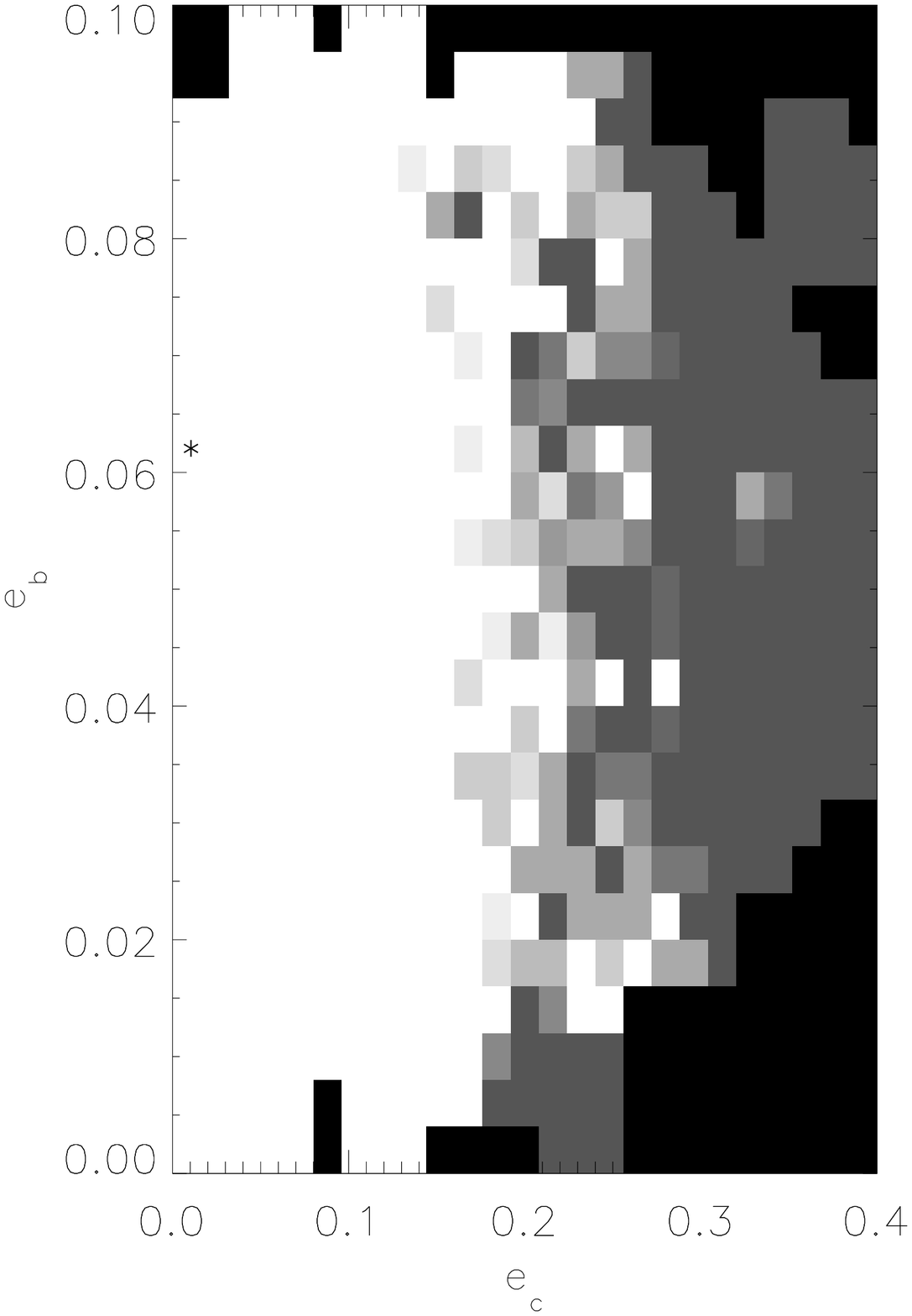}
\figcaption[47uma.ecc.grey.ps]{\label{fig:asymptotic}
\small{The stability map for 47UMa. The relevant orbital elements are $e_b$
and $e_c$. For this system the value of $e_c$ determines
stability. The current best fit to this system, as of 1 Nov 2002, is
marked by the asterisk (Laughlin, Chambers, \& Fischer 2002). Note
that although this system appears to lie far from the edge, the
observational error for $e_c$ is $\pm$0.115.}} \medskip

In Figs. 35-37 we examine several possible orbital evolutions for
47UMa. Some sample simulations from the suite of 1000 are listed in
Table 9. First, in \Fig 35, is a stable regular configuration. Over
50\% of all configurations for this system are regular. In this
case the eccentricities, and inclinations show very small oscillations
(~10\%) and $\Lambda$ librates with an amplitude of $45^o$.

\medskip
\epsfxsize=8truecm
\epsfbox{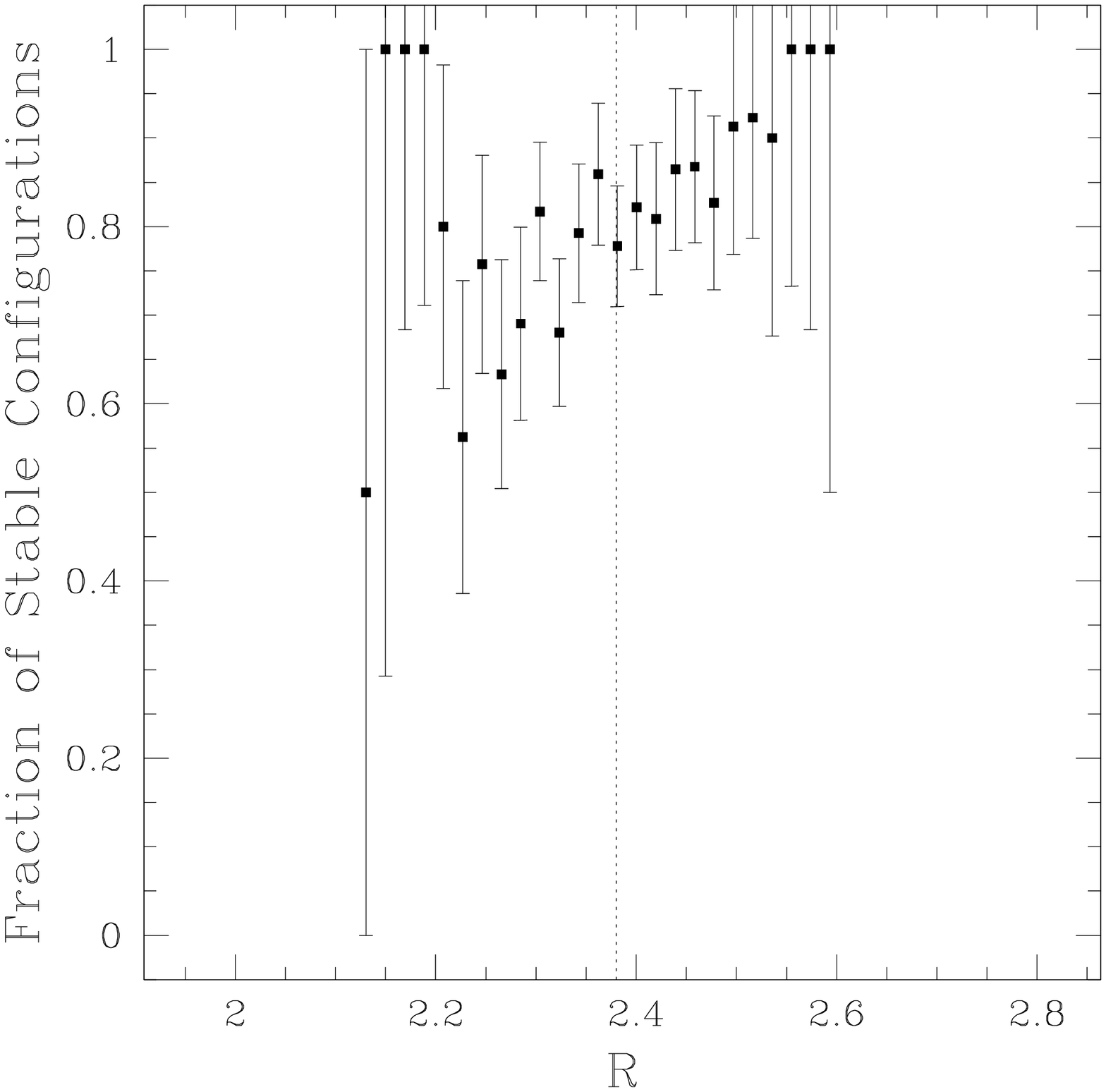}
\figcaption[47uma.res.dyn.ps]{\label{fig:asymptotic}
\small{The stability of 47UMa as a function of $R$. This system lies very
close to the 5:2 mean motion resonance. There is some indication of
more stability exterior to 5:2, and less interior, but the statistics
are not robust enough to confirm this. The dashed line represents the
current best fit to the system.}}
\medskip

\begin{figure*}
\psfig{file=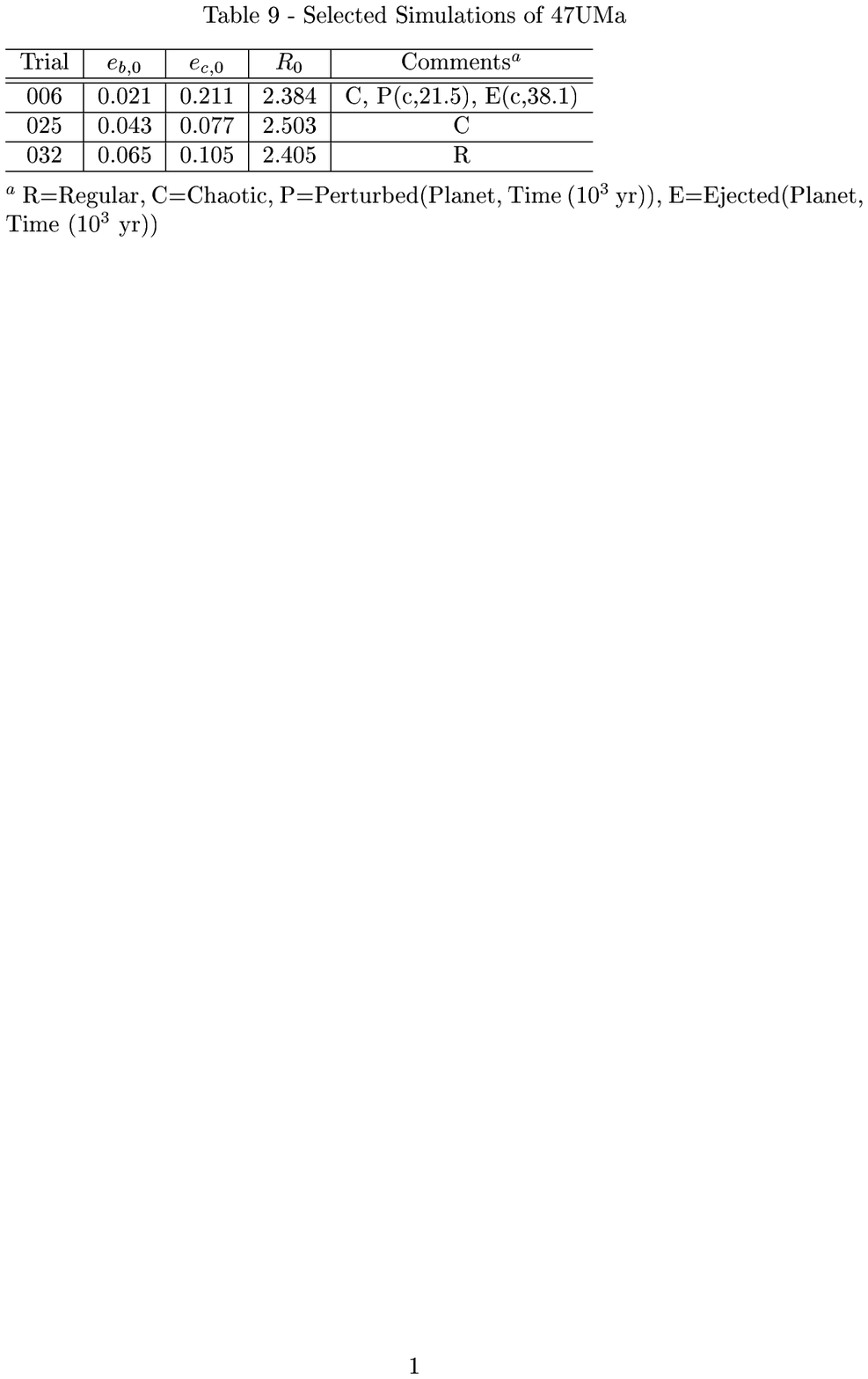,width=19.truecm}
\end{figure*}

\medskip
\epsfxsize=8truecm
\epsfbox{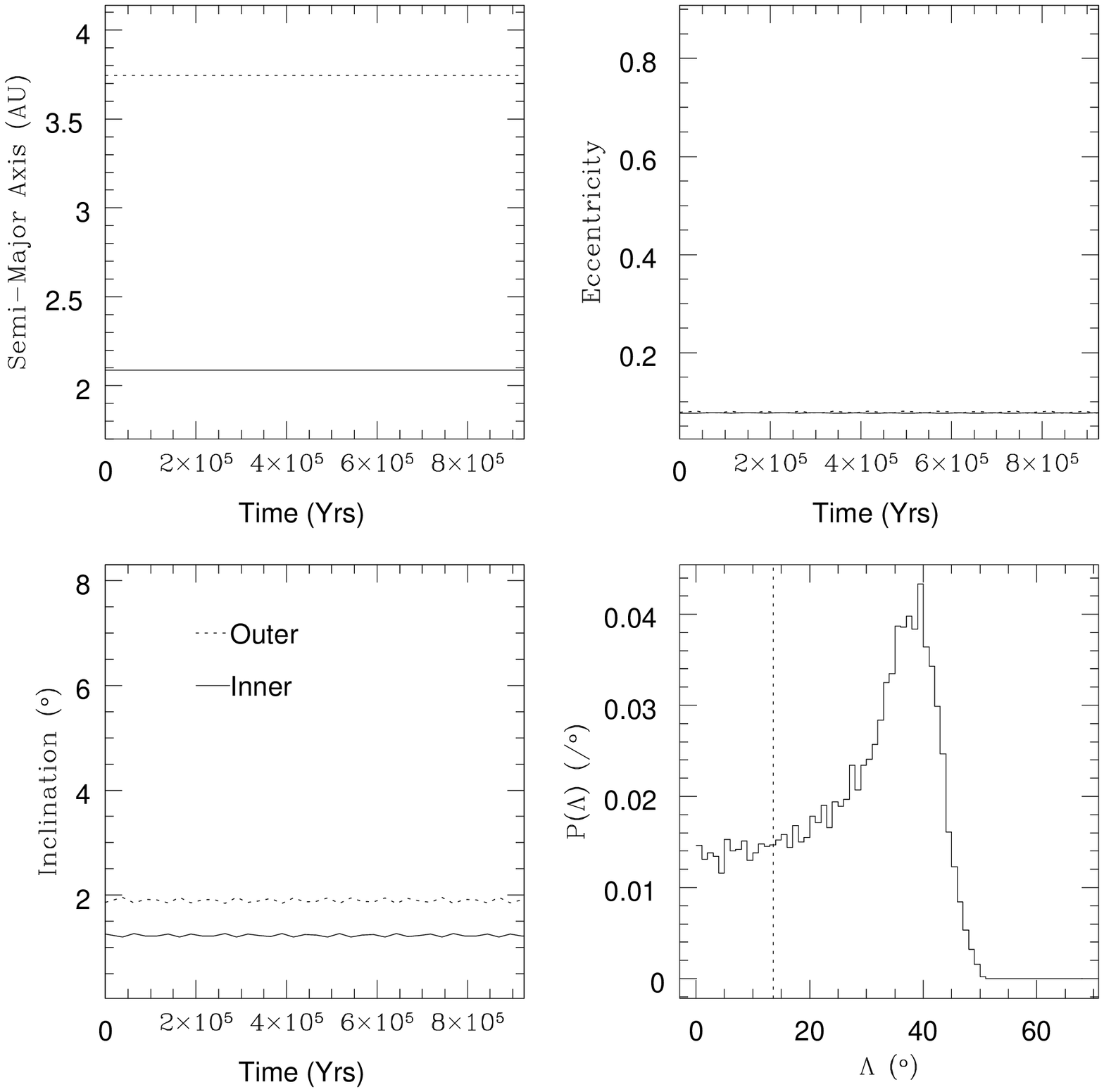}
\figcaption[47uma_032.smooth.ps]{\label{fig:asymptotic}
\small{The orbital evolution of 47UMa-032, a stable, regular configuration
of 47UMa. The data are smoothed on 25,000 year intervals.{\it Top
Left}: The semi-major axes show no hint of perturbations. {\it Top
Right}: The eccentricities vary on a 4800 year timescale for the
duration of the simulation. {\it Bottom Left}: This system never
deviates more than $5^o$ from coplanarity. The inclinations oscillate
on a 4400 year period. {\it Bottom Right}: In this configuration
$\Lambda$ is aligned, but librates with an amplitude of $45^o$.}}
\medskip

As with all other systems, there are chaotic, but stable examples of
47UMa. One such system, 47UMa-025, is examined in \Fig 36. The
semi-major axes of this configuration show no signs of perturbations,
but the eccentricities and inclinations show clear signs of chaos,
though no perturbation is very large. As with stable, chaotic
configurations of other systems, circulation appears to be the
dominant mode of $\Lambda$ evolution. However, this system is actually
librating about an anti-aligned orientation, with only occasional
circulation. The amplitude of this libration is quite variable,
sometimes as high as $165^o$.

In \Fig 37, we present an example of the ejection of planet
c. This system evolves regularly for 10,000 years, then experiences
25,000 years of chaotic evolution, culminating in the ejection of the
outer planet. For this configuration $\Lambda$ circulates during the
first 30,000 years, both regular and chaotic epochs, but with
different angular velocities in each stage. During the final 10,000
years, however, when $e_c$ becomes very large, $\Lambda$ becomes
locked at $155^o$.

\medskip
\epsfxsize=8truecm
\epsfbox{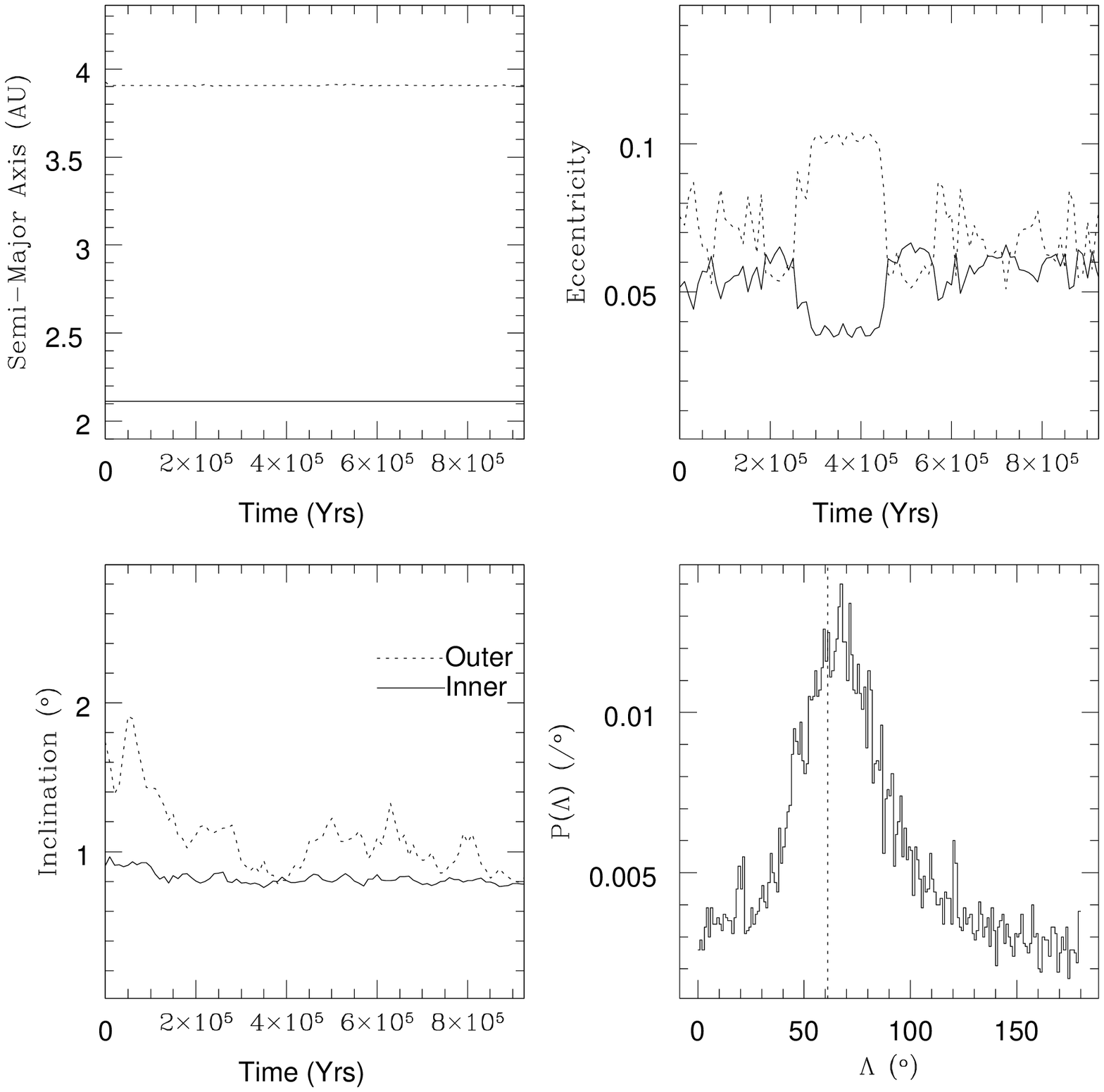}
\figcaption[47uma_025.bw.ps]{\label{fig:asymptotic}
\small{The orbital evolution of 47UMa-025, a stable, chaotic configuration
of 47UMa. {\it Top Left}: The semi-major axes show no hint of
perturbations. {\it Top Right}: The eccentricities vary by only 25\%
for the duration of the simulation, but show clear coupled, chaotic
behavior. {\it Bottom Left}: This nearly coplanar simulation also
shows obvious chaos in the evolution of the inclinations. {\it Bottom
Right}: This $\Lambda$ distribution function is libration about
anti-alignment with an amplitude which varies between $100^o$ and
$165^0$, but also with occasional circulation.}}
\medskip

\medskip
\epsfxsize=8truecm
\epsfbox{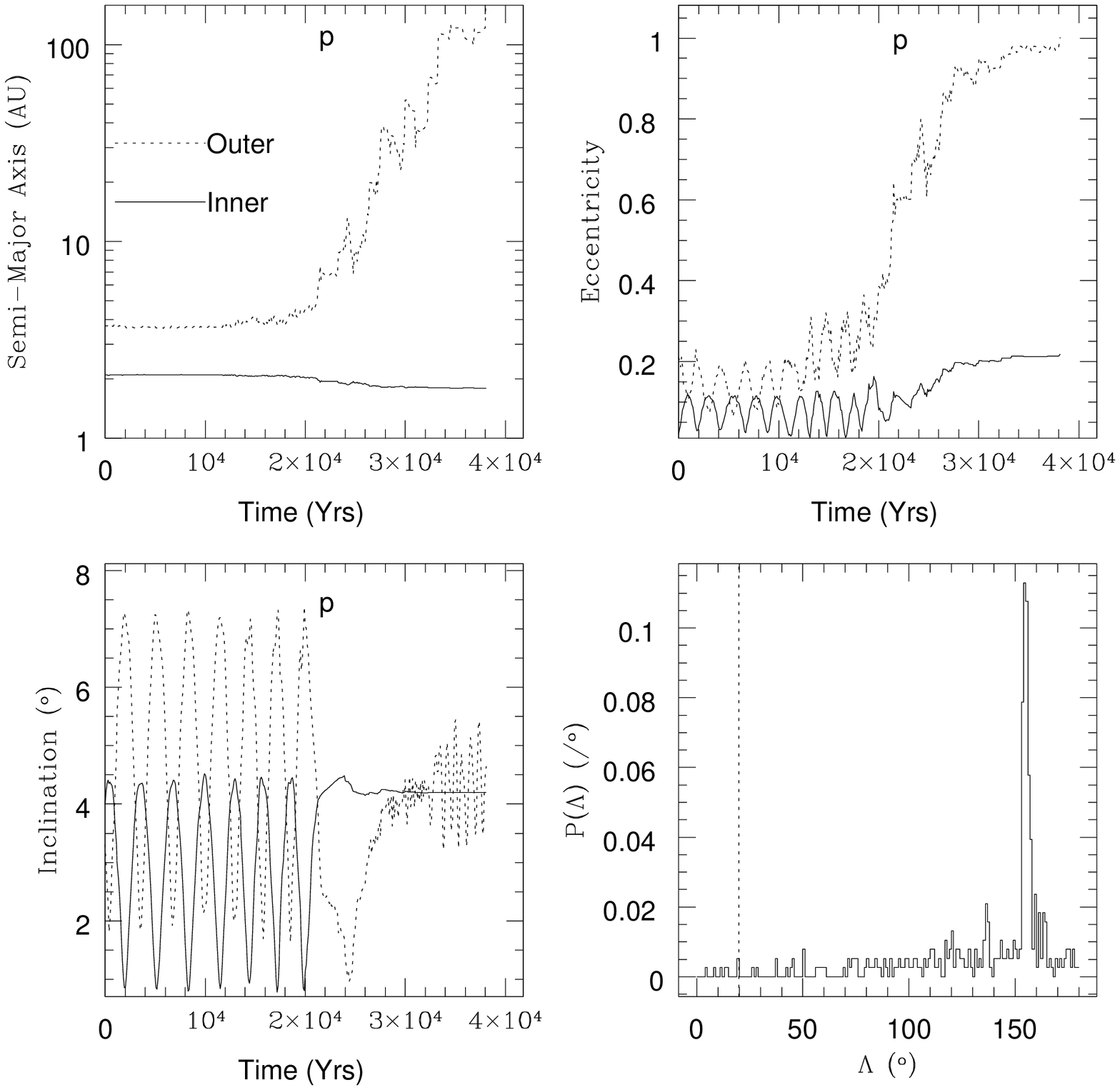}
\figcaption[47uma_006.bw.ps]{\label{fig:asymptotic}
\small{The orbital evolution of 47UMa-006, a chaotic, unstable
configuration of 47UMa. The $p$ marks the time that our perturbation
criterion is met. {\it Top Left}: The semi-major axes evolve
quiescently for 10,000 years, then $e_c$ begins its slow trek to
upward of 200AU. The system is perturbed, however, until 21,500
years. Note that the y-axis is logarithmic in this example. {\it Top
Right}: After 10,000 years, the system suddenly becomes chaotic,
eventually pushing $e_c$ to unity in 40,000 years. {\it Bottom Left}:
As with the eccentricities, the inclinations evolve regularly for
10,000 years, but then become chaotic. {\it Bottom Right}: In this
configuration $\Lambda$ circulates, but eventually becomes locked at
$155^o$ for the final 10,000 years.}}
\medskip

\begin{figure*}
\psfig{file=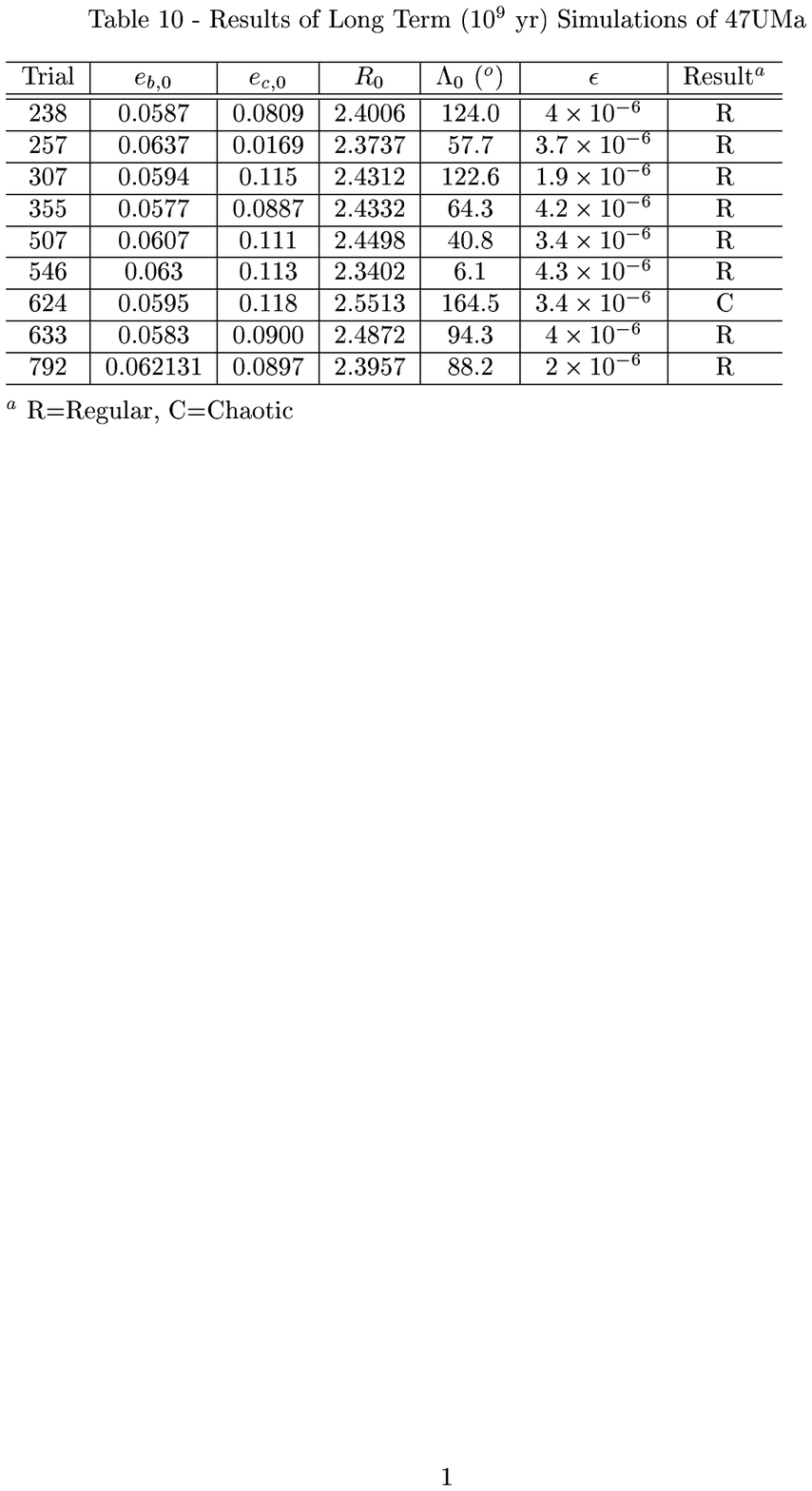,width=19.truecm}
\end{figure*}

\medskip 
\epsfxsize=8truecm 
\epsfbox{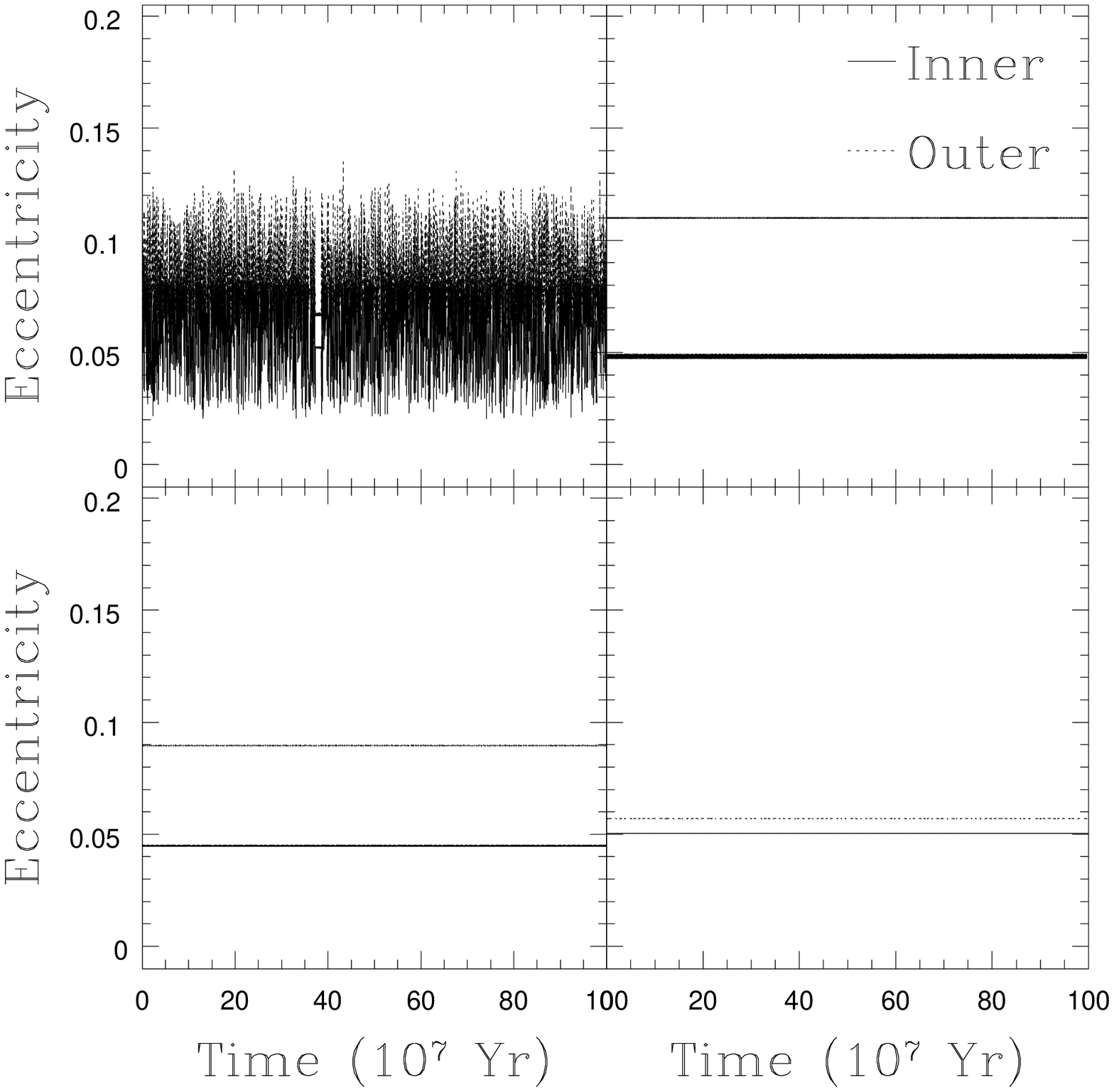}
\figcaption[47uma.long.bw.ps]{\label{fig:asymptotic} \small{Eccentricity
evolution of 4 long term simulations of 47UMa. {\it Top Left}:
Evolution of 47UMa-624. {\it Top Right}: Evolution of 47UMa-307. {\it
Bottom Left}: Evolution of 47UMa-238. {\it Bottom Right}: Evolution of
47UMa-257. Most systems are regular for the duration of the
simulation. However the top right is clearly chaotic, yet it still
survives for $10^9$ years.}} \medskip

Long term simulations of 47UMa were integrated for $10^9$ years. Table
10 is a summary of all long term simulations for 47UMa. \Fig 38
shows the eccentricity evolution of 4 systems. The top right panel of
this figure is fascinating. The system is very chaotic for the first
350 million years, then enters a short ($\sim$20 million years) period
of quiescence, only to return again to a similar chaotic state. The
other configurations all appear regular. \Fig 39, plots the
$\Lambda$ evolution. In 47UMa no secular resonance locking
occurs. However these plots confirm the results of Laughlin, Chambers,
\& Fischer (2002). They show that for low values of $e_c$ ($<0.1$) the
system should librate in an aligned configuration, but above 0.1 the
system should be anti-aligned. The chaotic case, as expected, has a
flat distribution function. 

\medskip
\epsfxsize=8truecm
\epsfbox{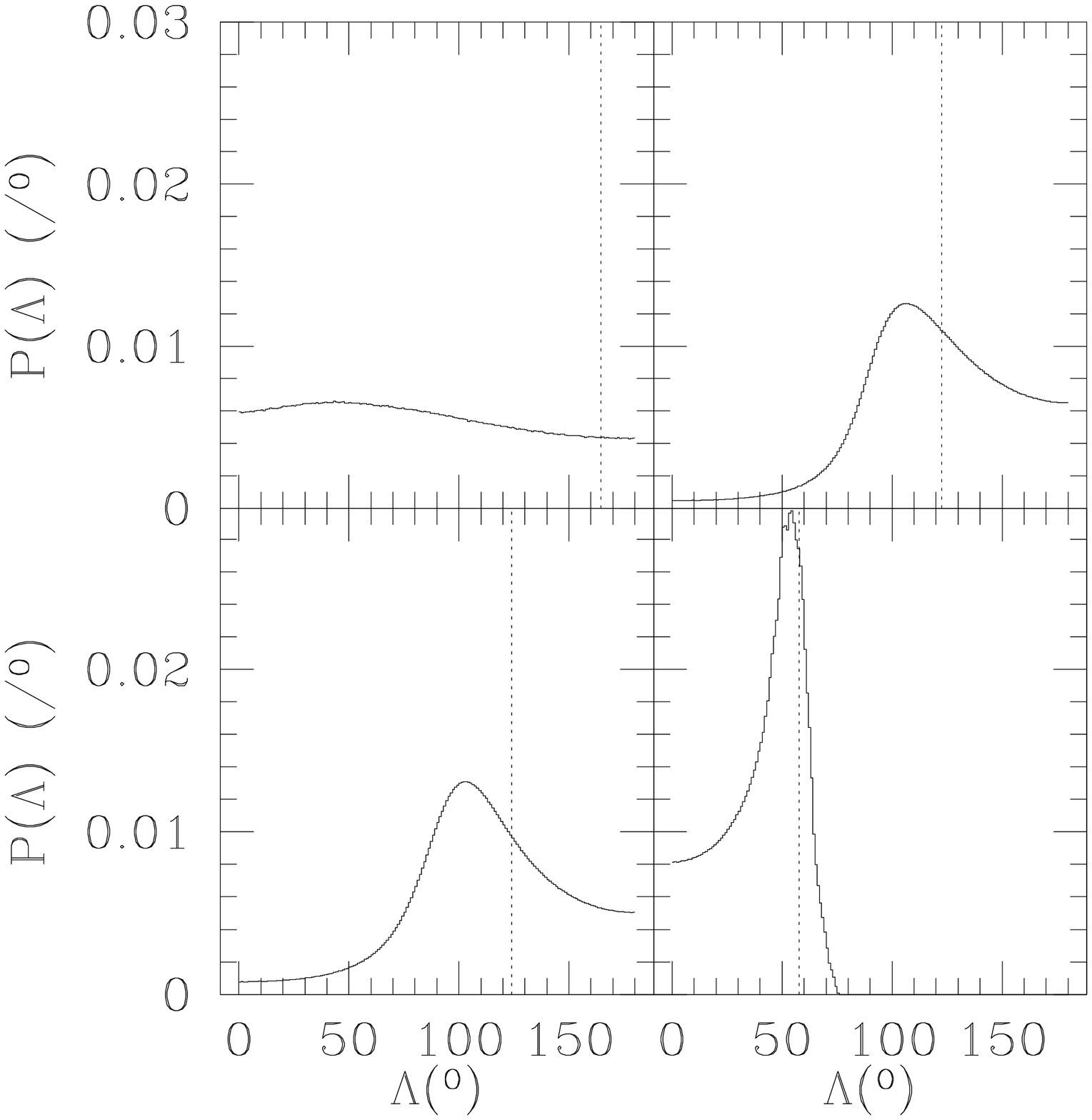}
\figcaption[47uma.lamprob.ps]{\label{fig:asymptotic}
\small{The distribution of $\Lambda$ for 4 stable long term simulations of
47UMa. The chaotic system (top left) shows a nearly flat distribution
in $\Lambda$. This suggests that $\Lambda$} is behaving chaotically as
well. Two regular systems (top right, and bottom left) show libration
about anti-parallel configurations, whereas the bottom right librates
about $\Lambda = 0$.}
\medskip

\subsection{Gas Giants}
Perhaps the most interesting aspect of the new
ESPS is their comparison to the SS. The procedure outlined in $\S$2
permits a comparison of the ESPS and the SS. We must first, however,
determine how to vary the initial conditions of the gas giants. As can
be seen in Table 6, the ``error'' associated with each planet is
arbitrary.  We have given a spread in initial conditions that is
roughly similar to the percent error as listed in the ESPS. For
example the periods are allowed to vary by approximately 10\%, but the
eccentricities have a standard deviation of 0.1 for all planets. This
procedure will allow us to create a stability map for the SS, but will
make a comparison of the probabilities of survival less meaningful.

The outer SS consists of 4 gas giants located between 5.2 and
40AU. The gas giants are on much more circular orbits than the ESPS
planets (Saturn has the largest eccentricity at just over
0.05). Because of the large semi-major axes, $\tau_{SS}$ corresponds
to $3.32\times 10^6$ years. Compared to known ESPS the gas giants are
relatively low-mass planets. In fact Uranus and Neptune could not be
detected via the Doppler method at the precision level currently
achieved (see $\S$4.4).

Chaos in the SS is well documented (e.g. Sussman \& Wisdom 1988, Sussman \&
Wisdom 1992, Varadi \etal 1999, Lecar \etal 2001). In fact Varadi
\etal (1999) show that the Jupiter-Saturn system lies very near
chaotic regions. They vary the semi-major axes of these two planets
slightly (less than 1\%) and find that this is enough to identify
broad chaotic regions. Below we show that by enlarging this variation,
the system moves into total instability; ejections are inevitable.

For the gas giants 85.3\%$\pm$4.3\% of the trials were unperturbed for
3.3$\times 10^6$ years. In Paper I, we integrated 32 gas giant
configurations. Of these 81\% survived. As in $\S$4.1 we again recover
the results of Paper I. In this system Jupiter was never ejected; it
always removed the other planets from the system. Saturn was ejected in
14\% of the simulations, Uranus, (the least massive) 62\%, and
Neptune, 24\%. In the SS therefore, ejection rate is tightly coupled
to mass, as was observed in the other ESPS. It therefore seems that
the SS behaves like other interacting systems.

\Fig 40 is the instability rate. Most configurations survive for
$10^5$ years. Once again it appears the perturbation rate does not
fall to zero, and we note that this means that we have not found all
the unstable systems.  The last bin in this plot contains only 20,000
years worth of data. It is therefore unclear if the ejection rate
might still be rising with larger time. Should that be the case that
our choice for $\tau_{SS}$ is too small, and suggests that we have not
identified all unstable configurations.  \Fig 41 is the stability
map for the gas giants. The gas giants show a plateau as in 47UMa and
in $\upsilon$ And, however the edge is much less dramatic.  The actual
values for our gas giants shows that our system lies quite far from
the stability edge. We also note that the stability plateau shows many
depressions, and the abyss contains many spires. This apparent
difference between the SS and other interacting systems may result
from not determining all unstable configurations. Perhaps longer
integrations would sharpen the edge, and broaden the instability
abyss. Dynamically speaking the major difference between the SS and
other interacting systems is that the SS has a much broader range of
orbital times. Jupiter orbits 13 times more quickly that
Neptune. Perhaps instability is more relevant on timescale based on
the orbit of Neptune (see $\S6$).  However these features may also
arise from the system's proximity to the 5:2 resonance, the so called
``Great Inequality''.  We address this possibility in $\S$4.4.

\medskip
\epsfxsize=8truecm
\epsfbox{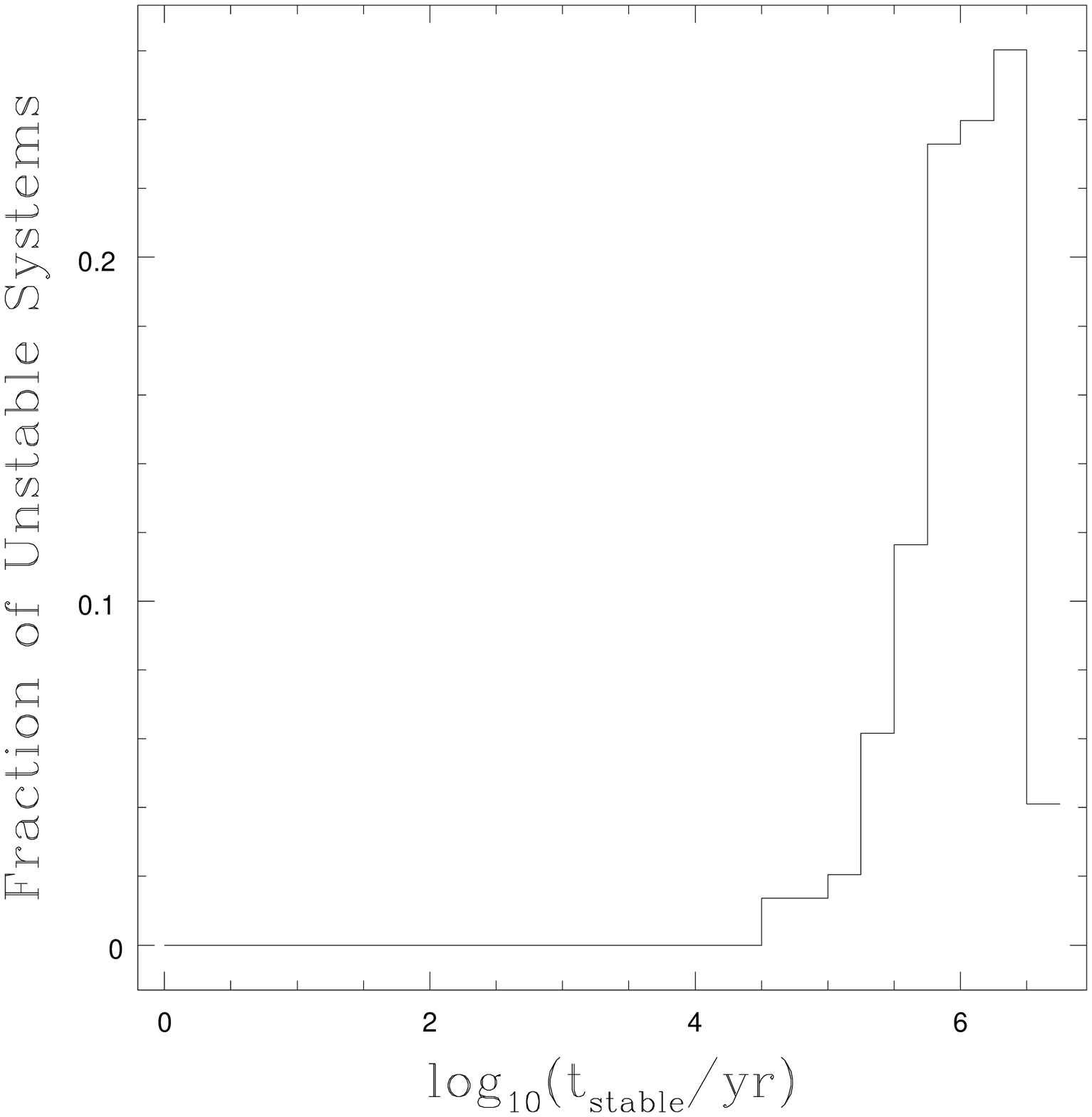}
\figcaption[ss.ejrate.ps]{\label{fig:asymptotic}
\small{Rate of instability in the SS. Instability requires at least 30,000
years to develop, and continues through $\tau_{SS}$.}}
\medskip

Several example simulations are shown in Figs.\ 42-47. The initial
conditions and outcomes of these simulations are shown in Table 11. A
regular, stable example is shown in \Fig 42. The eccentricities and
inclinations are a linear superposition of eigenmodes. Although
$\Lambda$ begins nearly anti-aligned, it quickly moves into a large
amplitude librational mode. The circulation, which is typically an
indicator of chaos, appears to not influence the eccentricity
evolution.

The best fit to $e_J$, $e_S$ and $R$ is simulation SS-183. The orbital
evolution of this system is shown in \Fig 43. Note, however, that
$\Lambda_0$ differs substantially from its standard value of
$68.5^o$. This difference is responsible for the chaotic evolution of
$e_J$ and $e_S$. Curiously, though, the evolution of $e_U$ and $e_N$
are regular. Unlike the eccentricities, all the inclinations evolve
regularly. The nodes of Uranus and Neptune librate about
anti-alignment, but with occasional circulation. Note that in $e$ and
$i$ Uranus oscillates from 2 modes, whereas Neptune experiences 3
modes.

\medskip
\epsfxsize=8truecm
\epsfbox{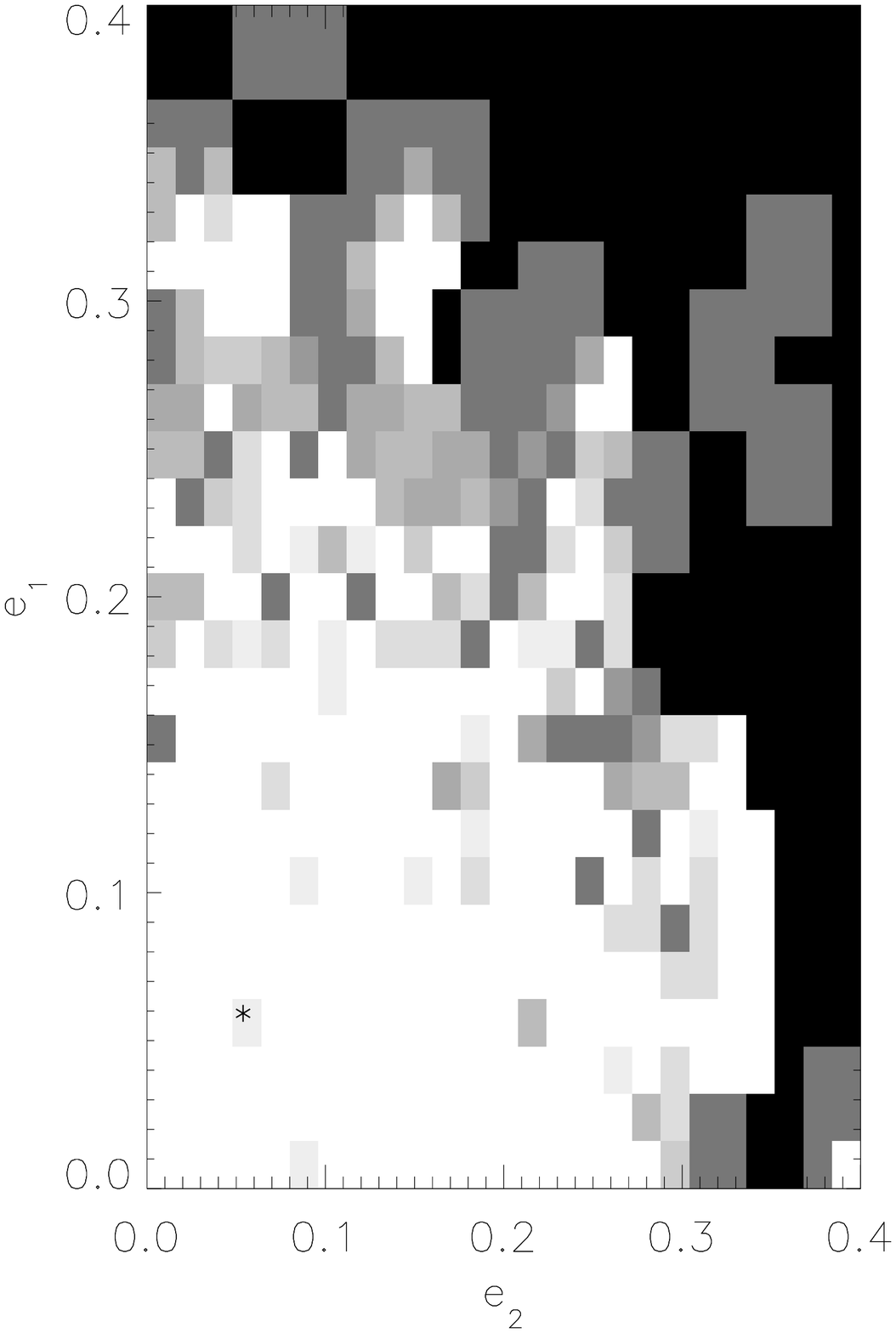}
\figcaption[ss.ecc.grey.ps]{\label{fig:asymptotic}
\small{Stability map for the gas giants. In eccentricity space, the SS lies
near a small depression. The edge in the gas giant system is not
nearly as clean as in other interacting systems. This may because our
choice of $\tau$ is too low to find most unstable configurations.}}
\medskip

\begin{figure*}
\psfig{file=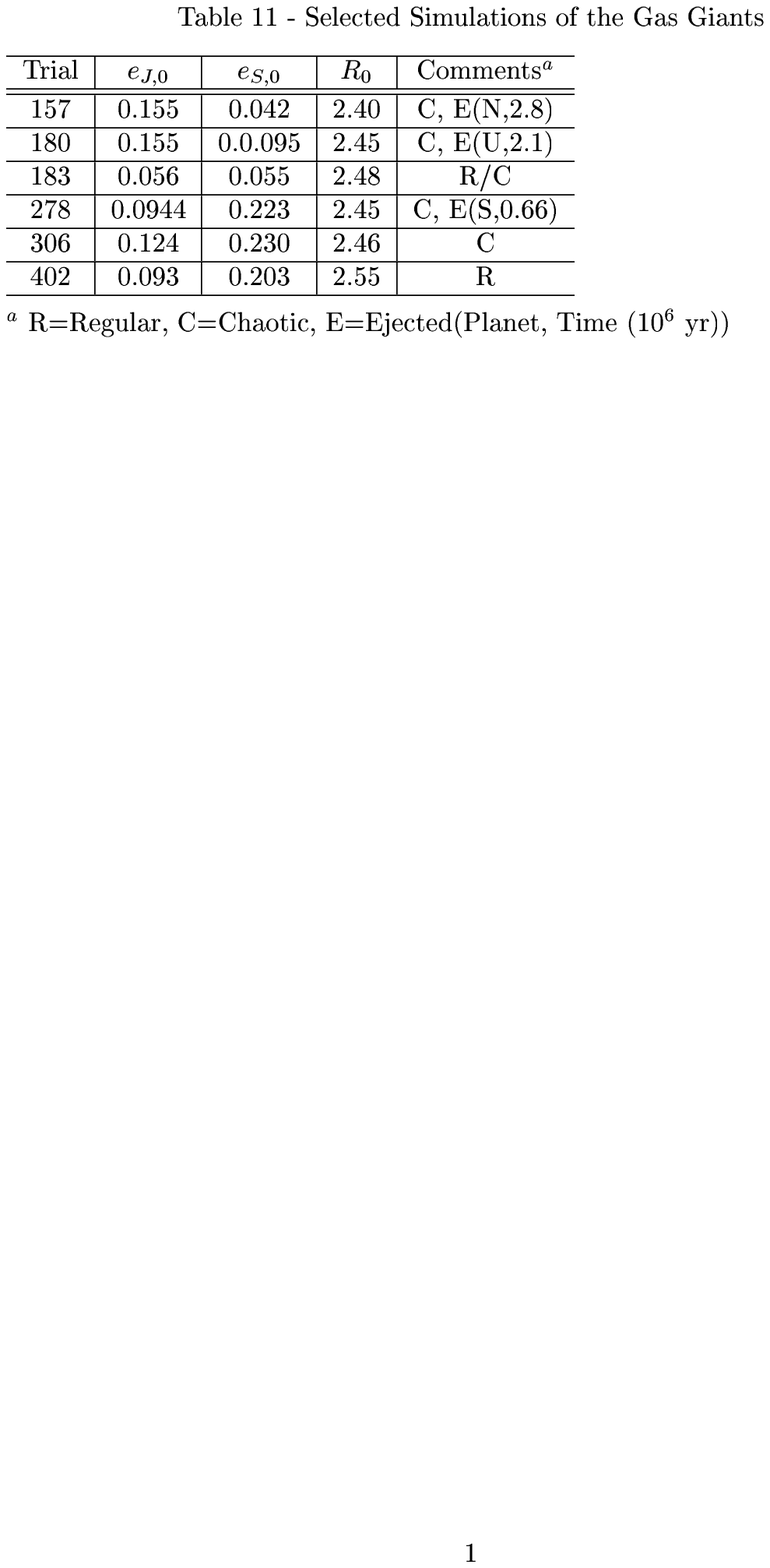,width=19.truecm}
\end{figure*}

\medskip
\epsfxsize=8truecm
\epsfbox{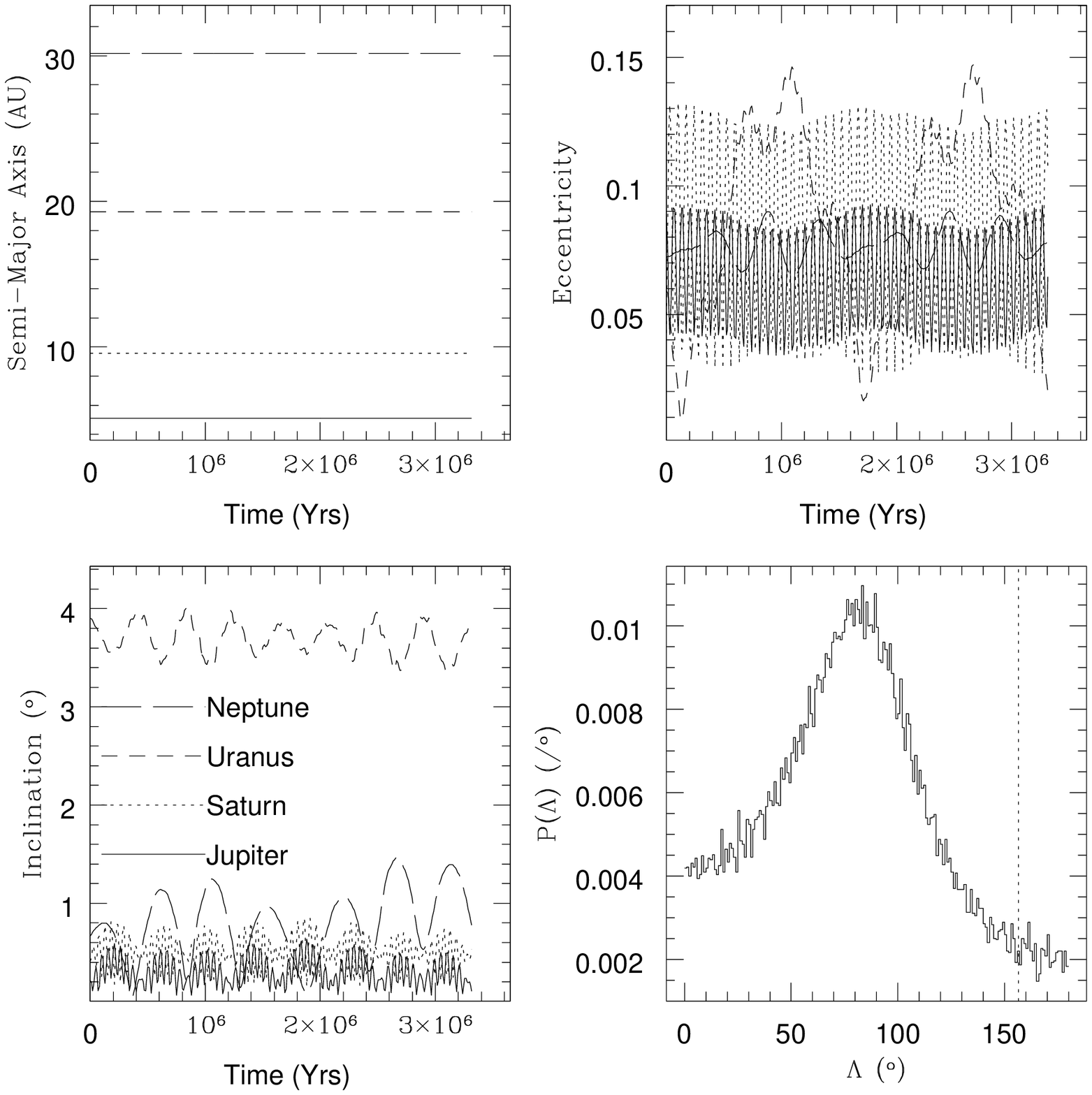}
\figcaption[ss_402.bw.ps]{\label{fig:asymptotic}
\small{The orbital evolution of SS-402, a stable, regular configuration of
the gas giants. The 4 planetary bodies in this system result in more
complicated dynamics as the smaller 2 planets often experience several
modes of oscillation. {\it Top Left}: The semi-major axes do not change for
the duration of the simulation. {\it Top Right}: The eccentricities
are combinations of sinusoids. Jupiter and Saturn both oscillate on a 70,000 year period. Uranus and Neptune vary on a 1.5 million year timescale. Their motion however also has 4 million year mode, due to to beating with the oscillations from Jupiter and Saturn. {\it Bottom Left}: All the inclinations evolve
as a function of two modes. Jupiter and Saturn oscillate on a 40,000
year cycle. Uranus and Neptune oscillate on a 500,000 year period. As
with eccentricity we observe beating between these 2 frequencies in
all planets. {\it Bottom
Right}: $\Lambda$ switches between libration with a $80^o$ amplitude,
and circulation.}}
\medskip

\medskip
\epsfxsize=8truecm
\epsfbox{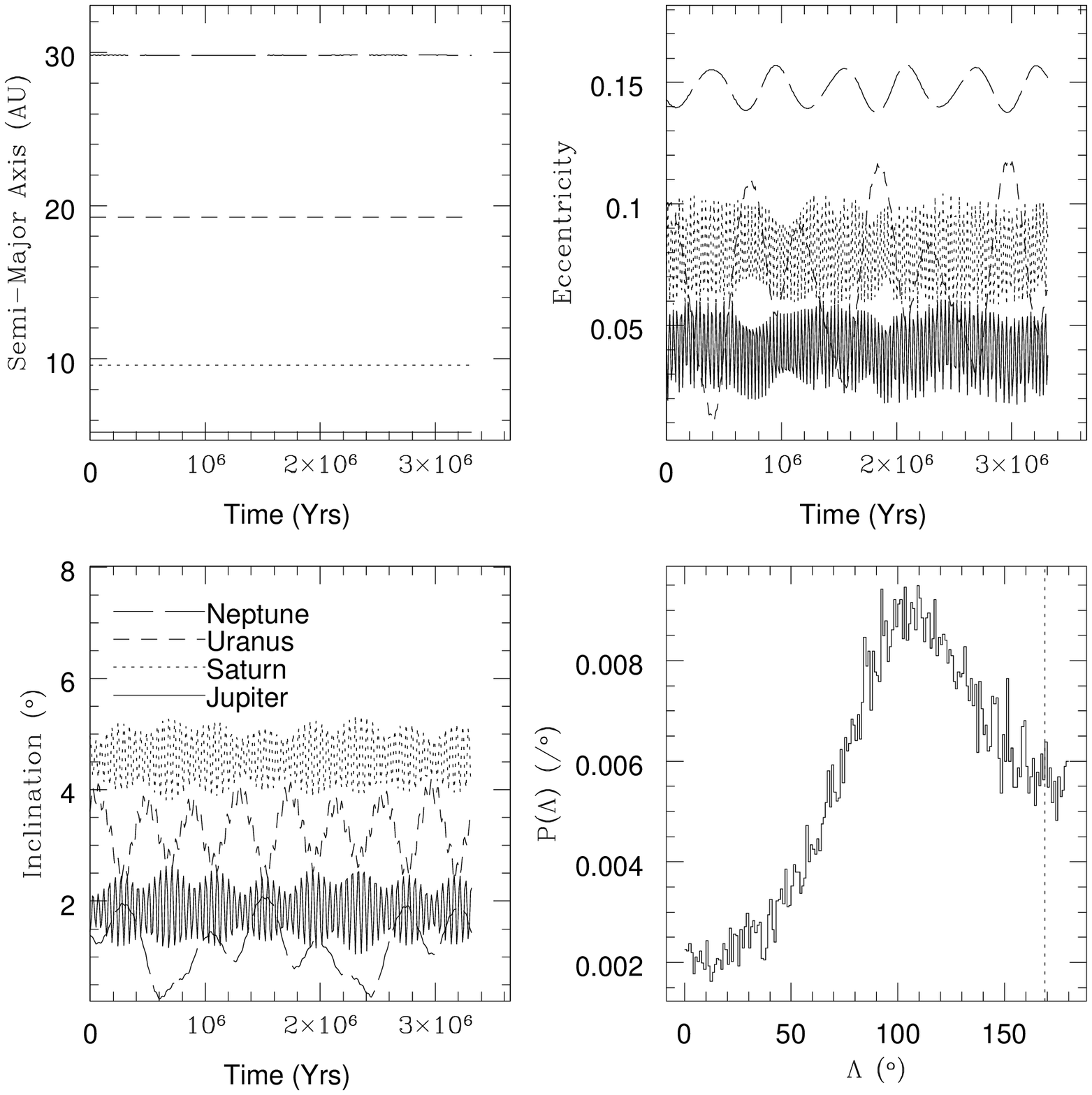}
\figcaption[ss_183.bw.ps]{\label{fig:asymptotic}
\small{The orbital evolution of SS-183, a stable configuration of the gas
giants in which some elements evolve regularly, others
chaotically. {\it Top Left}: The semi-major axes do not change for the
duration of the simulation. {\it Top Right}: The eccentricities of
Jupiter and Saturn evolve chaotically, but Neptune and Uranus appear
to be regular. {\it Bottom Left}: All inclinations evolve regularly,
although the number of modes is different, Jupiter, Saturn, and
Neptune have 2 modes, but Uranus has 3. {\it Bottom Right}: $\Lambda$
shows the typical distribution function of libration about
anti-alignment and circulation. This alternating results in the
chaotic evolution of $e_J$ and $e_S$ through $e-\varpi$ coupling.}}
\medskip

In \Fig 44 a fully chaotic, yet stable configuration is shown. The
semi-major axes show obvious signs of encounters, but never change by
more than 20\%. The eccentricities are very chaotic, but rarely reach
0.3. The inclinations, too, are extremely chaotic, but never surpass
$12^o$. The double peaked $\Lambda$ distribution function is another
clear indicator of chaos.

\medskip
\epsfxsize=8truecm
\epsfbox{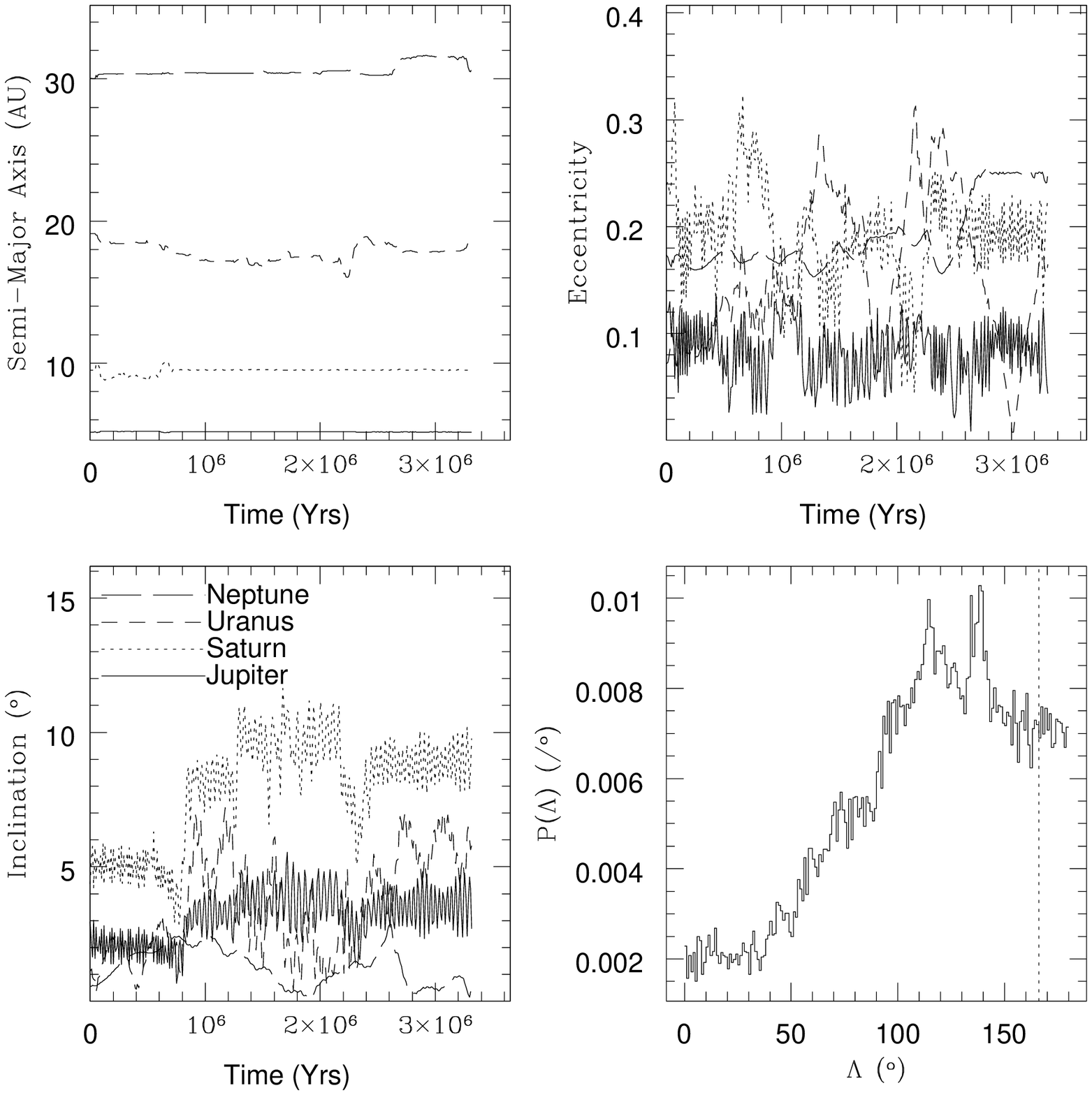}
\figcaption[ss_306.bw.ps]{\label{fig:asymptotic}
\small{The orbital evolution of SS-306, a stable, chaotic configuration of
the gas giants. {\it Top Left}: The semi-major axes begin migrating
immediately, but a change greater than 20\% does not occur for the
duration of the simulation. {\it Top Right}: All eccentricities
undergo chaotic oscillations, but the amplitudes are small. No
eccentricity ever reaches 0.35. {\it Bottom Left}: The inclinations
also evolve chaotically with low level oscillations. The system
remains close to coplanarity, as $i_S$ never exceeds $12^o$. {\it
Bottom Right}: This $\Lambda$ distribution is clearly chaotic as the 2
peaks and circulation demonstrate.}}
\medskip

In \Fig 45 we present an example of the ejection of Saturn. Although the
semi-major axes show obvious signs of perturbations, it is only during the
last 50,000 years that any planet's semi-major axis changes by more than
10\%. The eccentricities and inclinations do not show large fluctuations
until Saturn is quickly ejected after 65,000 years.

\medskip
\epsfxsize=8truecm
\epsfbox{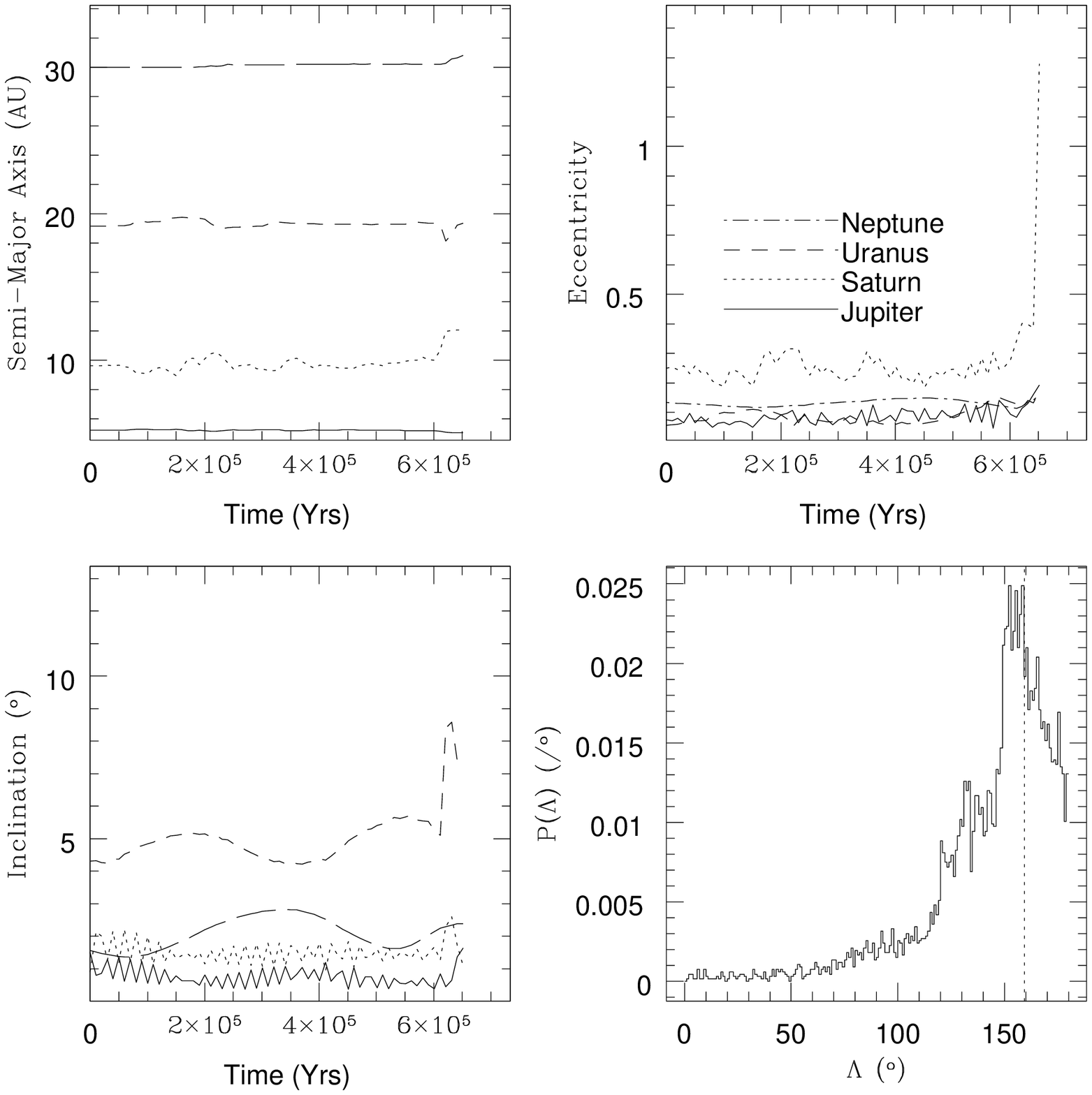}
\figcaption[ss_278.bw.ps]{\label{fig:asymptotic}
\small{The orbital evolution of SS-278, the ejection of Saturn. {\it Top
Left}: The semi-major axes deviate from the initial values by less
than 10\% until the final 50,000 years of integration. Even then $a_S$
only increases by 15\%. {\it Top Right}: All eccentricities are
chaotic, but the values remain close to their initial conditions until
Saturn is thrown from the system. {\it Bottom Left}: The inclinations
appear to evolve regularly for 35,000 years until Jupiter and Saturn
become chaotic. Uranus becomes chaotic after 60,000 years, but Neptune
remains regular for the duration of the simulation. {\it Bottom
Right}: The $\Lambda$ distribution for this configuration is quite
unusual and also indicates the system is chaotic.}}
\medskip

The ejection of Uranus is shown in \Fig 46. The semi-major axes of
Jupiter and Saturn remain nearly constant throughout the simulation, but
Neptune and Uranus are clearly interacting. The eccentricities are
chaotic, but remain near the starting values, except for Uranus, which
steadily increases until it is ejected after $2\times 10^6$ years. The
inclinations are also chaotic. Jupiter and Saturn appear to have 1 normal
mode, and 1 chaotic mode, while Neptune and Uranus are fully chaotic. The
longitudes of periastron tend to remain near alignment, but the 3 peaks
clearly bely the chaos in the system.

\medskip
\epsfxsize=8truecm
\epsfbox{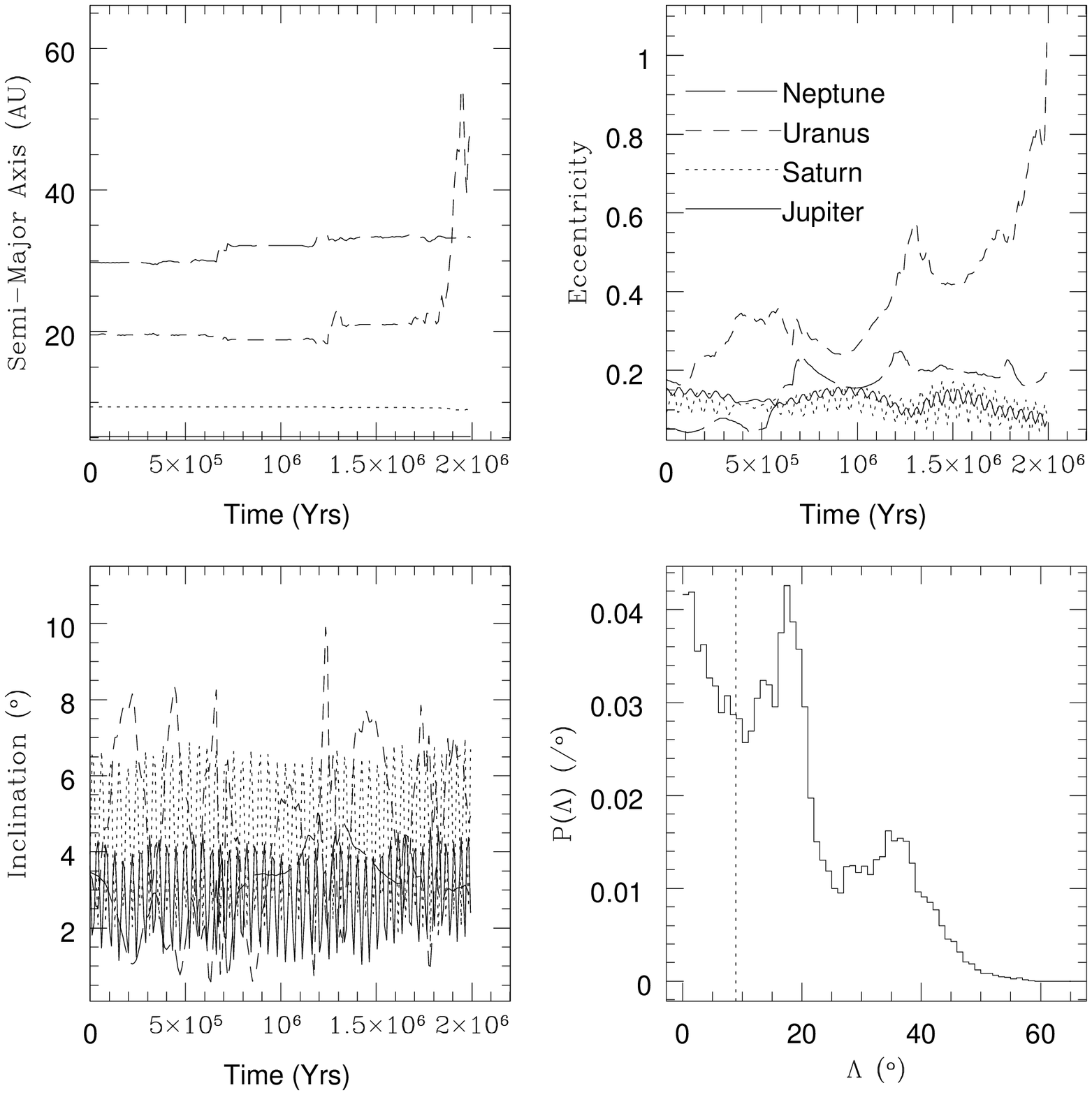}
\figcaption[ss_180.bw.ps]{\label{fig:asymptotic}
\small{The orbital evolution of SS-180, the ejection of Uranus. {\it Top
Left}: The semi-major axis of Jupiter does not change, and Saturn
changes by less than 0.1AU at the very end of the simulation. Uranus
and Neptune are clearly interacting, but the fluctuations of $e_U$ and
$E_N$ are small until $1.75\times 10^6$ years. {\it Top Right}: While
the eccentricities of Jupiter, Saturn, and Neptune remain low,
Uranus's eccentricity gradually grows until it is ejected after
$2\times 10^6$ years. {\it Bottom Left}: The inclinations all also
show chaos, but Jupiter and Saturn appear to have a regular 3000 year
mode superimposed on small chaotic fluctuations. {\it Bottom Right}:
The $\Lambda$ distribution for this configuration is quite unusual and
also indicates the system is chaotic. Note, however, that the system
is experiencing a generic form of libration, as $\Lambda$ never
exceeds $60^o$.}}
\medskip

Finally we plot the evolution of a configuration that ejects Neptune
in \Fig 47: SS-157. After 750,000 years Uranus becomes the most
distant planet. After Uranus and Neptune have a close approach which
flings Neptune from the system, Uranus drops to a 15AU orbit. It is
likely that it would then encounter Saturn, and another planet would
be removed from the system. Despite the large variations in semi-major
axes, the eccentricities remain relatively calm. Jupiter and Saturn in
particular appear nearly regular. In inclination, they in fact are
regular. For all the unstable cases shown here, this configuration's
$\Lambda$ shows the least chaos. Although there is some flipping
between a librational and a circulation mode. $\Lambda$ also
demonstrates that the chaos between Neptune and Uranus does not affect
the evolution of Jupiter and Saturn's orbits.

These last three figures reveal why the stability map for the SS is
different than for other interacting systems. Sometimes, even when $e_J$
and $e_S$ are low, the smaller bodies in the system can interact with each
other violently. Although the eccentricity of Jupiter and Saturn are
generally the most important factor in system stability. The larger the
values of any eccentricity, the more likely is instability.

We ran no long term simulations on the SS. The orbital elements for
the SS are well determined, and many long term simulations have
already been performed on this system (see Duncan \& Quinn 1993, Lecar
\etal 2001).

\medskip
\epsfxsize=8truecm
\epsfbox{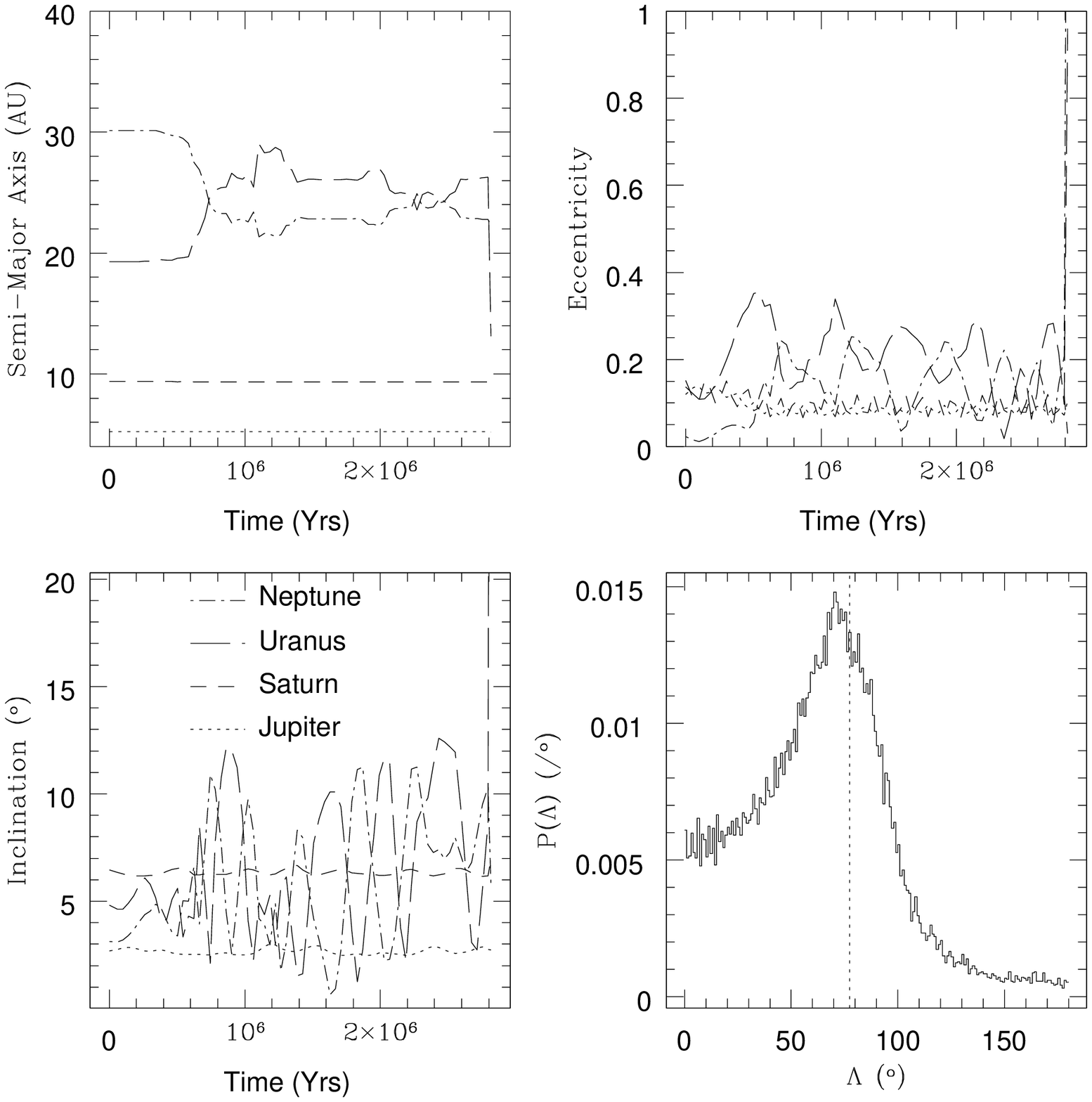}
\figcaption[ss_157.smooth.ps]{\label{fig:asymptotic}
\small{The orbital evolution of SS-157, the ejection of Neptune. These
data are smoothed over a 40,000 year interval. {\it Top
Left}: After 750,000 years Neptune and Uranus cross orbits, and Uranus
becomes the most distant planet. After ejecting Neptune Uranus drops
to a 15AU orbit which would most likely encounter Saturn relatively
quickly and result in another ejection. Jupiter and Saturn remain on
regular orbits. {\it Top Right}: All
eccentricities remain low for the duration of the simulation until
Neptune is removed after $2.8\times 10^6$ years. This is true even
while Uranus and Neptune are in 1:1 resonance at 750,000 years and 2.2
million years. Jupiter and Saturn experience sinusoidal oscillations
throughout the simulation. {\it Bottom Left}: The inclinations of Jupiter and
Saturn are regular but Uranus and Neptune are clearly chaotic. Note
that the encounter that throws Neptune from the system sends Uranus
into a highly inclined orbit. {\it
Bottom Right}: For this configuration $\Lambda$ is generally librating
in an aligned state with an amplitude of $70^o$, but with some
circulation.}}
\medskip

\subsection{Jupiter and Saturn}
As mentioned above Uranus and Neptune do not provide enough reflex
velocity, $K$, motion in the star to be observable by current
technology ($K\sim 3$m/s). Should any planet of Uranus or Neptune mass
exist in the observed ESPS they would not be detected. Therefore we
followed the same procedure with just Jupiter and Saturn. This suite
of simulations can also provide clues as to how other ESPS will behave
if they have additional, distant companions.

Not surprisingly this 3-body system is more stable than the 5-body
system, as $96.3\pm 2.4$\% of the trials remained stable. In this
system Jupiter was ejected 6.9\% of the time, and Saturn 93.1\%.  It
seems as though instability is being passed through Saturn and into
the smaller planets. We can therefore apply this result to the other
ESPS. This Jupiter-Saturn system is analogous to the 47UMa system. The
mass ratio of the planets are about the same, as is $R$. The only
substantial difference is that the masses are higher and the orbits
closer in 47UMa. However, our simulations of 47UMa used
$e_c$=0.1. This is about twice as high as $e_{Saturn}$. This decrease
in eccentricity is clearly important as the Jupiter-Saturn system is
substantially more stable than 47UMa. This again demonstrates that the
eccentricities of the planets are key in determining stability.

\medskip
\epsfxsize=8truecm
\epsfbox{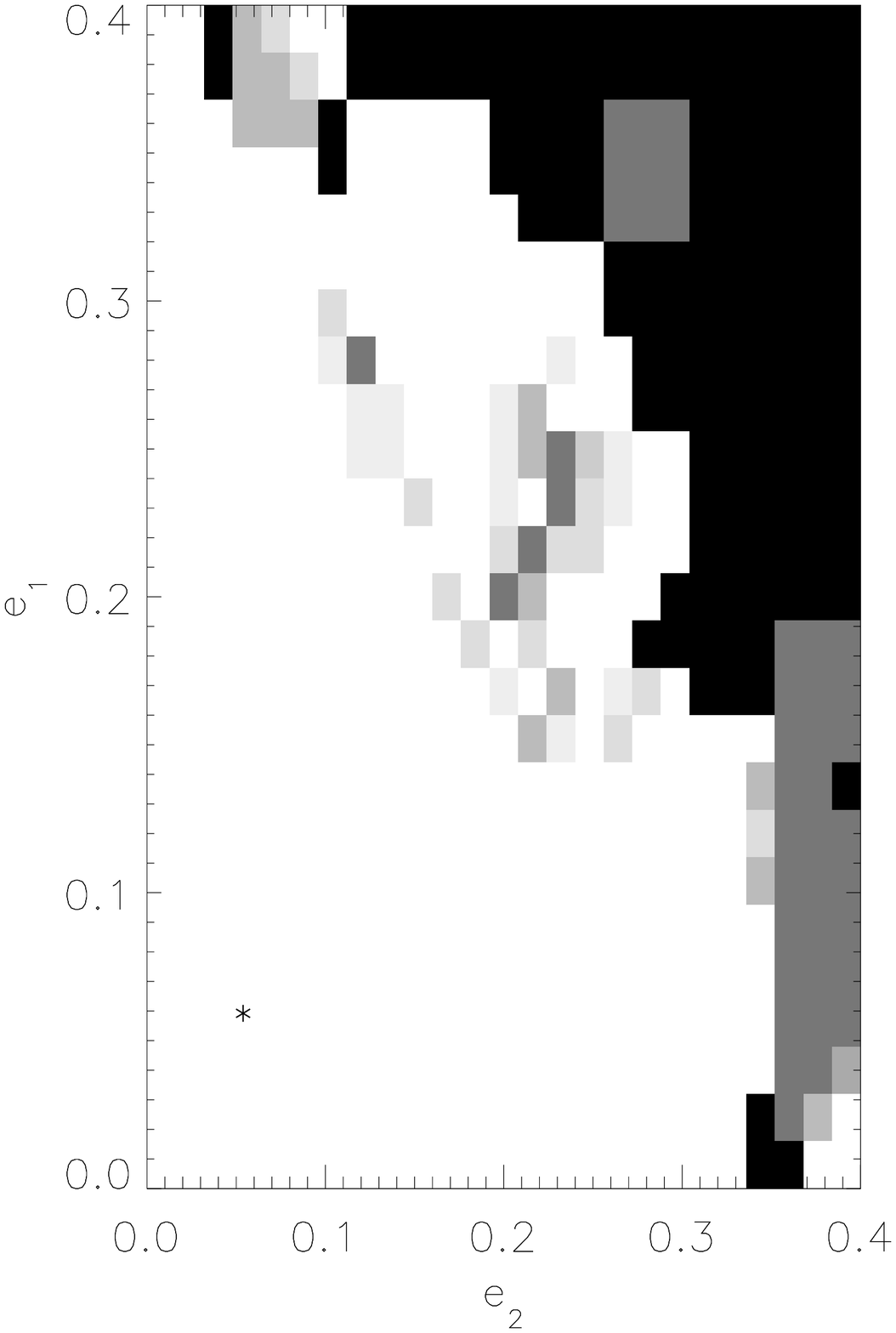}
\figcaption[jupsat.ecc.grey.ps]{\label{fig:asymptotic}
\small{Stability map for the Jupiter-Saturn system. This map shows some
similarities to that for the gas giants (see \Fig 44). The line
demarcating the plateau follows approximately the same diagonal
line. As well we still see a few depressions in the plateau.}}
\medskip

\Fig 48 is the stability map for the Jupiter-Saturn system. This plot is
similar to \Fig 41. There is a boundary at approximately the same
location in eccentricity space, however the drop is not so sharp, nor as
deep as in the gas giant system. Further an additional plateau rises at
larger eccentricities. This last phenomenon is not observed in any other system
in this paper.

As mentioned in $\S$4.3, stability might be correlated with the 5:2
resonance. \Fig 49 shows stability as a function of R for the
SS. There is a hint that as the system moves out of this third order
resonance, stability increases, but the errors on these distant
configurations are too large to confirm this possibility. In 47UMa
$R=2.38$, while in the SS, $R=2.48$. Although there are no
statistically meaningful points in \Figs 34 and 49, the same trend
appears in both. Namely that interior to 5:2 instability is more
prevalent. As has been shown throughout this paper, the eccentricities
determine the overall stability, and the statistics are too poor to
claim any trend with $R$ in either system.

Comparing the gas giants with Jupiter-Saturn provides us with an excellent
opportunity to explore completeness. As mentioned above, the Jupiter-Saturn
system is similar to 47UMa, yet the stability maps are quite
different. The Jupiter-Saturn system is the only system examined with no
instability abyss, and it is the only system we know to be incomplete. 

\medskip
\epsfxsize=8truecm
\epsfbox{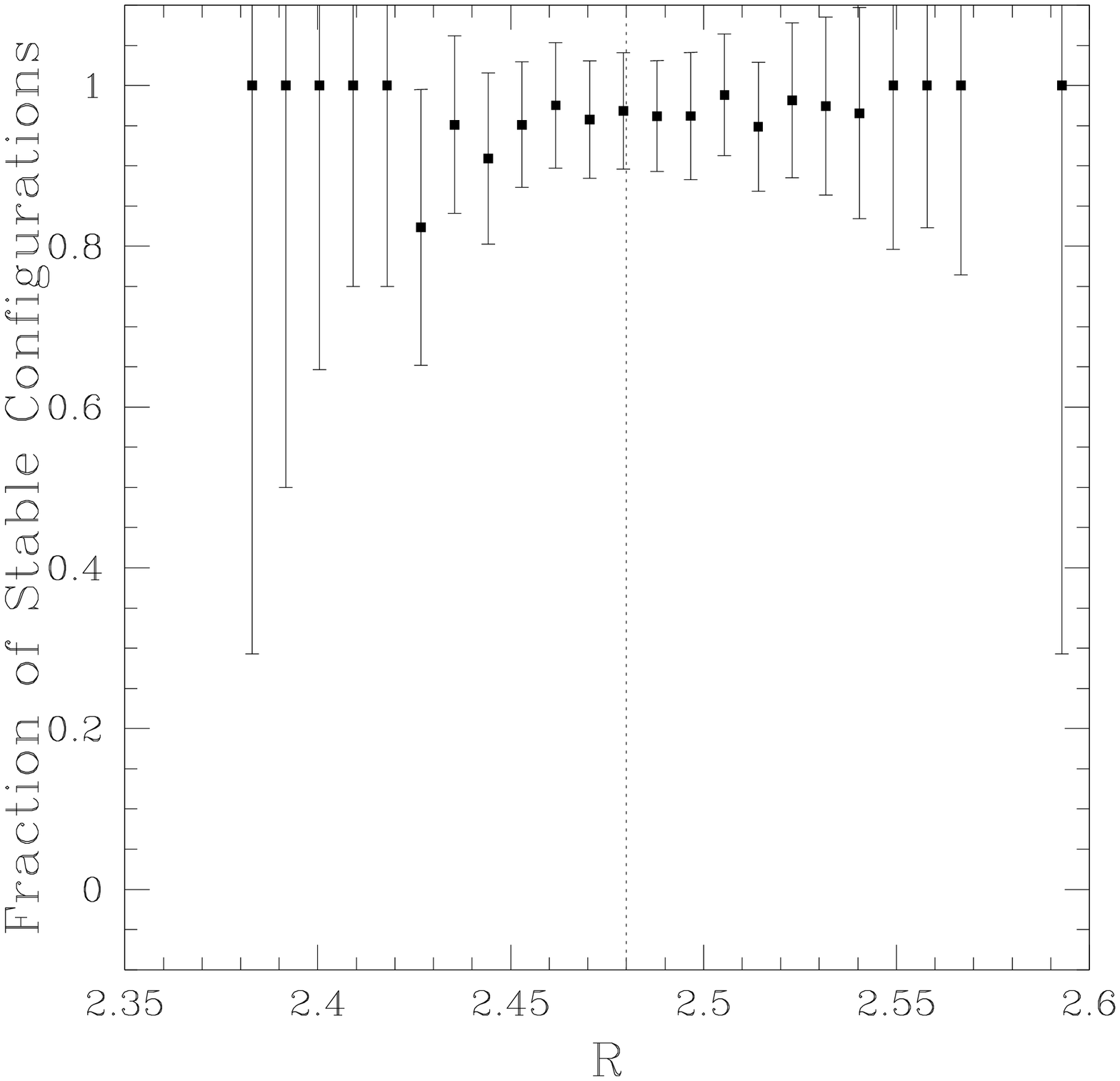}
\figcaption[jupsat.res.dyn.ps]{\label{fig:asymptotic}
\small{Stability as a function of $R$ for the Jupiter-Saturn system. As in
47UMa there is a hint that instability increases inside the 5:2
resonance; no data point lies more than 1$\sigma$ from the mean rate
of stability (96\%). It therefore apears that this resonance has
minimal impact on the system. The dashed vertical line represents the
true value of $R$ in the SS.}}
\medskip

\section{Separated Systems} 
By the end of 2002 three separated systems had been announced:
HD83443\footnote{http://obswww.unige.ch/$\sim$udry/planet/hd83443\_syst.html},
HD168443 (Marcy \etal 2001b), and
HD74156\footnote{http://obswww.unige.ch/$\sim$udry/planet/hd74156.html}.
The values and errors for these systems are reproduced in Table
12. HD83443 consists of 2 Saturn mass planets in very tight
orbits. HD168443 consists of 2 very large companions ($m_1\ge
17M_{Jup}$, $m_2\ge 7.5M_{Jup}$). In fact planet c should be
considered a brown dwarf, and if the system is more inclined than 35
degrees planet b would also be a brown dwarf. For this system
$R=30.5$. HD74156 contains 2 bodies of slightly more than a Jupiter
mass, with $R=44.6$. We only examined HD168443 and HD83443. Evidence
is mounting that HD83443 is not a multiple system (Butler \etal 2002),
therefore we stopped the simulations on this system after 847 trials
had been completed.  For HD168443 and HD83443, all simulations
survived to $\tau$. HD74156 has a larger $R$, and smaller masses than
HD168443. It therefore seems highly doubtful that any simulation of
HD74156 would produce an unstable configuration.

\begin{figure*}
\psfig{file=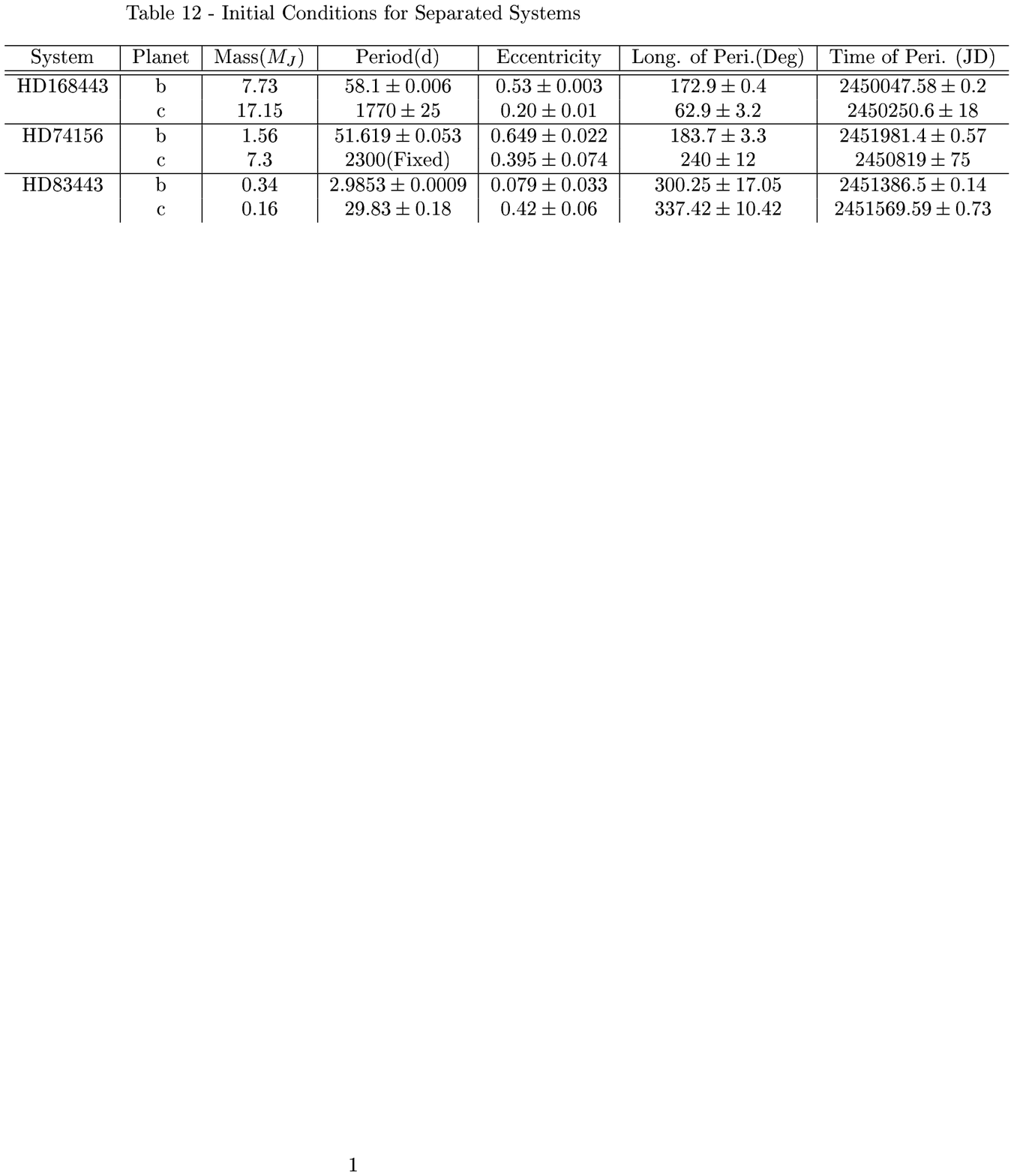,width=19.truecm}
\end{figure*}

Although all simulations were stable, the dynamics of HD168443 are
still interesting. The eccentricities and inclinations of this system
show a weak planet-planet interaction. Although no evidence of chaos
is evident, the planets apparently are close enough that they feel
each other. This system may be fully stable, but it appears to lie
close to the boundary between interacting and separated systems. We
hypothesize that this proximity to interacting systems is due to the
large planetary masses in this system.

\section{Summary} 
In this paper we have described the dynamics of three different
morphological classifications of planetary systems. We find that the
systems in each of the classifications have similar stable regions. In resonant
systems very small stability zones exist in phase space, and stability
is tightly coupled with $R$ and, to a lesser degree, $e_1$, where
$e_1$ is the eccentricity of the most massive companion. In
interacting systems, the zones are larger, but are correlated with
$e_1$ and $e_2$, where $e_2$ is the eccentricity of the second most
massive companion. In these systems we see a correlation with
eccentricity and the location of the most massive planet. Large
interior planets are almost impossible to eject (47UMa, the gas
giants), whereas large exterior planets can be ejected sometimes
($\upsilon$ And). Separated systems are completely stable as observed.

Table 13 summarizes the results of this paper.  In this table
$f_{stable}$ is the fraction of configurations that were stable, and
$f_j, j=1,2,3,...$ is the fraction of unstable systems which
perturbed/ejected that planet. Note that the subscripts correspond to
mass, not semi-major axis (1 being the most massive companion). We
therefore strengthen the theory suggested in Paper I: all interacting
planetary systems lie near the edge of stability.

\begin{figure*}
\psfig{file=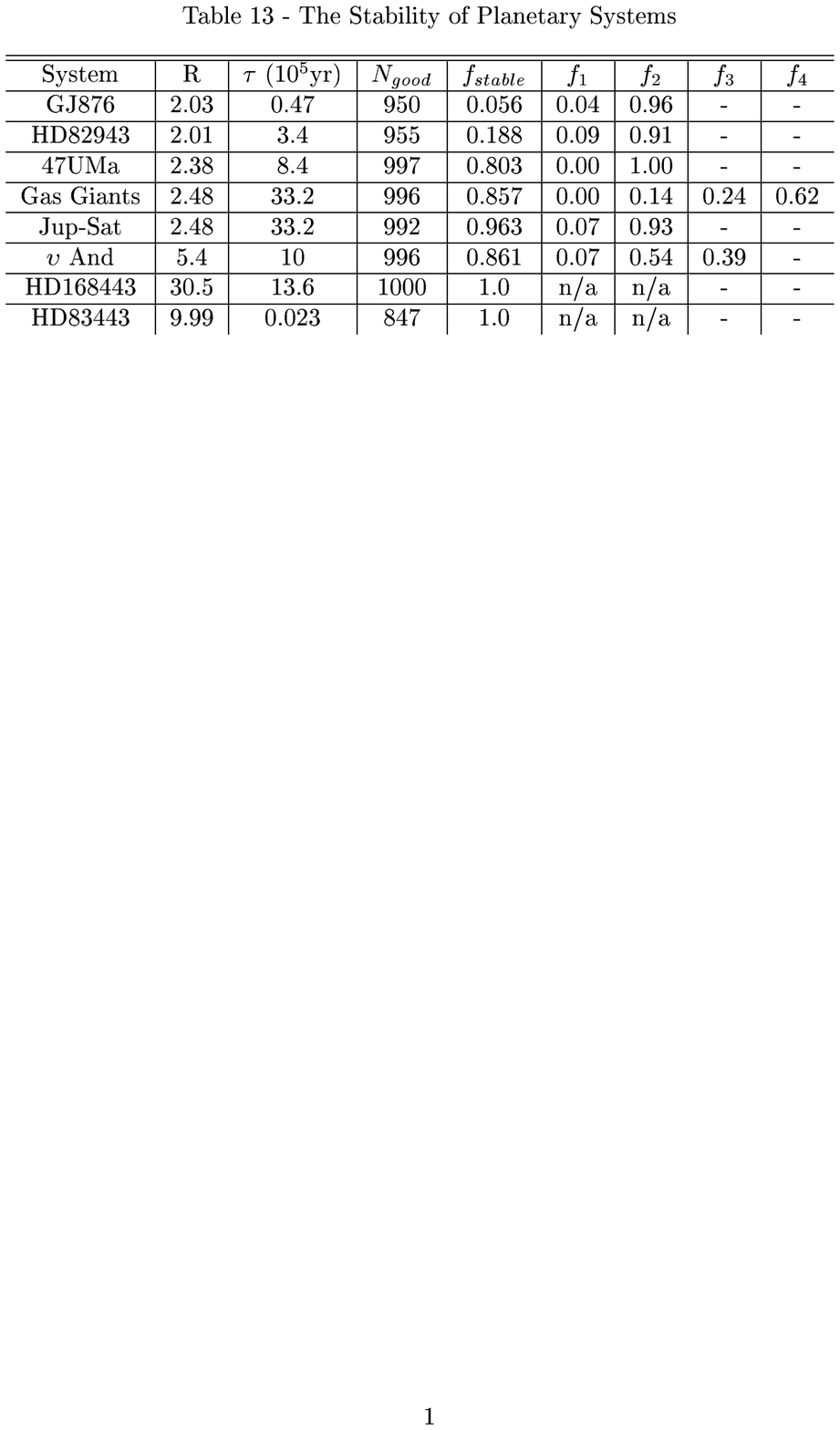,width=19.truecm}
\end{figure*}

There are some obvious similarities among the systems. One
particularly intriguing result is the similarity in $f_{stable}$
between systems in the same classification. Resonant systems have
survival probabilities less than 20\%, whereas interacting systems lie
close to 80\% and the separated systems are completely stable. Of the
interacting systems, 47UMa stands out as being far from the stability edge.

The choice of $\tau = 2.8\times 10^5 P_1$ appears to identify most
unstable configurations. For both resonant systems $\tau < 10^6$ yrs,
but since the simulations were integrated to $10^6$ yrs we may
estimate the usefulness of this arbitrary value. For both GJ876 and
HD82943 approximately 2-3\% of configurations ejected a planet after
$\tau$. For the $\upsilon$ And system, 1.5\% of unstable systems were
ejected in the last bin of \Fig 20. For 47UMa the rate was over
10\%, and for the SS the rate was 4\%. However as was noted in
$\S$4.3, this low rate for the SS may be the result of poor sampling
in the last bin. Although all these systems reached a maximum
ejection rate before $\tau$, the nonzero rate at $\tau$ demonstrates that our
choice for $\tau$ was slightly too short. In GJ876 some ejections
occurred right up to $10^6$ yrs, but by $10^4$ yrs
(0.25$\tau_{GJ876}$) over 90\% of unstable cases had been
identified. The situation is nearly the same for HD82943; 90\% of
unstable cases were identified by 30,000 years (0.1$\tau_{HD82943}$),
but ejections continued for $10^6$ years. Given these statistics, a
better choice for $\tau$ would be $\tau = 10^6 P_{outer}$. However, it is
important to note that instability can arise after this, as is shown
in \Figs 9, 18, and 30. The simulations
presented here clearly demonstrate the unpredictable behavior of
chaotic systems; no choice of $\tau$ would identify all unstable
configurations. One should therefore note that all the global results
presented here are upper limits. The probability of survival and
extent of stable phase space are smaller than what is shown here.

Long term simulations ($\ge 10^8$ orbits) show that all systems have
regular configurations on this timescale. However only one simulated
configuration of GJ876 showed this behavior. Some configurations may
show a large degree of chaos for up to $10^9$ years (see \Fig 38, top
left), eject a planet after an arbitrarily long period of time (\Fig
30, bottom right), in addition to quiescent, regular evolution.
Regular orbits tend to librate about $\Lambda$=0, but this is not
necessary for stability (see \Figs 10, 19, 31, and 39).

\section{Discussion}
Ideally this research provides insights into planet formation. In
particular, the current distribution of orbits may give clues to the
formation scenario. Two features of ESPS are particularly interesting:
the apsidal alignments and the large eccentricities. As $e$ and
$\varpi$ are coupled, these two phenomena are likely the result of the
same mechanism. There are two generic ways to pump up eccentricities:
adiabatically, or impulsively, with respect to the secular timescale
($\gtrsim 10^5$yr). Our variation of orbital elements provides a
unique view into the effects of these mechanisms on the dynamics and
stability of actual planetary systems. Several groups have examined
this problem, and in this section we interpret our results in the
context of theirs.

Of adiabatic scenarios a remnant planetary disk is the most likely
candidate (Chiang \& Murray 2002; Goldreich \& Sari 2003). For at
least the $\upsilon$ And system, a remnant disk external to planet d
can provide a mechanism to pump up $e_c$ and $e_d$ to their current
observed values (Chiang \& Murray 2002). This method also predicts
libration of $\Lambda$ about 0, which is observed in this
work. Conversely an impulsive force may also drive eccentricities to
values significantly higher than zero (Malhotra 2002).  The impulsive
scenario also perturbs $\Lambda$. In adiabatic schemes the libration
amplitude is small, whereas in impulsive cases it can be quite large
($>45^o$). Throughout this paper we have shown configurations with
libration amplitudes larger than $45^o$, i.e. Figs.\ 5 and 39, and
smaller than $45^o$, i.e. Figs.\ 10, 31.  This work therefore finds
examples of systems which may result from either mechanism.

This impulsive scenario is difficult to reconcile with resonant
systems. These systems most likely form as a result of resonance
capture during the orbital migration epoch of planet formation
(Snellgrove, Papaloizou, \& Nelson 2001), which assumes adiabatic
migration. This phenomenon seems qualitatively similar to the external
disk model of Chiang \& Murray. Their model, based on torques produced
by Lindblad and corotation resonances, is very similar to planetary
migration. However current orbital migration theory predicts that a
Jupiter mass planet at 5AU in a plausible minimum mass solar nebula
should migrate on a timescale of order 2500 years (Tanaka, Takeuchi,
\& Ward 2002) to 5000 years (Lufkin \etal 2003), which is a factor of
5 to 10 times shorter than the typical secular timescale for planetary
systems. This suggests the planetary disk model might actually be
impulsive, but only marginally so. However Lufkin \etal also point out
that the migration might be very impulsive in heavier
disks. Understanding the rates of migration will be a major step
toward resolving this issue of high eccentricities and apsidal
alignment. All we can say now is that we are too limited in the number
of resonant and interacting systems to determine if their
eccentricities result from similar processes.

The results presented here, coupled with Malhotra (2002) support the
theory that the eccentricities of planets in interacting systems
result from planet-planet scatterings. This possibility has been
investigated substantially (Rasio \& Ford 1996; Ford, Havlickova, \&
Rasio 2001; Weidenschilling \& Marzari 2002; Malhotra 2002). The
proximity of these systems to the edge of stability might imply that
planet formation is an efficient process.  Perhaps too efficient. As
planets form, they are constantly perturbing each other with ever
greater force. It is well known that ejections are common during
planet formation. In fact some research predicts that the ejection of
a fifth terrestrial planet may be needed to explain the period of
heavy bombardment in the SS (Chambers, Lissauer, \& Morbidelli
2001). So it is not too surprising that we find systems near
instability because they form in an unstable state and eject massive
bodies until they arrive in resonance, or reach the stability
plateau. Some work has shown that if planet formation is very
efficient (i.e. initially 10 Jupiter mass planets) then the subsequent
scattering and ejections can produce distributions of $a$ and $e$ that
are similar to those observed (Adams \& Laughlin, 2003). Clearly this
scenario is appealing, and will be verified in the next several years
as simulations become more sophisticated and more multiple planet
systems are detected.

Beyond the origin of large eccentricities and apsidal alignment, we
find some inconsistencies in the theory of the origin of the very
short period ($P\lesssim 10$days) planets. As mentioned in $\S$2 and
$\S$4.1, the effects of tidal circularization were not included in
this model. For $\upsilon$ And the timescale for circularization is 80
million years (Trilling 2000), but we see that the eccentricity of
$\upsilon$ And b can oscillate on $10^5$ year timescales with an
amplitude of 0.3. At this point is is unclear if the tidal damping
will always overwhelm the perturbations of other companions. It could
be that we have detected $\upsilon$ And b at a point in time in which
its orbit is nearly circular. Perhaps we will discover planetary
systems in which a close planet is being perturbed by external
companions and the tidal circularization cannot compensate. However it
seems more likely that the circularization is a stronger effect as we
have yet to detect any planet inside the circularization radius on an
eccentric orbit. Future numerical work should resolve this issue.

We observe that, in general, there is good agreement between our results
and those performed with MEGNO. The sizes of stable regions appear to be
overestimated in those papers, but that is most likely due their choice
for $\tau$, usually $10^4$ years. The shortcomings of MEGNO are most clearly demonstrated in the
long term integrations of 47UMa. The top left of \Fig 38 shows a
stable, but chaotic system which persists for $10^9$ years. The uniqueness
of systems such as this is unknown, but understanding the dynamics of
chaotic, yet long-lasting, systems could yield new insights into planetary
dynamics. Our own SS is another example of a system that displays weak
chaos, yet can survive for very long periods of time (Laskar 1994).

This research is the first to examine the origin of high
eccentricities and apsidal alignment with known systems. Most other
work in this field is purely hypothetical. That type of research has
the benefit of being unconstrained by statistics; they may integrate
as many systems as they wish, with arbitrary initial conditions. This
work, conversely, is the first attempt to coherently and consistently
compare known systems in order to understand their dynamics and
origins. At this point, with so few known systems, the two methods are
complimentary, but as we discover more ESPS, the method described in
this paper will become more valuable as it uses true ESPS as a
starting point.

\section{Conclusions}
We have shown that this type of experiment can indeed constrain the
observed orbital elements of planetary systems. Further we see that in
almost all interacting and resonant systems the current best fits to
the system place them near the boundary between stability and
instability. The fact that no system is completely unstable implies
that the observations of these systems are reliable, and that the
errors in the system are probably conservative. That is, all
systematics have been removed, and statistical fluctuations are being
overestimated.

Note that our estimates of the instability of these systems is in some
sense an underestimate because of the possible presence of yet
undetected lower mass companions. For example, it may turn out that
47UMa may have an undetected planet that would put it closer to the
edge. On the other hand, the very existence of these systems shows
that they are not unstable. As unsettling as it may be, it seems a
large fraction of planetary systems, including our own, lie
dangerously close to instability. As more and different types of
systems are detected we will discover if all planetary systems are on
the edge.

This method has shown that, dynamically, the SS is not a unique
system. In fact, it lies in the middle stability category. Some
systems lie nearer instability, others further. As the radial velocity
searches continue, and astrometric searches begin, a SS analogue
(circular orbits, large semi-major axes) will undoubtedly be
discovered and we will finally be able to determine how the SS fits in
with other planetary systems.  But this experiment has shown that,
with regards to its (close) proximity to unstable regions, the SS is a
typical planetary system.

Recently more systems were announced; HD38529 (Fischer \etal 2003), a
separated system, HD12661 (Fischer \etal 2003), an interacting system,
and 55Cnc a 3 planet system with interior planets in 3:1 resonance and
a distant companion (Marcy \etal 2002). The planets in HD38529 have
masses less than HD168443, and comparable values for $R$, therefore it
seems likely that they are fully stable. 55Cnc, however, might
demonstrate different dynamics and edges as it is in a different mean
motion resonance. Future planetary systems will most likely fall into
the categories outlined in this paper. The results presented here
suggest that $f_{stable}$ for 55Cnc would lie between resonant and
coupled systems. HD12661 is very similar to $\upsilon$ And, so we
expect this system to show similar edges, probabilities, and dynamics.

Future work will address many of the issues brought up in $\S$7. If
planet formation is an efficient phenomenon, then we might suspect
that additional companions lie in separated systems. Further we should
be able to determine that the eccentricity distribution of ESPS result
from a late scattering event. We also need to determine the origin of
the edges presented here. A mathematical relationship probably exists
which would make the classification of planetary systems trivial. The
categories as defined here may be the result of small number
statistics. Two systems, HD169830 and HD37124, in which $R\approx 10$
have been announced (Mayor \etal 2003, Butler \etal 2003). These
system may reveal the boundary between interacting and separated
systems. An analysis of these two systems and 55Cnc will help sharpen
our classification of planetary systems.

Future work, both observational and theoretical, must address these
issues.  These systems as they are observed now reflect their
histories, and hence provide us with the best path to unlocking the
secrets of planet formation. As more and more observations of these
planetary systems, additional planetary systems, and (hopefully)
protoplanets are
made, numerical studies such as this, and those cited here, should
provide a deeper understanding of planet formation and dynamics.

\section{Acknowledgments}
The authors wish to thank Chance Reschke for his help in completing
these simulations. This manuscript was greatly clarified after an edit
by Derek C. Richardson. We would also like to thank Renu Malhotra, and
Nader Haghighipour for useful discussions and suggestions. This work
was funded by grants from NASA, NAI, and the NSF, and was simulated on
computers donated by the University of Washington Student Technology
Fund.

%\end{multicols}

\end{document}